%% file: main.tex
\pdfoutput=1
\documentclass[11pt]{lmcs}
\usepackage{amsmath,amsthm,amssymb,stmaryrd,xspace,pdfsync,mathrsfs,verbatim,bbm}
\usepackage{xspace}
\usepackage{dutchcal}
\usepackage[mathscr]{euscript}
\usepackage{enumitem}

\usepackage{hyperref}
\usepackage{microtype}
\usepackage{tikz}
\usepackage[english]{babel}
\usetikzlibrary{decorations.text,arrows,decorations.pathreplacing,shapes,fit,shadows,calc,patterns,intersections}
\definecolor{my1}{cmyk}{0,.6,0,0}
\definecolor{my2}{cmyk}{.3,.0,.0,.0}

\usepackage[scaled=1.1]{dsserif}

\theoremstyle{plain}

\newtheorem{theorem}{Theorem}[section]
\newtheorem{corollary}[theorem]{Corollary}
\newtheorem{proposition}[theorem]{Proposition}
\newtheorem{example}[theorem]{Example}
\newtheorem{lemma}[theorem]{Lemma}
\newtheorem{fct}[theorem]{Fact}
\newtheorem{remark}[theorem]{Remark}

\input{macros.tex}

\begin{document}

%%
%% The "title" command has an optional parameter,
%% allowing the author to define a "short title" to be used in page headers.
\title[The amazing mixed polynomial closure]{The amazing mixed polynomial closure and its applications to two-variable first-order logic}

%%
%% The "author" command and its associated commands are used to define
%% the authors and their affiliations.
%% Of note is the shared affiliation of the first two authors, and the
%% "authornote" and "authornotemark" commands
%% used to denote shared contribution to the research.
\author{Thomas~Place}
\address{LaBRI, Universit\'e de Bordeaux}
\email{tplace@labri.fr}

%%
%% By default, the full list of authors will be used in the page
%% headers. Often, this list is too long, and will overlap
%% other information printed in the page headers. This command allows
%% the author to define a more concise list
%% of authors' names for this purpose.
\renewcommand{\shortauthors}{Thomas Place}

%%
%% The abstract is a short summary of the work to be presented in the
%% article.
\begin{abstract}
	Polynomial closure is a standard operator. It takes as input a class of regular languages and builds a new one. In this paper, we investigate three restrictions: left (\ldeto), right (\rdeto) and mixed polynomial closure (\mdeto). The first two were known while \mdeto is new. We look at three decision problems that one may associate to each class \Cs: membership (decide if an input regular language belongs to \Cs), separation (decide if two input regular languages can be separated by a third one in \Cs) and covering (which generalizes separation to arbitrarily many inputs). We prove that \ldeto, \rdeto and \mdeto preserve the decidability of membership under mild hypotheses on the input class, and the decidability of covering under much stronger hypotheses. 
	
	We apply our results to natural hierarchies that are built from a single input class by applying \ldeto, \rdeto and \mdeto recursively. We prove that these hierarchies can actually be defined using almost exclusively \mdeto. We also consider quantifier alternation hierarchies for \emph{two-variable} first-order logic (\fod) and prove that one can climb them using \mdeto. This result is generic in the sense that it holds for most standard choices of signatures. We use it to prove that for most of these choices, membership is decidable for all levels in the hierarchy. Finally, we prove that separation and covering are decidable for the hierarchy of two-variable first-order logic equipped with only the linear order (\fodw).
\end{abstract}

%%
%% The code below is generated by the tool at http://dl.acm.org/ccs.cfm.
%% Please copy and paste the code instead of the example below.
%%

%%
%% Keywords. The author(s) should pick words that accurately describe
%% the work being presented. Separate the keywords with commas.

%% A "teaser" image appears between the author and affiliation
%% information and the body of the document, and typically spans the
%% page.
%\begin{teaserfigure}
%  \includegraphics[width=\textwidth]{sampleteaser}
%  \caption{Seattle Mariners at Spring Training, 2010.}
%  \Description{Enjoying the baseball game from the third-base
%  seats. Ichiro Suzuki preparing to bat.}
%  \label{fig:teaser}
%\end{teaserfigure}

%%
%% This command processes the author and affiliation and title
%% information and builds the first part of the formatted document.
\maketitle

\section{Introduction}
\label{sec:intro}
\input{intro.tex}

\section{Preliminaries}
\label{sec:prelims}
\input{prelims.tex}

\section{Operators}
\label{sec:polc}
\input{polc.tex}

\section{Framework}
\label{sec:frame}
\input{framework.tex}

\section{Algebraic characterizations}
\label{sec:carac}
\input{carac.tex}

\section{Deterministic hierarchies}
\label{sec:deth}
\input{deth.tex}

\section{Two-variable first-order logic}
\label{sec:logic}
\input{logic.tex}

\section{Covering framework: \ratms}
\label{sec:ratms}
\input{ratms.tex}

\section{Covering for left and right polynomial closure}
\label{sec:cov}
\input{cov.tex}

\section{Covering for mixed polynomial closure}
\label{sec:covm}
\input{covm.tex}

\section{Conclusion}
\label{sec:conc}
\input{conc.tex}

%%
%% The next two lines define the bibliography style to be used, and
%% the bibliography file.
\bibliographystyle{alpha}
\bibliography{main}

\end{document}

%% file: macros.tex
\newcommand{\veps}{\ensuremath{\varepsilon}\xspace}
\newcommand{\nat}{\ensuremath{\mathbb{N}}\xspace}
\newcommand{\efgame}{Ehrenfeucht-Fraïssé\xspace}

\newcommand{\Bs}{\ensuremath{\mathscr{B}}\xspace}
\newcommand{\Cs}{\ensuremath{\mathscr{C}}\xspace}
\newcommand{\Ds}{\ensuremath{\mathscr{D}}\xspace}

\newcommand{\Fs}{\ensuremath{\mathscr{F}}\xspace}
\newcommand{\Gs}{\ensuremath{\mathscr{G}}\xspace}

\newcommand{\Is}{\ensuremath{\mathscr{I}}\xspace}

\newcommand{\Ps}{\ensuremath{\mathscr{P}}\xspace}

\newcommand{\Hb}{\ensuremath{\mathbf{H}}\xspace}

\newcommand{\Kb}{\ensuremath{\mathbf{K}}\xspace}
\newcommand{\Lb}{\ensuremath{\mathbf{L}}\xspace}

\newcommand{\Ub}{\ensuremath{\mathbf{U}}\xspace}
\newcommand{\Vb}{\ensuremath{\mathbf{V}}\xspace}

\newcommand{\frI}{\ensuremath{\mathbb{I}}\xspace}

\newcommand{\frP}{\ensuremath{\mathbb{P}}\xspace}

\newcommand{\frS}{\ensuremath{\mathbb{S}}\xspace}

\newcommand{\wsuit}{well-suited\xspace}

\newcommand{\vari}{prevariety\xspace}

\newcommand{\varis}{prevarieties\xspace}

\def\inv{^{-1}}

\newcommand{\Jrel}{\ensuremath{\mathrel{\mathscr{J}}}\xspace}

\newcommand{\Rrel}{\ensuremath{\mathrel{\mathscr{R}}}\xspace}
\newcommand{\Lrel}{\ensuremath{\mathrel{\mathscr{L}}}\xspace}

\newcommand{\Jord}{\ensuremath{\leqslant_{\mathscr{J}}}\xspace}

\newcommand{\Rord}{\ensuremath{\leqslant_{\mathscr{R}}}\xspace}
\newcommand{\Lord}{\ensuremath{\leqslant_{\mathscr{L}}}\xspace}

\newcommand{\Jords}{\ensuremath{<_{\mathscr{J}}}\xspace}

\newcommand{\Rords}{\ensuremath{<_{\mathscr{R}}}\xspace}
\newcommand{\Lords}{\ensuremath{<_{\mathscr{L}}}\xspace}

\newcommand{\su}{\ensuremath{\textup{SU}}\xspace}
\newcommand{\at}{\ensuremath{\textup{AT}}\xspace}
\newcommand{\pt}{\ensuremath{\textup{PT}}\xspace}

\newcommand{\abg}{\ensuremath{\textup{AMT}}\xspace}

\newcommand{\md}{\ensuremath{\textup{MOD}}\xspace}

\newcommand{\stzer}{\ensuremath{\textup{ST}}\xspace}
\newcommand{\canec}{\ensuremath{\sim_\Cs}\xspace}
\newcommand{\caned}{\ensuremath{\sim_\Ds}\xspace}

\newcommand{\canedo}{\ensuremath{\sim_{\Ds_1}}\xspace}
\newcommand{\canedt}{\ensuremath{\sim_{\Ds_2}}\xspace}

\newcommand{\etam}[1]{\ensuremath{\eta_{#1}}\xspace}
\newcommand{\etac}{\etam{\Cs}}

\newcommand{\canm}[1]{\ensuremath{N_{#1}}\xspace}
\newcommand{\canc}{\canm{\Cs}}

\newcommand{\caneld}{\ensuremath{\sim_{\ldet{\Ds}}}\xspace}
\newcommand{\canerd}{\ensuremath{\sim_{\rdet{\Ds}}}\xspace}

\newcommand{\bool}[1]{\ensuremath{Bool(#1)}\xspace}
\newcommand{\pol}[1]{\ensuremath{Pol(#1)}\xspace}
\newcommand{\bpol}[1]{\ensuremath{BPol(#1)}\xspace}

\newcommand{\upol}[1]{\ensuremath{UPol(#1)}\xspace}

\newcommand{\sfp}[1]{\ensuremath{\textup{SF}(#1)}\xspace}

\newcommand{\upolo}{\ensuremath{UPol}\xspace}

\newcommand{\booln}{\ensuremath{Bool}\xspace}
\newcommand{\poln}{\ensuremath{Pol}\xspace}
\newcommand{\bpoln}{\ensuremath{BPol}\xspace}

\newcommand{\ldet}[1]{\ensuremath{LPol(#1)}\xspace}
\newcommand{\rdet}[1]{\ensuremath{RPol(#1)}\xspace}
\newcommand{\adet}[1]{\ensuremath{APol(#1)}\xspace}

\newcommand{\idet}[1]{\ensuremath{\ldet{#1} \cap \rdet{#1}}\xspace}
\newcommand{\jdet}[1]{\ensuremath{\ldet{#1}\! \vee\! \rdet{#1}}\xspace}
\newcommand{\mdet}[1]{\ensuremath{MPol(#1)}\xspace}
\newcommand{\ldeto}{\ensuremath{LPol}\xspace}
\newcommand{\rdeto}{\ensuremath{RPol}\xspace}

\newcommand{\mdeto}{\ensuremath{MPol}\xspace}

\newcommand{\ldetp}[2]{\ensuremath{LP_{#1}(#2)}\xspace}
\newcommand{\rdetp}[2]{\ensuremath{RP_{#1}(#2)}\xspace}

\newcommand{\idetp}[2]{\ensuremath{\ldetp{#1}{#2} \cap \rdetp{#1}{#2}}\xspace}
\newcommand{\jdetp}[2]{\ensuremath{\ldetp{#1}{#2}\! \vee\! \rdetp{#1}{#2}}\xspace}
\newcommand{\ldetn}[1]{\ldetp{n}{#1}}
\newcommand{\rdetn}[1]{\rdetp{n}{#1}}

\newcommand{\idetn}[1]{\idetp{n}{#1}}
\newcommand{\jdetn}[1]{\jdetp{n}{#1}}

\newcommand{\eqlp}[1]{\ensuremath{\mathrel{\rhd_{#1}}}\xspace}

\newcommand{\eqlep}[1]{\eqlp{\eta,#1}}
\newcommand{\eqlek}{\eqlep{k}}
\newcommand{\eqrp}[1]{\ensuremath{\mathrel{\lhd_{#1}}}\xspace}

\newcommand{\eqrep}[1]{\eqrp{\eta,#1}}
\newcommand{\eqrek}{\eqrep{k}}
\newcommand{\eqmp}[1]{\ensuremath{\mathrel{\bowtie_{#1}}}\xspace}

\newcommand{\eqmep}[1]{\eqmp{\eta,#1}}
\newcommand{\eqmek}{\eqmep{k}}
\newcommand{\eqmap}[1]{\eqmp{\alpha,#1}}
\newcommand{\eqmak}{\eqmap{k}}

\newcommand{\posxp}[3]{\ensuremath{\poschar_{\xvar}(#1,#2,#3)}\xspace}
\newcommand{\poslp}[3]{\ensuremath{\poschar_{\rhd}(#1,#2,#3)}\xspace}
\newcommand{\posrp}[3]{\ensuremath{\poschar_{\lhd}(#1,#2,#3)}\xspace}
\newcommand{\posmp}[3]{\ensuremath{\poschar_{\bowtie}(#1,#2,#3)}\xspace}

\newcommand{\poslak}[1]{\poslp{\alpha}{k}{#1}}
\newcommand{\posrak}[1]{\posrp{\alpha}{k}{#1}}
\newcommand{\posmak}[1]{\posmp{\alpha}{k}{#1}}
\newcommand{\posxek}[1]{\posxp{\eta}{k}{#1}}
\newcommand{\poslek}[1]{\poslp{\eta}{k}{#1}}
\newcommand{\posrek}[1]{\posrp{\eta}{k}{#1}}
\newcommand{\posmek}[1]{\posmp{\eta}{k}{#1}}

\newcommand{\qtlp}[1]{\ensuremath{\preceq_{\eta,#1}}\xspace}
\newcommand{\qtlkn}{\qtlp{k,n}}
\newcommand{\eqtlp}[1]{\ensuremath{\cong_{\eta,#1}}\xspace}
\newcommand{\eqtlkn}{\eqtlp{k,n}}

\newcommand{\infix}[3]{\ensuremath{#1(#2,#3)}\xspace}
\newcommand{\suffix}[2]{\infix{#1}{#2}{|#1|+1}}
\newcommand{\prefix}[2]{\infix{#1}{0}{#2}}

\newcommand{\xvar}{\ensuremath{\mathbcal{x}}\xspace}

\newcommand{\poschar}{\textup{P}}

\newcommand{\pos}[1]{\ensuremath{\poschar(#1)\xspace}}
\newcommand{\posc}[1]{\ensuremath{\poschar_{\mathbcal{c}}(#1)\xspace}}

\newcommand{\mods}{\ensuremath{\mathit{MOD}}\xspace}

\newcommand{\fo}{\ensuremath{\textup{FO}}\xspace}

\newcommand{\fod}{\ensuremath{\fo^2}\xspace}

\newcommand{\fodw}{\ensuremath{\fod(<)}\xspace}
\newcommand{\fodws}{\ensuremath{\fod(<,+1)}\xspace}

\newcommand{\fodwm}{\ensuremath{\fod(<,\mods)}\xspace}

\newcommand{\fodwsm}{\ensuremath{\fod(<,+1,\mods)}\xspace}

\newcommand{\fodsi}[1]{\ensuremath{\sic{#1}^2}\xspace}
\newcommand{\fodsn}{\fodsi{n}}

\newcommand{\fodb}[1]{\ensuremath{\bsc{#1}^2}\xspace}
\newcommand{\fodbn}{\fodb{n}}

\newcommand{\sic}[1]{\ensuremath{\Sigma_{#1}}\xspace}

\newcommand{\bsc}[1]{\ensuremath{\Bs\Sigma_{#1}}\xspace}

\newcommand{\prefsig}[1]{\ensuremath{\frP_{#1}}\xspace}

\newcommand{\prefsigg}{\prefsig{\Gs}}

\newcommand{\infsig}[1]{\ensuremath{\frI_{#1}}\xspace}

\newcommand{\infsigc}{\infsig{\Cs}}

\newcommand{\infsigg}{\infsig{\Gs}}

\newcommand{\infsiggp}{\infsig{\Gs^+}}

\newcommand{\dclosp}[1]{\ensuremath{\mathord{\downarrow_{#1}}}\xspace}
\newcommand{\dclosr}{\dclosp{R}}

\newcommand{\imprint}{imprint\xspace}
\newcommand{\imprints}{imprints\xspace}

\newcommand{\Imprints}{Imprints\xspace}

\newcommand{\tame}{multiplicative\xspace}

\newcommand{\Ratms}{Rating maps\xspace}

\newcommand{\ratms}{rating maps\xspace}

\newcommand{\Nice}{Nice\xspace}
\newcommand{\nice}{nice\xspace}

\newcommand{\mratm}{multiplicative rating map\xspace}
\newcommand{\mratms}{multiplicative rating maps\xspace}

\newcommand{\Mratms}{Multiplicative rating maps\xspace}

\newcommand{\prin}[2]{\ensuremath{\Is[#1](#2)}\xspace}

\newcommand{\opti}[2]{\ensuremath{\Is_{#1}\left[#2\right]}\xspace}

\newcommand{\dopti}[1]{\opti{\Ds}{#1}}

\newcommand{\popti}[3]{\ensuremath{\Ps_{#1}[#2,#3]}\xspace}

	\newcommand{\lhtyp}[2]{\ensuremath{[#1]^{\smash{\rhd}}_{\smash{#2}}}\xspace}
\newcommand{\rhtyp}[2]{\ensuremath{[#1]^{\smash{\lhd}}_{\smash{#2}}}\xspace}
\newcommand{\mhtyp}[2]{\ensuremath{[#1]^{\smash{\bowtie}}_{\smash{#2}}}\xspace}

\newcommand{\pldetoptid}{\popti{\ldet{\Ds}}{\etac}{\rho}}

\newcommand{\prdetoptid}{\popti{\rdet{\Ds}}{\etac}{\rho}}

\newcommand{\pmdetoptid}{\popti{\mdet{\Ds}}{\etac}{\rho}}

\newcommand{\typ}[2]{\ensuremath{[#1]_{#2}}\xspace}
\newcommand{\ctype}[1]{\typ{#1}{\Cs}}
\newcommand{\dtype}[1]{\typ{#1}{\Ds}}
\newcommand{\atyp}[1]{\typ{#1}{\alpha}}

\newcommand{\gtyp}[1]{\typ{#1}{\gamma}}
\newcommand{\ftyp}[2]{\ensuremath{[#1]_{#2}}\xspace}

%% file: intro.tex
This paper is part of a research program whose aim is to investigate natural subclasses of the regular languages of finite words. We are interested in classes that are specified by a syntax (inspired by either regular expressions or logic), that one can use to describe their languages. For each class \Cs, we use three decision problems as means of investigation. First, \Cs-membership takes a regular language $L$ as input and asks if $L\in\Cs$. Second, \Cs-separation takes two regular languages $H,L$ as input and asks if there exists $K \in \Cs$ such that $H \subseteq K$ and \mbox{$K \cap L = \emptyset$}. Finally, \Cs-covering is a generalization of \Cs-separation to arbitrarily many input languages. The key idea is that in practice, obtaining algorithms for these problems requires techniques that cannot be developed without a solid understanding of~\Cs.

In the paper, we consider several \emph{operators}. Each of them defines a family of closely related classes. Let us clarify with logic. Each fragment of first-order logic (\fo) defines several classes: one per choice of \emph{signature} (\emph{i.e.}, the set of predicates that one may use in formulas). For instance, in the literature, several classes are associated to first-order logic itself by considering natural predicates such as the linear order ``$<$''~\cite{mnpfosf,schutzsf}, successor ``$+1$''~\cite{Beauquier_1991} or modular predicates ``$MOD$''~\cite{MIXBARRINGTON1992478}. Hence, a generic approach is desirable. This typically involves two steps. First, one characterizes the investigated fragment with an \emph{operator} $\Cs\mapsto Op(\Cs)$ on classes. For example, first-order logic is characterized \emph{star-free closure} which builds the least class \sfp{\Cs} containing its input class \Cs and closed under union, complement and concatenation. More precisely, it is known~\cite{mnpfosf,PZ:generic18} that if \Cs is a \emph{Boolean algebra closed under quotients} (we call this a \vari), there exists a signature \infsigc such that $\sfp{\Cs} = \fo(\infsigc)$. This captures most of the natural signature choices. The second step then consists in investigating the operator $\Cs \mapsto Op(\Cs)$ in a generic way: on has to identify hypotheses on \Cs which ensure the decidability of membership, separation or covering for $Op(\Cs)$. For example, \sfp{\Cs}-membership is decidable as soon as \Cs-separation is decidable~\cite{pzsfclos}. Finally, a similar results is known for \sfp{\Cs}-separation and \sfp{\Cs}-covering~\cite{pzsfclos} but it is restricted to special input \varis \Cs containing only \emph{group languages}. These are the languages recognized by a finite group, or equivalently by a permutation automaton (\emph{i.e.}, a complete, deterministic \emph{and} co-deterministic automaton). %This implies that separation is decidable for variants of \fo such as \fow or \fowm.

We investigate restrictions of \emph{polynomial closure}. Given an input \Cs, the class \pol{\Cs} contains the finite unions of marked products $K_0a_1K_1 \cdots  a_nK_n$ where \mbox{$K_0,\dots,K_n  \in \Cs$}. We look at variants defined by imposing semantic restrictions on the products. A marked product $K_0a_1K_1\cdots a_nK_n$ is \emph{unambiguous} if for each $w\in K_0a_1K_1\cdots a_nK_n$, the decomposition of $w$ witnessing this membership is \emph{unique}. This defines \emph{unambiguous polynomial closure} (\upolo) which is well-understood~\cite{Pin80,Pinambigu,pzupol}. We look at stronger restrictions.  For a marked product $K_0a_1K_1 \cdots a_nK_n$, we let $L_i = K_0 a_1K_1 \dots a_{i-1}K_{i-1}$ and $R_i = K_i a_{i+1}  \cdots K_{n-1}a_n K_n$ for all $i \leq n$. The whole marked product is \emph{left} (resp. \emph{right}) deterministic if for all $i \leq n$, $L_ia_iA^*$ (resp. $A^*a_iR_i$) is unambiguous. It is \emph{mixed deterministic} if for all $i \leq n$, either $L_ia_iA^*$ or $A^*a_i R_i$ is unambiguous. This leads to three operators: \emph{left, right} and \emph{mixed polynomial closure} (\ldeto, \rdeto and \mdeto). Historically, \ldeto and \rdeto are well-known. They were first investigated by Schützenberger~\cite{schul} and Pin~\cite{Pin80,jep-intersectPOL}. On the other hand, \mdeto is new. We first prove that these operators have robust properties which are similar to those of \upolo~\cite{pzupol}. First, we prove that if \Cs is a \vari, then so are \ldet{\Cs}, \rdet{\Cs} and \mdet{\Cs}. Moreover, we prove that if \Cs has decidable membership, then this is also the case for \ldet{\Cs}, \rdet{\Cs} and \mdet{\Cs}.

We also look at hierarchies of classes. In general, \ldet{\Cs} and \rdet{\Cs} are incomparable. Thus, given an input class \Cs, two hierarchies can be built. The first levels are \ldet{\Cs} and \rdet{\Cs}, then for all $n>1$, the levels \ldetn{\Cs} and \rdetn{\Cs} are defined as \ldet{\rdetp{n-1}{\Cs}} and \rdet{\ldetp{n-1}{\Cs}}. One may also define combined levels \idetn{\Cs} (the languages belonging to both classes) and \jdetn{\Cs} (the least Boolean algebra containing both classes). It follows from results of~\cite{pzupol} that the union of all levels is \upol{\Cs}. In the literature, this construction is well-known for a specific input class: the piecewise testable languages \pt~\cite{simonthm} (\emph{i.e.}, the Boolean combinations of marked products $A^*a_1A^* \cdots a_nA^*$). This hierarchy is strict and has characterizations based on algebra~\cite{twhiera,subDA} and logic~\cite{kwfo2alt2,kwfo2alt3}. While each hierarchy contains four kinds of levels, we prove that their construction process can be unified: each kind can be climbed using only \mdeto. For example, we show that $\mdet{\jdetp{n-1}{\Cs}} = \jdetn{\Cs}$ for all $n > 1$. %This makes the investigation of such hierarchies easier.

In the second part of the paper, we investigate the quantifier alternation hierarchies of \emph{two-variable} first-order logic (\fod). The fragment \fod contains the first-order formulas using at most two distinct reusable variables. For all $n \geq 1$, we let \fodbn as the set of all \fod formulas such that each branch in their parse trees contains at most $n$ blocks of alternating quantifiers ``$\exists$'' and ``$\forall$''. There are important classes associated to these fragments and several of them are prominent in the literature. Historically, the full logic \fod was first considered. It is known that membership is decidable for the variants \fodw and \fodws equipped with the linear order and successor~\cite{twfo2}, as well as for \fodwm equipped with modular predicates~\cite{DartoisP13}. For quantifier alternation, it is known that membership is decidable for all levels $\fodbn(<)$~\cite{kwfo2alt2,kwfo2alt3,ksfo2alt}, $\fodbn(<,+1)$~\cite{klfo2alts} and $\fodbn(<,+1,MOD)$~\cite{dpfo2}. While the arguments are connected, each of these results involves a tailored proof. In the paper, we develop a generic approach based on \mdeto and look at a family of signatures. Given a \vari \Gs containing only group languages, we associate a generic set of predicates \prefsigg. For every $L \in \Gs$, it contains a unary predicate $P_L(x)$: it  checks if the prefix preceding a given position belongs to $L$. We \mbox{consider} all signatures of the form $\{<,\prefsigg\}$ or $\{<,+1,\prefsigg\}$. This captures most of the natural examples such as $\{<\}$, $\{<,+1\}$, $\{<,MOD\}$, or $\{<,+1,MOD\}$ (we present other examples in this paper). We prove that if \frS is one of the two above kinds of signatures, the quantifier alternation hierarchy of $\fod(\frS)$ is climbed using \mdeto: $\fodb{n+1}(\frS) = \mdet{\fodbn(\frS)}$ for all $n \geq 1$. This also yields $\fod(\frS) = \upol{\fodb{1}(\frS)}$. We get a \emph{generic} language theoretic characterization of \fod and its quantifier alternation hierarchy which applies to many natural signature choices. Moreover, it follows from independent results~\cite{pz:csr} that if \frS is a signature built from a group \vari \Gs as above, then membership for $\fodb{1}(\frS)$ is decidable when \Gs-\emph{separation} is decidable. Hence, since this property is preserved by \mdeto, we are able to lift the decidability of membership to all levels $\fodb{n}(\frS)$ in this case. We reprove the aforementioned results and obtain new ones.

In the last part of the paper, we investigate separation and covering for \ldeto, \rdeto and \mdeto. We prove that if \Cs is a \emph{finite} \vari and \Ds is a \vari with decidable covering such that $\Cs\! \subseteq\! \Ds\! \subseteq\! \upol{\Cs}$, then covering is both decidable for \ldet{\Ds}, \rdet{\Ds} and \mdet{\Ds} as well. This is weaker than our results concerning membership as \Cs must be \emph{finite}. Yet, we detail a key application: the \vari \pt of \emph{piecewise testable languages}~\cite{simonthm}. While \pt is infinite, it is simple to verify that $\at \subseteq \pt \subseteq \upol{\at}$ where \at is the finite \vari of alphabet testable languages (\emph{i.e.}, the Boolean combinations of languages $B^*$ where $B$ is a sub-alphabet). Since \pt-covering is decidable~\cite{cmmptsep,pvzptsep,pzcovering}, a simple induction yields the decidability of covering for all classes that can be built recursively from \pt by applying \ldeto, \rdeto and \mdeto. This includes all levels \ldetn{\pt} and \rdetn{\pt}. Moreover, it is well-known that $\pt = \fodb{1}(<)$. Hence, this can be combined with our logical characterization of \mdeto by two-variable first-order logic to obtain the decidability of $\fodb{n}(<)$-covering for every $n \geq 1$. Let us point out that an alternate proof of this result was obtained recently using independent techniques~\cite{kkufo2}.

%\smallskip

We present the definitions and the mathematical tools that we shall use in Section~\ref{sec:prelims}. We properly define \poln, \ldeto, \rdeto and \mdeto in Section~\ref{sec:polc}. Then, in  Section~\ref{sec:frame}, we introduce a general framework that we shall use to manipulate them throughout the paper.  Section~\ref{sec:carac}, we present algebraic characterizations of \ldeto, \rdeto and \mdeto. They imply that all three of them preserve the decidability of membership. We investigate the language theoretic hierarchies that can be built with our operators in Section~\ref{sec:deth}. We turn to logic in Section~\ref{sec:logic} and use \mdeto to characterize quantifier alternation for \fod. Finally, Sections~\ref{sec:ratms}, \ref{sec:cov} and~\ref{sec:covm} are devoted to the separation and covering. This paper is the journal version of~\cite{pmixed}, it includes all proof arguments and the decidability results have been generalized to covering (only membership and separation were considered in~\cite{pmixed}).

%% file: prelims.tex
\subsection{Finite words and classes of languages} 

We fix an arbitrary finite alphabet $A$ for the whole paper. As usual, $A^*$ denotes the set of all words over $A$, including the empty word~\veps. We let $A^{+}=A^{*}\setminus\{\veps\}$. For $u,v \in A^*$, we write $uv$ the concatenation of $u$ and~$v$. If $w \in A^*$, we write $|w| \in \nat$ for its length. We also consider \emph{positions}. A word $w =a_1 \cdots a_{|w|} \in A^*$ is viewed as an \emph{ordered set $\pos{w} = \{0,1,\dots,|w|,|w|+1\}$ of $|w|+2$ positions}. A position $i$ such that $1 \leq i \leq |w|$ carries the label $a_i \in A$. We write $\posc{w} = \{1,\dots,|w|\}$ for this set of labeled positions.  On the other hand, the positions $0$ and $|w|+1$ are \emph{artificial} leftmost and rightmost positions which carry \emph{no label}. Finally, given a word $w= a_1\cdots a_{|w|} \in A^*$ and $i,j \in \pos{w}$ such that $i < j$, we write $\infix{w}{i}{j} = a_{i+1} \cdots a_{j-1} \in A^*$ (\emph{i.e.}, the infix obtained by keeping the letters carried by the positions \emph{strictly} between $i$ and $j$). Note that $\infix{w}{0}{|w|+1} = w$.

A \emph{language} is a subset of $A^*$. We lift the concatenation operation to languages: for $K,L \subseteq A^*$, we write $KL = \{uv \mid u \in K \text{\;and\;} v \in L\}$. All languages considered in this paper are \emph{regular}. These are the languages which can be defined by a finite automaton or a morphism into a finite monoid. We work with the latter definition  which we recall now.

\smallskip\noindent
{\bf Monoids and morphisms.}  A \emph{semigroup} is a pair $(S,\cdot)$ where $S$ is a set and ``$\cdot$'' is an associative multiplication on $S$.  It is standard to abuse terminology and make the binary operation implicit: one simply says that ``$S$ is a semigroup''. A \emph{monoid} $M$ is a semigroup whose multiplication has a neutral element denoted by ``$1_M$''. An idempotent of a semigroup $S$ is an element $e \in S$ such that $ee = e$. We write $E(S) \subseteq S$ for the set of all idempotents in $S$. It is standard that when $S$ is \emph{finite}, there exists $\omega(S) \in \nat$ (written $\omega$ when $S$ is understood) such that $s^\omega$ is idempotent for every $s \in S$.

Clearly, $A^*$ is a monoid whose multiplication is concatenation (\veps is the neutral element). Thus, given a monoid $M$, we may consider morphisms $\alpha: A^* \to M$. For the sake of avoiding clutter, we shall adopt the following notation. Given $w \in A^*$, we write $\atyp{w}  \subseteq A^*$ for the language $\atyp{w} = \alpha\inv(\alpha(w)) = \{u \in A^* \mid \alpha(u) = \alpha(w)\}$. A language $L \subseteq A^*$ is \emph{recognized} by such a morphism $\alpha$ when there exists $F \subseteq M$ such that $L = \alpha\inv(F)$. It is well-known that a language is regular if and only if it can be recognized by a morphism into a \emph{finite} monoid.

\begin{remark}
	Since the only infinite monoid that we consider is $A^*$, we implicitly assume that every arbitrary monoid $M,N,\dots$ that we consider is finite from now on.
\end{remark}

We also consider the standard Green relations that one may associate to each monoid~$M$. Given $s,t \in M$, we write $s \Rord t$ if there exists $r \in M$ such that $s = tr$. Moreover, $s \Lord t$ if there exists $q \in M$ such that $s = qt$. Finally, $s \Jord t$ if there exist $q,r \in M$ such that $s = qtr$. One may verify that these are preorders. We write \Rrel, \Lrel and \Jrel for the equivalences associated to \Rord, \Lord and \Jord (\emph{e.g} $s \Rrel t$ when $s \Rord t$ and $t \Rord s$). Finally, we write \Rords, \Lords and \Jords for the strict variants of these preorders (\emph{e.g} $s \Rords t$ when $s \Rord t$ and $s \neq t$). We shall need the following standard lemma concerning the Green relations of \emph{finite} monoids.

\begin{lemma} \label{lem:jlr}
	Let $M$ be a finite monoid and $s,t \in M$. If $s \Rord t$ and $t \Jord s$, then $s \Rrel t$. Symmetrically, if $s \Lord t$ and $t \Jord s$, then $s \Lrel t$.
\end{lemma}

\begin{proof}
	By symmetry, we only prove the first property. Assume that $s \Rord t$ and $t \Jord s$. We show that $s \Rrel t$. Since we already know that $t \Rord s$, this amounts to proving that $s \Rord t$. Since $t \Rord s$, we have $x \in M$ such that $sx = t$. Since $s \Jord t$, we have $y,z \in M$ such that $ytz = s$. This yields $s = ysxz = y^\omega s (xz)^\omega = y^\omega s (xz)^\omega (xz)^\omega = s (xz)^\omega$. Therefore, $s = sx (zx)^{\omega-1} z = t (zx)^{\omega-1} z$ and we get $s \Rord t$, completing the proof. 
\end{proof}

\smallskip\noindent
{\bf Classes.} A \emph{class of languages} \Cs is a set of languages. A \emph{lattice} is a class which is closed under both union and intersection, and containing the languages $\emptyset$ and $A^*$. Moreover, a \emph{Boolean algebra} is a lattice closed under complement. Finally, a class \Cs is \emph{quotient-closed} when for every $L \in \Cs$ and every $u \in A^*$, the following properties hold:
\[
u^{-1}L \stackrel{\text{def}}{=}\{w\in A^*\mid uw\in L\} \text{\quad and\quad} Lu^{-1} \stackrel{\text{def}}{=}\{w\in A^*\mid wu\in L\}\text{\quad both belong to \Cs}.
\]
Finally, a class \Cs is a \emph{\vari} when it is a quotient-closed Boolean algebra containing only \emph{regular languages}. In the paper, we investigate several operators on classes of languages. An operator is a mapping $\Cs \mapsto Op(\Cs)$ which builds a new class $Op(\Cs)$ from an arbitrary input class \Cs. In practice, we shall restrict ourselves to input classes that are \varis.

\smallskip
\noindent
{\bf Group languages.} We define particular classes: the group \varis. In the sequel, they will serve as key input classes for our operators. A \emph{group} is a monoid $G$ such that every $g \in G$ has an inverse $g\inv \in G$, \emph{i.e.}, such that $gg\inv = g\inv g = 1_G$. A language $L$ is a \emph{group language} if it is recognized by a morphism $\alpha: A^* \to G$ into a \emph{finite group $G$}. Finally, a \emph{group \vari} is a \vari \Gs which contains group languages only.

 We also consider ``extensions'' of the group \varis. One may verify that $\{\veps\}$ and $A^+$ are \emph{not} group languages. This motivates the following definition: given a class \Cs, the \emph{\wsuit extension of \Cs}, written $\Cs^+$, is the class consisting of all languages of the form $L \cap A^+$ or $L \cup \{\veps\}$ where $L \in \Cs$ (while the definition makes sense for ever class \Cs, we only use it when \Cs is a group \vari). The following fact can be verified from the definition.

\begin{fct} \label{fct:wsvari}
	Let \Cs be a \vari. Then, $\Cs^+$ is a \vari containing $\{\veps\}$ and $A^+$.
\end{fct}

\subsection{Membership, separation and covering}

We look at three decision problems. Each of them depends on an arbitrary class \Cs and are used as mathematical tools for analyzing \Cs. %Intuitively, obtaining an algorithm for one of these three problems requires a solid understanding of \Cs.

The \emph{\Cs-membership problem} is the simplest one. It takes as input a single regular language~$L$ and asks whether $L\in \Cs$. The second problem, \emph{\Cs-separation}, is more general. Given three languages $K,L_1,L_2$, we say that $K$ \emph{separates} $L_1$ from $L_2$ if we have $L_1 \subseteq K$ and $L_2 \cap K = \emptyset$. Given a class of languages \Cs, we say that $L_1$ is \emph{\Cs-separable} from $L_2$ if some language in \Cs separates $L_1$ from $L_2$. Observe that when \Cs is not closed under complement, the definition is not symmetrical: it is possible for $L_1$ to be \Cs-separable from $L_2$ while $L_2$ is not \Cs-separable from $L_1$. The separation problem associated to a given class \Cs takes two regular languages $L_1$ and $L_2$ as input and asks whether $L_1$ is \Cs-separable from $L_2$.

\begin{remark}
	The \Cs-separation problem generalizes \Cs-membership. A regular language belongs to $\Cs$ if and only if it is \Cs-separable from its complement, which is also regular.
\end{remark}

We do not consider separation directly and look at a third, even more general problem: \Cs-covering. A \emph{cover of a language $L$} is a \emph{finite} set of languages \Kb such that $L \subseteq \bigcup_{K \in \Kb} K$. Additionally, \Kb is a \emph{\Cs-cover} if every $K \in \Kb$ belongs to \Cs. Moreover, given two finite sets of languages \Kb and \Lb, we say that \Kb is \emph{separating} for \Lb if for every $K\in\Kb$, there exists $L\in\Lb$ such that $K \cap L = \emptyset$. Finally, given a language $L_1$ and a finite set of languages~$\Lb_2$, we say that the pair $(L_1,\Lb_2)$ is \emph{\Cs-coverable} if there exists a \Cs-cover of $L_1$ which is separating~for~$\Lb_2$.

The \Cs-covering problem is  defined as follows. Given as input a regular language $L_1$ and a finite set of regular languages $\Lb_2$, it asks whether the pair $(L_1,\Lb_2)$ \Cs-coverable. Covering generalizes separation if the class~\Cs is a lattice (see~\cite[Theorem~3.5]{pzcovering} for the proof).

\begin{lemma} \label{lem:covsep}
	Let \Cs be a lattice and $L_1,L_2$ be two languages. Then $L_1$ is \Cs-separable from $L_2$ if and only if $(L_1,\{L_2\})$ is \Cs-coverable.
\end{lemma}

\subsection{\texorpdfstring{\Cs-morphisms}{C-morphisms}}
 
Consider a \vari \Cs. A \emph{\Cs-morphism} is a \emph{surjective} morphism \mbox{$\eta: A^*\to N$} such that every language recognized by $\eta$ belongs to \Cs. This notion serves as a key mathematical tool in the paper. First, we use it for the membership problem.

Given a regular language $L$, one may associate a canonical morphism recognizing $L$. Let us briefly recall the definition. We associate a relation $\equiv_L$ on $A^*$ to $L$. Given $u,v \in A^*$, we have $u \equiv_L v$ if and only if $xuy \in L \Leftrightarrow xvy \in L$ for every \mbox{$x,y \in A^*$}. It can be verified that $\equiv_L$ is a congruence of $A^*$ and, since $L$ is regular, that it has finite index. Therefore, the map \mbox{$\alpha: A^* \to {A^*}/{\equiv_L}$} which associates its $\equiv_L$-class to each word is a morphism into a finite monoid. It is called the \emph{syntactic morphism of $L$} and it can be computed from any representation of $L$. The following standard result connects it to \Cs-membership (see \emph{e.g.} \cite[Proposition~2.11]{pzupol2} for a proof).

\begin{proposition} \label{prop:synmemb}%
	Let \Cs be a \vari. A regular language belongs to \Cs if and only if its syntactic morphism is a \Cs-morphism.
\end{proposition}

%\begin{proof}
%	We fix a regular language $L$ for the proof. Moreover, we let $\alpha_L: A^* \to M_L$ as its syntactic morphism. Since $L$ is recognized by $\alpha_L$, it is immediate that if $\alpha_L$ is a \Cs-morphism, then $L \in \Cs$. We prove the converse implication. Assume that $L \in \Cs$. We show that $\alpha\inv(F) \in \Cs$ for every $F \subseteq M$. Since \Cs is closed under union, it suffices to consider the case when $F = \{s\}$ for some $s \in M$. By definition of the syntactic morphism, $\alpha\inv(s)$ is an equivalence class of $\equiv_L$. By definition of $\equiv_L$, we know that given $w,w' \in A^*$, we have $w \equiv_L w'$ if and only if $w \in x\inv L y\inv \Leftrightarrow w' \in x\inv L y\inv$ for every $x,y \in L$. Hence, since $\equiv_L$ has finite index (this is because $L$ is regular), there finitely many languages $x\inv L y\inv$ for $x,y \in A^*$ and every $\equiv_L$-class is a finite Boolean combination of such languages. Since $L \in \Cs$ and \Cs is a quotient-closed, we have  $x\inv L y\inv \in \Cs$ for every $x,y \in A^*$. Hence, since \Cs is~a Boolean algebra, every $\equiv_L$-class belongs to \Cs which complete the proof.
%\end{proof}

By Proposition~\ref{prop:synmemb}, getting an algorithm for \Cs-membership boils down to finding a procedure which decides if some input morphism $\alpha: A^* \to M$ is a \Cs-morphism. This is how we approach the question in this paper. We shall also use \Cs-morphisms as mathematical tools in proof arguments. In this context, we shall use the following statement which is a simple corollary of Proposition~\ref{prop:synmemb} (see \cite[Proposition~2.12]{pzupol2} for a proof).

\begin{proposition} \label{prop:genocm}
	Let \Cs be a \vari and consider finitely many languages $L_1,\dots,L_k \in \Cs$. There exists a \Cs-morphism $\eta: A^* \to N$ such that $L_1,\dots,L_k$ are recognized by $\eta$.
\end{proposition}

%\begin{proof}
%	For every $i \leq k$, we let $\alpha_i: A^* \to M_i$ as the syntactic morphism of $L_i$. We know from Proposition~\ref{prop:synmemb} that $\alpha_i$ is a \Cs-morphism. Let $M =M_1\times \cdots \times M_n$ be the monoid equipped with the componentwise multiplication and $\alpha: A^* \to M$ the morphism defined by $\alpha(w) = (\alpha_1(w),\dots, \alpha_n(w))$ for every $w \in A^*$. Clearly, the languages $L_1,\dots,L_n$ are all recognized by $\alpha$. Moreover, for every $\bar{s} = (s_1,\dots,s_n) \in M$, it is immediate that $\alpha\inv(\bar{s}) = \alpha_1\inv(s_1) \cap \cdots \cap \alpha_n\inv(s_n)$. Hence, $\alpha\inv(\bar{s}) \in \Cs$ by closure under intersection. It follows that every language recognized by $\alpha$ belongs to \Cs. Hence, it suffices to define $\eta$ as the surjective restriction of $\alpha$ to get the desired \Cs-morphism.
%\end{proof}

We complete the presentation with a lemma which considers the classes that are \emph{group} \varis and their \wsuit extensions (see \cite[Lemmas~2.14 and 2.15]{pzupol2} for a proof).

\begin{lemma} \label{lem:gmorph}
	Let \Gs be a group \vari  and $\eta: A^* \to N$ a morphism. If $\eta$ is a \Gs-morphism, then $N$ is a group. Moreover, if $\eta$ is $\Gs^+$-morphism, then $\eta(A^+)$ is a group.
\end{lemma}

%\begin{proof}
%	We treat the case when $\eta$ is $\Gs^+$-morphism. The other one is handled with a similar argument which is left to the reader. Let $G = \alpha(A^+)$. We show that $G$ is a group. By definition, it suffices to prove that there is only one idempotent in $G$. Hence, we consider two idempotents $e,f \in G$ and show that $e = f$. Let $u,v \in A^+$ such that $\eta(u) = e$ and $\eta(v) = f$. By hypothesis we have $\eta\inv(e) \in \Gs^+$. This yields $L \in \Gs$ such that either $\eta\inv(e) = L \cup \{\veps\}$ or $\eta\inv(e) = L \cap A^+$. Since $L$ is group language, we have a morphism $\beta: A^* \to H$ into a group $H$ recognizing $L$. Let $p = \omega(H)$. Since $e$ is idempotent, we have $\alpha(u^p) = e$ and since $u^p \in A^+$, this yields $u^p \in L$. Moreover, since $H$ is a group, we have $\beta(u^p) = 1_H = \beta(v^p)$. Hence, $v^p \in L$ since $\beta$ recognizes $L$. Since $v^p \in A^+$, it follows that $\eta(v^p) = e$. Finally, since $\eta(v) = f$ which is an idempotent, we also have $\eta(v^p) = f$ and we get $e = f$ as desired.
%\end{proof}

\subsection{Canonical relations}

For each class \Cs and each morphism $\alpha: A^*\to M$, we define two relations on $M$. They were first introduced in~\cite{pzupol2,PZ:generic18}. We shall use them to formulate \emph{generic} algebraic characterizations of the operators $\Cs \mapsto Op(\Cs)$ that we consider: they depend on \Cs through these relations. %We refer the reader to~\cite[Section~5]{pzupol2} for the proofs of the statements.

\medskip
\noindent
{\bf \Cs-pairs.} Let \Cs be a class and $\alpha: A^* \to M$ a morphism. A pair $(s,t) \in M^2$ is a \emph{\Cs-pair} (for $\alpha$) if and only if $\alpha\inv(s)$ is \emph{not} \Cs-separable from $\alpha\inv(t)$. The \Cs-pair relation is not very robust.  First, it is reflexive when $\alpha$ is surjective (a nonempty language cannot be separated from itself). It is also symmetric if \Cs is closed under complement but \emph{not} transitive in general. If \Cs is a \vari, we have the following lemma proved in~\cite[Lemma~5.11]{pzupol2}.

\begin{lemma}\label{lem:cmorph}
	Let \Cs be a \vari and $\alpha: A^* \to M$ a morphism. The following holds:
	\begin{itemize}
		\item For every \Cs-morphism $\eta: A^* \to N$ and every \Cs-pair $(s,t) \in M^2$ for $\alpha$, there exist $u,v \in A^*$ such that $\eta(u) = \eta(v)$, $\alpha(u) = s$ and $\alpha(v) = t$.
		\item There exists a \Cs-morphism $\eta: A^* \to N$ such that for all $u,v \in A^*$, if $\eta(u) = \eta(v)$, then $(\alpha(u),\alpha(v))$ is a \Cs-pair for $\alpha$.
	\end{itemize}
\end{lemma}

%\begin{proof}
%	For the first assertion, let $\eta: A^* \to N$ be a \Cs-morphism and let $(s,t) \in M^2$ be a \Cs-pair for $\alpha$. Let $F = \eta(\alpha\inv(s)) \subseteq N$. By hypothesis on $\eta$, we have $\eta\inv(F) \in \Cs$. Moreover, $\alpha\inv(s) \subseteq \eta\inv(F)$. Therefore, since $(s,t)$ is a \Cs-pair, we get $\eta\inv(F) \cap \alpha\inv(t) \neq \emptyset$. This yields $v \in A^*$ such that $\eta(v) \in F$ and $\alpha(v) = t$. Finally, since $\eta(v) \in F$, the definition of $F$ yields $u \in A^*$ such that $\eta(u) = \eta(v)$ and $\alpha(u) = s$, concluding the proof of the first~assertion.
%	
%	We turn to the second assertion. Let $P \subseteq M^2$ be the set of all pairs $(s,t) \in M^2$ which are \emph{not} \Cs-pairs. For every $(s,t) \in P$, there exists $K_{s,t} \in \Cs$ separating $\alpha\inv(s)$ from $\alpha\inv(t)$. Proposition~\ref{prop:genocm} yields a \Cs-morphism $\eta: A^* \to N$ recognizing all languages $K_{s,t}$ for $(s,t) \in P$. It remains to prove that for every $u,v \in A^*$, if $\eta(u) = \eta(v)$, then $(\alpha(u),\alpha(v))$ is a \Cs-pair. We prove the contrapositive. Assume that $(\alpha(u),\alpha(v))$ is a \emph{not} a \Cs-pair, we show that $\eta(u) \neq \eta(v)$. By hypothesis, $(\alpha(u),\alpha(v)) \in P$, which means that $K_{\alpha(u),\alpha(v)} \in \Cs$ is defined and separates $\alpha\inv(\alpha(u))$ from $\alpha\inv(\alpha(v))$. Thus, $u \in K_{\alpha(u),\alpha(v)}$ and $v \not\in K_{\alpha(u),\alpha(v)}$. Since $K_{\alpha(u),\alpha(v)}$ is recognized by $\eta$, this implies $\eta(u) \neq \eta(v)$.
%\end{proof}

Moreover, a key property is that if \Cs is a \vari, the \Cs-pair relation is compatible with multiplication. We refer the reader to~\cite[Lemma~5.12]{pzupol2} for the proof.

\begin{lemma}\label{lem:mult}
	Let \Cs be a \vari and $\alpha: A^* \to M$ a morphism. If $(s_1,t_1),(s_2,t_2) \in M^2$ are \Cs-pairs, then $(s_1s_2,t_1t_2)$ is a \Cs-pair as well.
\end{lemma}

%\begin{proof}
%	Lemma~\ref{lem:cmorph} yields a \Cs-morphism $\eta: A^* \to N$ such that for all $u,v \in A^*$, if $\eta(u) = \eta(v)$, then $(\alpha(u),\alpha(v))$ is a \Cs-pair. Let $(s_1,t_1),(s_2,t_2) \in M^2$ be \Cs-pairs. Since $\eta$ is a \Cs-morphism, it follows from Lemma~\ref{lem:cmorph} that there exist $u_i,v_i \in A^*$ for $i = 1,2$ such that $\eta(u_i) = \eta(v_i)$, $\alpha(u_i) = s_i$ and $\alpha(v_i) = t_i$. Clearly, it follows that $\eta(u_1u_2) = \eta(v_1v_2)$, $\alpha(u_1u_2) = s_1s_2$ and $\alpha(v_1v_2) = t_1t_2$. Hence, $(s_1s_2,t_1t_2)$ is a \Cs-pair by definition of $\eta$.
%\end{proof}

\newcommand{\canoac}{\ensuremath{\preceq_{\Cs,\alpha}}\xspace}
\newcommand{\qanoac}{\ensuremath{\leq_{\Cs,\alpha}}\xspace}
\newcommand{\canaeq}{\ensuremath{\sim_{\Cs,\alpha}}\xspace}
\newcommand{\canabeq}{\ensuremath{\sim_{\Cs,\beta \circ \alpha}}\xspace}

\noindent
{\bf Canonical equivalence.} Consider a class \Cs and a morphism $\alpha: A^* \to M$. We define an equivalence \canaeq on $M$. Let $s,t \in M$. We write $s \canaeq t$ if and only if $s \in F \Leftrightarrow t \in F$ for all $F \subseteq M$ such that $\alpha\inv(F) \in \Cs$. It is immediate by definition that \canaeq is an equivalence. For the sake of avoiding clutter, we shall abuse terminology when the morphism $\alpha$ is understood and write \canec for \canaeq. Additionally, for every element $s\in M$, we write $\ctype{s} \in {M}/{\canec}$ for the $\canec$-class of $s$. Observe that by definition, computing \canaeq boils down to computing the sets $F \subseteq M$ such that $\alpha\inv(F) \in \Cs$, \emph{i.e.} to \Cs-membership.

\begin{fct} \label{fct:eqmemb}
	Let \Cs be a \vari with decidable membership. Given as input a morphism $\alpha: A^* \to M$, one may compute the equivalence \canaeq on $M$.
\end{fct}

We now connect our two relations in the following lemma proved in~\cite[Lemma~5.16]{pzupol2}.

\begin{lemma} \label{lem:transclos}
	Let \Cs be a \vari and $\alpha: A^* \to M$ be a morphism. The equivalence \canaeq on $M$ is the reflexive transitive closure of the \Cs-pair relation associated to $\alpha$.
\end{lemma}

Moreover, \emph{when $\alpha$ is surjective}, the equivalence \canaeq is a congruence of the monoid $M$. We refer the reader to~\cite[Lemma~5.18]{pzupol2} for the proof.

\begin{lemma} \label{lem:caquot}
	Let \Cs be a \vari and $\alpha: A^* \to M$ be a surjective morphism. Then, \canaeq is a congruence of $M$.
\end{lemma}

%\begin{proof}
%	We fix $s_1,t_1,s_2,t_2 \in M$ such that $s_1 \canec t_1$ and $s_2\canec t_2$. We prove that $s_1s_2 \canec t_1t_2$. Let $F \subseteq M$ such that $\alpha\inv(F) \in \Cs$. We show that $s_1s_2 \in F \Leftrightarrow t_1t_2 \in F$. By symmetry, we only prove the left to tight implication. Hence, we assume that $s_1s_2 \in F$. Let $u,v \in A^*$ such that $\alpha(u) = s_1$ and $\alpha(v) = t_2$ (this is where we use the hypothesis that $\alpha$ is surjective). Let $X = \{q \in M \mid s_1q \in F\}$. We have $s_2 \in X$ and $\alpha\inv(X) = \{w \in A^*\mid uw \in \alpha\inv(F)\}$ which belongs to \Cs by closure under quotients. Hence, since $s_2 \canec t_2$, we get $t_2 \in X$ which yields $s_1t_2 \in F$. Let $Y = \{r \in M \mid rt_2 \in F\}$. We know that $s_1 \in Y$ since $s_1t_2 \in F$. Moreover, we know that $\alpha\inv(Y) = \{w \in A^*\mid wv \in \alpha\inv(F)\}$ which belongs to \Cs by closure under quotients. Hence, since $s_1 \canec t_1$, we get $t_1 \in Y$ which yields $t_1t_2 \in F$ as desired.
%\end{proof}

In view of Lemma~\ref{lem:caquot}, when $\alpha: A^* \to M$ is surjective, the map $\ctype{\cdot}: M \to M/{\canec}$ which associates its \canec-class to each element in $M$ is a morphism. It turns out that the composition $\ctype{\cdot} \circ \alpha: A^*\to {M}/{\canec}$ is a \Cs-morphism. See~\cite[Lemma~5.19]{pzupol2} for the proof.

\begin{lemma} \label{lem:smult}
	Let \Cs be a \vari and $\alpha: A^* \to M$ be a surjective morphism. The languages recognized by $\ctype{\cdot}  \circ \alpha: A^*\to {M}/{\canec}$ are exactly those which are simultaneously in \Cs and recognized by $\alpha$. 
\end{lemma}

%
%\begin{proof}
%	The languages recognized by $\ctype{\cdot} \circ \alpha$ are of the form $\alpha\inv(F)$ where $F$ is a \canec-class. Hence, it suffices to prove that for every $F \subseteq M$, $\alpha\inv(F) \in \Cs$ if and only if $F$ is a \canec-class. We fix $F \subseteq M$ for the proof. Assume first that $\alpha\inv(F) \in \Cs$. Let $s \in F$ and $t \in M$ such that $s \canec t$. By definition of \canec and since $\alpha\inv(F) \in \Cs$, we have $t \in F$. Hence, $F$ is a \canec-class as desired. Conversely, assume that $F$ is a \canec-class. Let $s\in F$. By definition, we have $s \not\canec r$ for every $r\not\in F$. Hence, there exists a set $F_{s,r} \subseteq M$ such that $\alpha\inv(F_{s,r}) \in \Cs$, $s \in F_{s,r}$ and $r \not\in F_{s,r}$. It is now immediate that, $F = \bigcup_{s \in F} \bigcap_{r \not\in F} F_{s,r}$. Hence, since inverse image commutes with Boolean operation, we obtain, $\alpha\inv(F) = \bigcup_{s \in F} \bigcap_{r \not\in F} \alpha\inv(F_{s,r})$. This yields $\alpha\inv(F) \in \Cs$ since \Cs is a Boolean algebra.
%\end{proof}

%% file: polc.tex
We introduce the operators that we investigate in this paper. We first recall the definition of standard polynomial closure. Then, we define four \emph{semantic} restrictions

\subsection{Polynomial closure}

Given finitely many languages $L_0,\dots,L_n \subseteq A^*$, a \emph{marked product} of $L_0,\dots,L_n$ is a product of the form $L_0a_1L_1 \cdots a_n L_n$ where $a_1,\dots,a_n \in A$. Note that a single language $L_0$ is a marked product (this is the case $n = 0$). In the case $n = 1$ (\emph{i.e.}, there are two languages), we speak of \emph{marked concatenations}. 

The \emph{polynomial closure} of a class \Cs, denoted by \pol{\Cs}, is the class containing all \emph{finite unions} of marked products $L_0a_1L_1\cdots a_nL_n$ such that $L_0,\dots, L_n \in \Cs$. If \Cs is a \vari, \pol{\Cs} is a quotient-closed lattice (this is due to Arfi~\cite{arfi87}, see also~\cite{jep-intersectPOL,PZ:generic18} for recent proofs). On the other hand, \pol{\Cs} need not be closed under complement. Hence, it is natural to combine \poln with another operator. The Boolean closure of a class \Ds, denoted by \bool{\Ds}, is the least Boolean algebra containing \Ds. Finally, we write \bpol{\Cs} for \bool{\pol{\Cs}}. The following proposition is standard (see~\cite[Theorem~29]{PZ:generic18} for example).

\begin{proposition} \label{prop:bpvar}
	If \Cs is a \vari, then so is \bpol{\Cs}.
\end{proposition}

We do not investigate \bpoln itself. Yet, we use the classes \bpol{\Cs} as inputs for the operators that we do investigate. More precisely, we are mainly interested in all input classes of the form \bpol{\Gs} and \bpol{\Gs^+} where \Gs is a group \vari. They will be important for logical applications  (we detail this point in Section~\ref{sec:logic}).  In this context, we shall use the following result of~\cite{pz:csr} concerning membership for the classes \bpol{\Gs} and \bpol{\Gs^+}.

\begin{theorem}[\cite{pz:csr}] \label{thm:bpgm}
	Let \Gs be a group \vari with decidable separation. Then, membership is decidable for \bpol{\Gs} and \bpol{\Gs^+}.
\end{theorem}

\begin{remark} \label{rem:bpgm1}
	Theorem~\ref{thm:bpgm} is based on generic algebraic characterizations of the classes \bpol{\Gs} and \bpol{\Gs^+}. More precisely, it is shown that a regular language belongs to \bpol{\Gs} (resp. \bpol{\Gs^+}) if and only if its syntactic morphism satisfies a specific equation which depends on its \Gs-pairs. Since computing \Gs-pairs boils down to \Gs-separation, this is why membership for \bpol{\Gs} and \bpol{\Gs^+} is tied to separation for \Gs.
\end{remark}

\begin{remark} \label{rem:bpgm2}
	Actually, it is known that when a group \vari \Gs has decidable separation, then \bpol{\Gs} and \bpol{\Gs^+} have decidable separation and covering~\cite{pzconcagroup,pzlevelone}.  This is based on different techniques and we shall not use these results in the paper. 
\end{remark}

\subsection{Deterministic restrictions}

We define weaker variants of \poln by restricting the marked products with specific semantic conditions and the finite unions to \emph{disjoint} ones. %(\emph{i.e.}, the languages in the union must be pairwise disjoint). %The four of them are  restrictions of the \emph{polynomial closure} operator which we first define.

Consider a marked product $K_0a_1K_1\cdots a_nK_n$. Moreover, for each $i$ such that $1 \leq i \leq n$, let $L_i = K_0a_1K_1 \cdots a_{i-1}K_{i-1}$ (in particular, $L_1 = K_0$) and $R_i = K_ia_{i+1}K_{i+1} \cdots a_{n}K_n$ (in particular, $R_n = K_n$). We say that,
\begin{itemize}
	\item $K_0a_1K_1\cdots a_nK_n$ is \emph{left deterministic} if and only if for all $i \leq n$, we have $L_i \cap L_ia_iA^* = \emptyset$.
	\item $K_0a_1K_1\cdots a_nK_n$ is \emph{right deterministic} if and only if for all $i \leq n$, we have $R_i \cap A^*a_iR_i = \emptyset$.
	\item $K_0a_1K_1\cdots a_nK_n$ is \emph{mixed deterministic} if and only if for all $i \leq n$, either $L_i \cap L_ia_iA^* = \emptyset$ or $R_i \cap A^*a_iR_i = \emptyset$.
	\item $K_0a_1K_1\cdots a_nK_n$ is \emph{unambiguous} if and only if for every word $w \in K_0a_1K_1\cdots a_nK_n$, there exists a \emph{unique} decomposition $w = w_0a_1w_1 \cdots a_nw_n$ with $w_i \in K_i$ for $1 \leq i \leq n$.
\end{itemize}
These notions depend on the product itself and not only on the resulting language. For example, the product $A^*aA^*$ is not unambiguous and $(A \setminus \{a\})^* a A^*$ is left deterministic. Yet, they evaluate to the same language. Clearly, left/right deterministic products are also mixed deterministic. One may also verify that mixed deterministic products are unambiguous. 

%\begin{lemma} \label{lem:duna}
%	Every marked product which is left, right or mixed deterministic is also unambiguous.
%\end{lemma}

%\begin{proof}
%	We fix a mixed deterministic marked product $L_0a_1L_1 \cdots a_n L_n$ for the proof. We prove that it is unambiguous. We fix $w_i,w'_i \in L_i$ for every $i \leq n$. Assuming that $w_0a_1w_1 \cdots a_n w_n = w'_0a_1w'_1 \cdots a_n w'_n$, we show that $w_i = w'_i$ for every $i \leq n$. By definition, for $1 \leq i \leq n$, the marked concatenation $(L_0a_1L_1 \cdots a_{i-1}L_{i-1})a_i(L_ia_{i+1}L_{i+1} \cdots a_{n}L_n)$ is either left \emph{or} right deterministic. Hence, it is also unambiguous which implies that $w_0a_1w_1 \cdots a_{i-1}w_{i-1} = w'_0a_1w'_1 \cdots a_{i-1}w'_{i-1}$ for $1 \leq i \leq n$. The case $i = 1 $ exactly says that $w_0 = w'_0$ and it then follows from a simple induction that $w_i = w'_i$ for every $i \leq n$.
%\end{proof}

\begin{remark} \label{rem:hdet:conv}
	A mixed deterministic product needs not be left or right deterministic. Let $L_1 = (ab)^+$, $L_2 = c^+$ and $L_3 = (ba)^+$. The product $L_1cL_2cL_3$ is mixed deterministic since $L_1 \cap L_1cA^* = \emptyset$ and $L_3 \cap A^*cL_3 = \emptyset$. However, it is neither left deterministic nor right deterministic. Similarly, a unambiguous product need not be mixed deterministic. If  $L_4 = (ca)^+$, the product $L_1aL_4$ is unambiguous but it neither left nor right deterministic.
\end{remark}

%\begin{remark}
%	Clearly, a left (resp. right, mixed) \Cs-deterministic marked product is also left (resp. right, mixed) deterministic. However, the converse is not true in general. For example, if \Cs is the trivial \varie \stzer (recall that $\stzer(A) = \{\emptyset,A^*\}$ for every alphabet $A$), no marked product can be \Cs-deterministic.
%\end{remark} 

The \emph{left polynomial closure} of a class \Cs, written \ldet{\Cs}, contains the \emph{finite disjoint unions} of \emph{left deterministic marked products} $L_0a_1L_1 \cdots a_nL_n$ such that $L_0, \dots,L_n \in \Cs$. By ``disjoint'' we mean that the languages in the union must be pairwise disjoint. The \emph{right polynomial closure} of \Cs (\rdet{\Cs}), the \emph{mixed polynomial closure} of \Cs (\mdet{\Cs}) and the \emph{unambiguous polynomial closure} of \Cs (\upol{\Cs}) are defined analogously by replacing the ``left deterministic'' requirement on marked products by the appropriate one. The following lemma can be verified from the definition.

\begin{lemma} \label{lem:opcomp}
	Let \Cs be a class. Then, we have $\ldet{\Cs} \subseteq \mdet{\Cs}$, $\rdet{\Cs} \subseteq \mdet{\Cs}$ and $\mdet{\Cs} \subseteq \upol{\Cs} \subseteq \pol{\Cs}$.
\end{lemma}

The operators \ldeto, \rdeto and \upolo are standard. See for example~\cite{schul,Pin80,Pinambigu}. In particular, they admit the following alternate definition (see~\cite{jep-intersectPOL} for a proof).

\begin{lemma} \label{lem:dleast}
	Let \Cs be a class. Then, \ldet{\Cs} (resp. \rdet{\Cs}, \upol{\Cs}) is the least class containing \Cs which is closed under disjoint union and left deterministic (resp. right deterministic, unambiguous) marked concatenation.
\end{lemma}

On the other hand, \mdeto is new. It is arguably the key notion of the paper. In particular, the application to two-variable first-order logic is based on it (see Section~\ref{sec:logic}). Unfortunately, it is less robust than the other operators: no result similar to Lemma~\ref{lem:dleast} is known for \mdeto. In particular, it is not idempotent: in general \mdet{\Cs} is \emph{strictly included} in \mdet{\mdet{\Cs}}. Actually several of our results are based on this fact. This is because  a mixed product of mixed products is \emph{not} a mixed product itself in general.

\begin{example}
	Let $A = \{a,b,c\}$, $L_0 = b^+$, $L_1 = a^+$ and $K = (a+b+c)^+$. Clearly, $L_0bL_1$ and $K$ are defined by mixed deterministic products. Also, if $L = L_0bL_1$, then $LcK$ is mixed deterministic. Yet, the combined product $L_0bL_1cK$ is not mixed deterministic itself. Indeed, the marked concatenation $(L_0)b(L_1cK)$ is neither left deterministic nor right deterministic.
\end{example}

Note that \upolo is well-understood. We shall use two key results from~\cite{pzupol2}. While this is not apparent on the definition, \upol{\Cs} has robust properties.

\begin{theorem}[\cite{pzupol,pzupol2}] \label{thm:pupol}
	If \Cs is a \vari, then so is \upol{\Cs}.
\end{theorem}

%Moreover, we have the following generic characterization of the \upol{\Cs}-morphisms.

\begin{theorem}[\cite{pzupol,pzupol2}]  \label{thm:cupol}
	Let \Cs be a \vari and $\alpha: A^* \to M$ a surjective morphism. The following are equivalent:
	\begin{enumerate}[label=\alph*)]
		\item $\alpha$ is a \upol{\Cs}-morphism.
		\item $s^{\omega+1} = s^\omega t s^\omega$ for all \Cs-pairs $(s,t) \in M^2$.
		\item $s^{\omega+1} = s^\omega t s^\omega$ for all $s,t \in M$ such that $s \canec t$.
	\end{enumerate}
\end{theorem}

By Fact~\ref{fct:eqmemb}, the equivalence \canec can be computed from $\alpha$ when \Cs-membership is decidable. Hence, by Proposition~\ref{prop:synmemb}, Theorem~\ref{thm:cupol} implies that \upol{\Cs}-membership is also decidable in this case. We prove similar results for \ldeto, \rdeto and \mdeto in Section~\ref{sec:carac}.

%% file: framework.tex
We introduce a framework designed to manipulate \ldeto, \rdeto and \mdeto in proof arguments. We first define equivalences relations over $A^*$. We then show that for every \vari \Cs, they characterize the languages within \ldet{\Cs}, \rdet{\Cs} and \mdet{\Cs} in terms of \Cs-morphisms. Here, we present a first application by generalizing Theorem~\ref{thm:pupol} to \ldeto, \rdeto and \mdeto.

\subsection{Preliminaries}

We first introduce terminology and results that we shall use to define and manipulate our equivalence relations. Given a \emph{surjective} morphism $\eta: A^* \to N$ and $k \in \nat$, we use the Green relations of $N$ to associate three sets of positions to every $w \in A^*$. Let $w = a_1 \cdots a_\ell \in A^*$ with $a_1,\dots,a_\ell \in A$. We define two sets $\poslek{w} \subseteq \posc{w}$ and $\posrek{w} \subseteq \posc{w}$ by induction on $k$. When $k = 0$, we define $\poslp{\eta}{0}{w}=\posrp{\eta}{0}{w}=\emptyset$. Assume now that $k \geq 1$ and let $i \in \posc{w}$. We let,
\begin{itemize}
	\item $i \in \poslek{w}$ if and only if there exists $j \in \poslp{\eta}{k-1}{w} \cup \{0\}$ such that $j < i$ and $\eta(\infix{w}{j}{i}a_i) \Rords \eta(\infix{w}{j}{i})$.
	\item $i \in \posrek{w}$ if and only if there exists $j \in \posrp{\eta}{k-1}{w} \cup \{|w|+1\}$ such that $i < j$ and $\eta(a_i\infix{w}{i}{j}) \Lords \eta(\infix{w}{i}{j})$.
\end{itemize}
Finally, we define $\posmek{w} = \poslek{w} \cup \posrek{w}$ for every $k \in \nat$. We complete the definition with a key lemma.   In practice, we often consider the three sets \poslak{w}, \posrak{w} and \posmak{w} in the special case when $\alpha: A^*\to M$ is a \upol{\Cs}-morphism. The lemma states that in the case, all three sets can be specified using only a \Cs-morphism: there exists a \Cs-morphism $\eta: A^* \to N$ and $k' \geq k$ such the sets are included in \poslp{\eta}{k'}{w}, \posrp{\eta}{k'}{w} and \posmp{\eta}{k'}{w}. The proof is based on Theorem~\ref{thm:cupol}. 

\begin{lemma} \label{lem:isdet}
	Let \Cs be a \vari and $\alpha: A^* \to M$ a \upol{\Cs}-morphism. For every $k \in \nat$ and $w \in A^*$, $\poslp{\alpha}{k}{w} \subseteq \poslp{\ctype{\cdot} \circ \alpha}{k|M|}{w}$ and $\posrp{\alpha}{k}{w} \subseteq \posrp{\ctype{\cdot} \circ \alpha}{k|M|}{w}$.
\end{lemma}

\begin{proof}
	We write $N = {M}/{\canec}$ and $\eta = \ctype{\cdot} \circ \alpha: A^* \to N$ for the proof. We show that $\poslak{w} \subseteq \poslp{\eta}{k|M|}{w}$ for all $w \in A^*$ and $k\in\nat$. The other inclusion is symmetrical and left to the reader. Let  $a_1,\dots,a_\ell \in A$ be the letters such that $w = a_1 \cdots a_\ell$. We use induction on $k$. If $k =0$, then $\poslp{\alpha}{0}{w} = \poslp{\eta}{0}{w} = \emptyset$. Assume now that $k \geq 1$ and let $i \in \poslak{w}$. We show that $i \in \poslp{\eta}{k|M|}{w}$. By definition, there is \mbox{$j \in \poslp{\alpha}{k-1}{w} \cup \{0\}$} such that $j < i$ and $\alpha(\infix{w}{j}{i}a_i) \Rords \alpha(\infix{w}{j}{i})$. By induction, we get $j \in \poslp{\eta}{(k-1)|M|}{w} \cup \{0\}$. Let $i_1,\dots,i_n \in \posc{w}$ be \emph{all} the positions in $w$ which satisfy $j<i_1< \cdots <i_n$ and $\alpha(\infix{w}{j}{i_h}a_{i_h}) \Rords \alpha(\infix{w}{j}{i_h})$ for $1 \leq h \leq n$. Note that $n \leq |M|$ by definition. Since $i \in \{i_1,\dots,i_n\}$ by hypothesis, it now suffices to prove that $i_1,\dots,i_n \in \poslp{\eta}{k|M|}{w}$. We write $i_0 = j$. For every $h$ such that $1 \leq h \leq n$, we prove that $\eta(\infix{w}{i_{h-1}}{i_h}a_{i_h}) \Rords \eta(\infix{w}{i_{h-1}}{i_h})$. Since we have $i_0 = j \in \poslp{\eta}{k|M| - |M|}{w} \cup \{0\}$ and $n \leq |M|$, this implies that $i_1,\dots,i_n\in \poslp{\eta}{k|M|}{w}$ by definition.
	
	We proceed by contradiction. Assume that there exists an index $1 \leq h \leq n$ such that $\eta(\infix{w}{i_{h-1}}{i_h}a_{i_h}) \Rrel \eta(\infix{w}{i_{h-1}}{i_h})$. We write $u = \infix{w}{j}{i_{h-1}}a_{i_{h-1}}$ and $v = \infix{w}{i_{h-1}}{i_h}$. Our contradiction hypothesis states that $\eta(va_{i_h}) \Rrel \eta(v)$. We get $y \in A^*$ such that $\eta(va_{i_h}y) = \eta(v)$. Moreover, $\alpha(uva_{i_h}) \Rords \alpha(uv) \Rrel \alpha(u)$ by definition of $i_1,\dots,i_n$. Hence, we get a word $z \in A^*$ such that $\alpha(uvz) = \alpha(u)$. Since $\eta(va_{i_h}y) = \eta(v)$, we have $\eta(va_{i_h}yz) = \eta(vz)$, \emph{i.e.} $\alpha(va_{i_h}yz) \canec \alpha(vz)$ by definition of $\eta$. Therefore, since $\alpha$ is a \upol{\Cs}-morphism, it follows from Theorem~\ref{thm:cupol} that $(\alpha(vz))^{\omega+1} = (\alpha(vz))^{\omega} \alpha(va_{i_h}yz) (\alpha(vz))^{\omega}$. We multiply on the left by $\alpha(u)$. Since $\alpha(uvz) = \alpha(u)$, we get $\alpha(u) = \alpha(u)\alpha(va_{i_h}yz) (\alpha(vz))^{\omega}$. Hence, we obtain $\alpha(uv) \Rord \alpha(uva_{i_h})$, contradicting the hypothesis that $\alpha(uva_{i_h}) \Rords \alpha(uv)$.
\end{proof}

\newcommand{\snp}[2]{\ensuremath{\sigma_{#1}(#2)}\xspace}
\newcommand{\snpe}[1]{\snp{\eta}{#1}}

We turn to a second independent notion that we shall use conjointly with the first one. Let $\eta: A^* \to N$ be a surjective morphism. Given a word $w = a_1 \cdots a_{\ell} \in A^*$ and a set $P \subseteq\posc{w}$, we use $\eta$ to associate a tuple in $N \times (A \times N)^{|P|}$ that we call the \emph{$\eta$-snapshot} of $(w,P)$. Let $m = |P|$ and $i_1 < \cdots < i_{m}$ be the positions such that $P = \{i_1,\dots,i_{m}\}$. Finally, we let $i_0=0$ and $i_{m+1}=|w|+1$. For $0 \leq h \leq m$, we let $s_h = \eta(\infix{w}{i_h}{i_{h+1}}) \in N$. The $\eta$-snapshot of $(w,P)$, denoted by $\snpe{w,P}$, is the following tuple:
\[
\snpe{w,P} = (s_0,a_{i_1},s_1,\dots, a_{i_{m}},s_{m}) \in N \times (A \times N)^{m}.
\]
We complete the definition with a result that will be useful when manipulating $\eta$-snapshots in proof arguments.

\begin{fct} \label{fct:snap}
	Let $\eta: A^* \to N$ be a surjective morphism, $w,w' \in A^*$, $P \subseteq \posc{w}$ and $P' \subseteq \posc{w'}$. Assume that $\sigma_\eta(w,P) = \sigma_\eta(w',P')$ and let $P_1,P_2 \subseteq P$ such that $P_1 \cup P_2=P$. There exist $P'_1,P'_2 \subseteq P'$ such that $P'_1 \cup P'_2 = P'$, $\sigma_\eta(w,P_1) = \sigma_\eta(w',P'_1)$ and $\sigma_\eta(w,P_2) = \sigma_\eta(w',P'_2)$.
\end{fct}

\begin{proof}
	Since $\sigma_\eta(w,P) = \sigma_\eta(w',P')$, we have $|P|=|P'|$. Hence, there exists a unique increasing bijection $f: P \to P'$ (by ``increasing'', we mean that $i < j \Rightarrow f(i) < f(j)$ for every $i,j \in P$). We let $P'_1 = f(P_1)$ and $P'_2 = f(P_2)$. Clearly, we have $P'_1 \cup P'_2 = P'$ since $P_1 \cup P_2 = P$. One may then verify using our hypothesis on $(w,P)$ and $(w',P')$ that $\sigma_\eta(w,P_1) = \sigma_\eta(w',P'_1)$ and $\sigma_\eta(w,P_2) = \sigma_\eta(w',P'_2)$.
\end{proof}

Finally, we connect these two notions to the operators \ldeto, \rdeto and \mdeto.

\begin{lemma} \label{lem:detprod}
	Let $\eta: A^* \to N$ be a morphism, $w \in A^*$ and $k \in \nat$. Let $P$ be the set \poslek{w} (resp. \posrek{w}, \posmek{w}) and $(s_0,a_1,s_1,\dots,a_n,s_n) = \sigma_\eta(w,P)$. Then, the marked product $\eta\inv(s_0)a_1\eta\inv(s_1) \cdots a_n\eta\inv(s_n)$ is left (resp. right, mixed) deterministic.
\end{lemma}

\begin{proof}
	We treat the case $P = \posmek{w}$ (the other two are similar and left to the reader). For each $h$ such that $1\leq h \leq n$, we let $U_h = \eta\inv(s_0)a_1\eta\inv(s_1) \cdots a_{h-1}\eta\inv(s_{h-1})$ and $V_h = \eta\inv(s_h)a_{h+1} \cdots \eta\inv(s_{n-1})a_n\eta\inv(s_n)$. We have to show that for each such $h$, either $U_h \cap U_ha_hA^* = \emptyset$ or $V_h \cap A^*a_hV_h = \emptyset$. Let \mbox{$i_1 < \cdots < i_n$} such that $\posmek{w} = \{i_1,\dots,i_n\}$ ($i_h$ has label $a_h$). By definition of \posmek{w}, we know that either $i_h \in \poslek{w}$ or $i_h\in\posrek{w}$ for $1 \leq h \leq n$. In the former case, one may prove that $U_h \cap U_ha_hA^* = \emptyset$ and in the latter case, one may prove that $V_h \cap A^*a_hV_h = \emptyset$. By symmetry, we only prove the former property. Let $h$ such that $1 \leq h \leq n$ and assume that $i_h \in \poslek{w}$. We use induction on the least number $m$ such that $i_h \in \poslp{\eta}{m}{w}$ to show that $U_h \cap U_ha_hA^* = \emptyset$.
	
	By definition, we get $j \in \poslp{\eta}{m-1}{w} \cup \{0\}$ such that $\eta(\infix{w}{j}{i_h}a_h) \Rords \eta(\infix{w}{j}{i_h})$. Let $q = \eta(\infix{w}{j}{i_h})$. Observe that $\eta\inv(q)a_hA^* \cap \eta\inv(q) = \emptyset$. Indeed, otherwise we get $x \in A^*$ such that $q = q\eta(a_h)\eta(x)$ which contradicts $q\eta(a_h) \Rords q$. This concludes the proof when $j = 0$. Since $q = \eta(\infix{w}{0}{i_h})$ in this case, one may verify that $U_h \subseteq \eta\inv(q)$. Hence, we get $U_h \cap U_ha_hA^* = \emptyset$. Assume now that $j \neq 0$. Hence, $j \in \poslp{\eta}{m-1}{w}$ which implies that $j = i_g$ for some $g \leq h$. By induction, $U_g \cap U_ga_gA^* = \emptyset$. We use contradiction to prove that $U_h \cap U_ha_hA^* = \emptyset$. Assume that there exists $u \in U_h \cap U_ha_hA^*$. Since $q = \eta(\infix{w}{i_g}{i_h})$, one may verify that $U_h \subseteq U_ga_g\eta\inv(q)$. Hence, we get $x,x' \in U_g$, $y,y' \in \eta\inv(q)$ and $z \in A^*$ such that $u = xa_gya_hz = x'a_gy'$. Since we have $U_g \cap U_ga_gA^* = \emptyset$, this yields $x = x'$. Thus, $ya_hz = y'$. This is a contradiction since $\eta\inv(q)a_hA^* \cap \eta\inv(q) = \emptyset$. 
\end{proof}

\subsection{Equivalence relations} We may now define our equivalences. Consider a surjective morphism $\eta: A^* \to N$. For every $k\in\nat$, we associate three equivalence relations \eqlek, \eqrek and \eqmek on $A^*$. Consider $u,v \in A^*$. We define,
\begin{itemize}
	\item $u \eqlek v$ if and only if $\snpe{u,\poslek{u}} = \snpe{v,\poslek{v}}$.
	\item $u \eqrek v$ if and only if $\snpe{u,\posrek{u}} = \snpe{v,\posrek{v}}$.
	\item $u \eqmek v$ if and only if $\snpe{u,\posmek{u}} = \snpe{v,\posmek{v}}$.
\end{itemize} 
It is immediate by definition that \eqlek, \eqrek and \eqmek are equivalence relations. Moreover, they have finite index. For example, consider \eqmek. By definition, the \eqmek-class of a word $w\in A^*$ is determined by the $\eta$-snapshot \snpe{w,\posmek{w}}. One may verify using induction on $k$ that $|\posmek{w}| \leq 2|N|^{k}$. Since this bound depends only on $\eta$ and $k$ (and not on $w$), it follows that there finitely many possible $\eta$-snapshot \snpe{w,\posmek{w}} for $w \in A^*$. Thus, \eqmek has finite index. For every $w \in A^*$, we shall write $\lhtyp{w}{\eta,k} \subseteq A^*$ for the \eqlek-class of $w$, $\rhtyp{w}{\eta,k} \subseteq A^*$ for the \eqrek-class of $w$ and $\mhtyp{w}{\eta,k} \subseteq A^*$ for the \eqlek-class of $w$. 

\begin{lemma} \label{lem:finiteind}
	If $\eta: A^* \to N$ is a surjective morphism and $k\in \nat$, then \eqlek, \eqrek and \eqmek are equivalences of finite index.
\end{lemma}

We complete the definition with a key technical lemma that we shall use whenever we need to prove that two words are equivalent for \eqlek, \eqrek or \eqmek.

\begin{lemma} \label{lem:eqb}
	Let $\eta: A^* \to N$ be a surjective morphism, $k \in \nat$ and $\xvar \in \{\rhd,\lhd,\bowtie\}$. Let $w,w' \in A^*$ and $P' \subseteq \posc{w'}$. If $\snpe{w,\posxek{w}} = \snpe{w',P'}$, then $P' = \posxek{w'}$.
\end{lemma}

\begin{proof}
	First, note that the case $\xvar = {\bowtie}$ is a corollary of the other two. Indeed, assume for now that they hold and that we have $\sigma_\eta(w,\posmek{w}) = \sigma_\eta(w',P')$. By definition, we know that $\posmek{w} = \poslek{w} \cup \posrek{w}$. Consequently, Fact~\ref{fct:snap} yields $P'_1,P'_2 \subseteq P'$ which satisfy $P' = P'_1 \cup P'_2$, $\sigma_\eta(w,\poslek{w}) = \sigma_\eta(w',P'_1)$ and $\sigma_\eta(w,\posrek{w}) = \sigma_\eta(w',P'_2)$. Hence, the cases when $\xvar \in \{\rhd,\lhd\}$ yield $P'_1 = \poslek{w'}$ and $P'_2 = \posrek{w'}$. We get $P' = \poslek{w'} \cup \posrek{w'} = \posmek{w'}$ as desired.
	
	We now treat the case when $\xvar= {\rhd}$ (the symmetrical case $\xvar = {\lhd}$ is left to the reader). Let $w,w' \in A^*$ and $a_1,\dots,a_m,b_1,\dots,b_n \in A$ such that $w = a_1 \cdots a_m$ and $w' = b_1\cdots b_{n}$. We assume that $\sigma_\eta(w,\poslek{w}) = \sigma_\eta(w',P')$ and prove that $P' = \poslek{w'}$. We have $|\poslek{w}| = |P'|$ by hypothesis. Hence, we may consider the unique increasing bijection $f: \poslek{w} \to P'$ (by ``increasing'', we mean that $i < j \Rightarrow f(i) < f(j)$ for all $i,j$). We extend it to the unlabeled positions $0$ and $|w|+1$ by defining $f(0) = 0$ and $f(|w|+1) = |w'|+1$. The following two properties can be verified from our hypotheses:
	\begin{enumerate}
		\item\label{itm:hdet:geq1} for all $i \in \poslek{w}$, we have $a_i = b_{f(i)}$ ($i$ and $f(i)$ have the same label), and,
		\item\label{itm:hdet:geq2} for all $i,j \in \poslek{w} \cup \{0,|w|+1\}$, if $i < j$, then $\eta(\infix{w}{i}{j}) = \eta(\infix{w'}{f(i)}{f(j)})$.
	\end{enumerate}
	First, we show that $P' \subseteq \poslek{w'}$. Let $h \leq k$. We use induction on $h$ to prove that for all $i \in \poslp{\eta}{h}{w}$, we have $f(i) \in \poslp{\eta}{h}{w'}$. Since $f$ is surjective, the case \mbox{$h = k$} yields \mbox{$P' \subseteq \poslek{w'}$}. Let $i \in \poslp{\eta}{h}{w}$. By definition, $h \geq 1$ and there is \mbox{$j \in \poslp{\eta}{h-1}{w} \cup \{0\}$} such that $j < i$ and $\eta(\infix{w}{j}{i}a_i) \Rords \eta(\infix{w}{j}{i})$. We have $f(j) < f(i)$ since $f$ is increasing. Moreover we know that $f(j)\in \poslp{\eta}{h-1}{w'} \cup \{0\}$ by induction. We know that $a_{i} = b_{f(i)}$ and $\eta(\infix{w}{j}{i}) = \eta(\infix{w'}{f(j)}{f(i)})$. Consequently, we obtain that $\eta(\infix{w'}{f(j)}{f(i)}b_{f(i)}) \Rords \eta(\infix{w'}{f(j)}{f(i)})$ which yields $f(i) \in \poslp{\eta}{h}{w'}$ as desired.
	
	We now prove that $\poslek{w'} \subseteq P'$. Let $h \leq k$. Using induction on $h$, we prove that for all $i' \in \poslp{\eta}{h}{w'}$, there is $i \in \poslp{\eta}{h}{w}$ such that $i' = f(i)$. This implies $\poslek{w'} \subseteq P'$ as desired. We fix $i'\in \poslp{\eta}{h}{w'}$. By definition, $h \geq 1$, and there exists $j' \in \poslp{\eta}{h-1}{w'} \cup \{0\}$ such that $j' < i'$ and $\eta(\infix{w'}{j'}{i'}b_{i'}) \Rords \eta(\infix{w'}{j'}{i'})$. Induction yields a position $j\in\poslp{\eta}{h-1}{w} \cup \{0\}$ such that $j' = f(j)$. Let $i_1,\dots,i_p$ be \emph{all} positions of $w$ such that $j < i_1 < \cdots < i_p$ and $\eta(\infix{w}{j}{i_\ell}a_{i_\ell}) \Rords \eta(\infix{w}{j}{i_\ell})$ for $1 \leq \ell \leq n$. Since we have $j \in \poslp{\eta}{h-1}{w} \cup \{0\}$, we get $i_1,\dots,i_n \in \poslp{\eta}{h}{w}$. Thus, it suffices to prove that $i' = f(i_\ell)$ for some $\ell \leq p$.  We proceed by contradiction. Assume that $i' \neq f(i_\ell)$ for $1 \leq \ell \leq p$. For the proof, we write $i_0 = j$ and $i_{p+1} = |w|+1$. Clearly, we have $i_0 < i_1 < \cdots < i_{p+1}$ which implies that $f(i_0) < f(i_1) < \cdots < f(i_{p+1})$. Hence, by hypothesis on $i'$ and since $f(i_0) = j'< i'$, there exists $\ell$ such that \mbox{$0 \leq \ell \leq n$} and $f(i_\ell) < i' < f(i_{\ell+1})$. By definition of $i_1,\dots,i_p$, we have $\eta(\infix{w}{j}{i_\ell}a_{i_\ell}) \Rrel \eta(\infix{w}{j}{i_{\ell+1}})$. Since \mbox{$j' = f(j)$}, we get $\eta(\infix{w'}{j'}{f(i_\ell)}b_{f(i_\ell)}) \Rrel \eta(\infix{w}{j'}{f(i_{\ell+1})})$. Therefore, since $f(i_\ell) < i' < f(i_{\ell+1})$, we get \mbox{$\eta(\infix{w'}{j'}{i'}) \Rrel \eta(\infix{w}{j'}{i'}b_{i'})$}. This is a contradiction since $\eta(\infix{w'}{j'}{i'}b_{i'}) \Rords \eta(\infix{w'}{j'}{i'})$.
\end{proof}

Lemma~\ref{lem:eqb} has an important consequence for the equivalences \eqlek, \eqrek and \eqmek. Indeed, we have the following immediate corollary.

\begin{corollary} \label{cor:eqbij}
	Let $\eta: A^* \to N$ be a surjective morphism, $k \in \nat$ and $\xvar \in \{\rhd,\lhd,\bowtie\}$. For every $w,w' \in A^*$, we have $w \mathrel{\xvar_{\eta,k}} w'$ if and only if there exists $P' \subseteq \posc{w'}$ such that  $\snpe{w,\posxek{w}} = \snpe{w',P'}$.
\end{corollary}

We use Corollary~\ref{cor:eqbij} to prove a first useful result concerning these equivalences: the three of them are congruences.

\begin{lemma} \label{lem:areparts}
	If $\eta: A^* \to N$ is a surjective morphism and $k\in \nat$, then \eqlek, \eqrek and \eqmek are congruences.
\end{lemma}

\begin{proof}
	We present a proof for \eqmek (the arguments for \eqlek and \eqrek are identical). Let $u_1,u_2,v_1,v_2 \in A^*$ such that $u_h \eqmek v_h$ for $h= 1,2$. We prove that $u_1u_2 \eqmek v_1v_2$. Let $P$ be the set of all positions $i \in \posc{u_1u_2}$ such that $i \in \posmek{u_1}$ or $i - |u_1| \in \posmek{u_2}$. Symmetrically, let $Q$ be the set of all positions $i \in \posc{v_1v_2}$ such that either $i \in \posmek{v_1}$ or $i - |v_1| \in \posmek{v_2}$. By hypothesis, $\sigma_\eta(u_h,\posmek{u_h}) = \sigma_\eta(v_h,\posmek{v_h})$ for $h =1,2$ which implies that $\sigma_\eta(u_1u_2,P) = \sigma_\eta(v_1v_2,Q)$ by definition. Also, one may verify from that $\posmek{u_1u_2} \subseteq P$. This yields $Q' \subseteq Q$ such that $\sigma_\eta(u_1u_2,\posmek{u_1u_2}) = \sigma_\eta(v_1v_2,Q')$ by Fact~\ref{fct:snap}. Hence, $u_1u_2 \eqmek v_1v_2$ as desired by Corollary~\ref{cor:eqbij}.
\end{proof}

\subsection{Application to \ldeto, \rdeto and \mdeto}

We are ready to characterize the classes built with \ldeto, \rdeto and \mdeto using these three equivalences. %We do so in the following proposition.

\begin{proposition} \label{prop:opcar}
	Let \Cs be a \vari and $L \subseteq A^*$. Then, we have $L \in \ldet{\Cs}$ (resp. $L \in \rdet{\Cs}$, $L \in \mdet{\Cs}$) if and only if there exist a \Cs-morphism $\eta: A^* \to N$ and $k \in \nat$ such that $L$ is a union of \eqlek-classes (resp. \eqrek-classes, \eqmek-classes).
\end{proposition}

\begin{proof}
	We present a proof argument for \mdet{\Cs} (the other cases are similar and left to the reader). Assume first that $L \in \mdet{\Cs}$. We exhibit a \Cs-morphism $\eta: A^* \to N$ and $k \in \nat$ such that $L$ is a union of \eqmek-classes. By definition of \mdet{\Cs}, there exists a \emph{finite} set \Hb of languages in \Cs and $m \geq 1$ such that $L$ is a finite disjoint union of mixed deterministic marked products of at most $m$ languages in \Hb. By definition, every unambiguous product of languages in \Hb belongs to \upol{\Cs}. Hence, since \upol{\Cs} is a \vari by Theorem~\ref{thm:pupol}, Proposition~\ref{prop:genocm} yields a \upol{\Cs}-morphism $\alpha: A^* \to M$ recognizing every unambiguous marked product of at most $m$ languages in \Hb. Consider the congruence \canec on $M$. We let $N = {M}/{\canec}$ and $\eta= \ctype{\cdot} \circ \alpha: A^* \to N$ and $k = |M|$.  Lemma~\ref{lem:smult} implies that $\eta$ is a \Cs-morphism. Moreover, since all $H \in \Hb$ belong to \Cs and are recognized by $\alpha$ (by definition), the lemma also implies that $\eta$ recognizes every $H \in \Hb$. It remains to prove that $L$ is a union of \eqmek-classes. For all $w,w' \in A^*$ such that $w \eqmek w'$, we prove that $w \in L \Leftrightarrow w' \in L$. By symmetry, we only prove one implication: assuming that $w \in L$, we prove that $w' \in L$.
	
	Since $w \in L$, the definitions of \Hb and $m$ yield $H_0,\dots, H_n \in \Hb$ and $a_1,\dots,a_n \in A$ such that $n+1 \leq m$, $w\in H_0a_1H_1 \cdots a_nH_n \subseteq L$ and $H_0a_1H_1 \cdots a_nH_n$ is mixed deterministic. It now suffices to prove $w' \in H_0a_1H_1 \cdots a_nH_n$. Since $w \in H_0a_1H_1 \cdots a_nH_n$, we get $w_j \in H_j$ for $0 \leq j \leq n$ such that $w = w_0a_1w_1 \cdots a_n w_n$. Let $P \subseteq \posc{w}$ be the set of all positions carrying the letters $a_1,\dots,a_n$. We prove that $P \subseteq \posmek{w}$. Let us first explain why this implies $w' \in H_0a_1H_1 \cdots a_nH_n$. Assume for now that $P \subseteq \posmek{w}$. Since $w \eqmek w'$, Fact~\ref{fct:snap} yields a set $P' \subseteq \posmek{w'}$ such that $\sigma_\eta(w,P) = \sigma_\eta(w',P')$. By definition of $P$, this exactly says that $w'$ admits a decomposition $w' = w'_0a_1w'_1 \cdots a_n w'_n$ such that $\eta(w'_j) = \eta(w_j)$ for every $j \leq n$. Since $H_0,\dots,H_n \in \Hb$ are recognized by $\eta$ and $w_j\in H_j$ for every $j \leq n$, this yields $w'_j \in H_j$ for every $j \leq n$. Therefore, we get $w' \in H_0a_1H_1 \cdots a_nH_n \subseteq L$ as desired.
	
	It remains to prove that $P \subseteq \posmek{w}$. Since \mbox{$\alpha: A^* \to M$} is a \upol{\Cs}-morphism and $k = |M|$, Lemma~\ref{lem:isdet} yields $\posmp{\alpha}{1}{w} \subseteq \posmek{w}$. We prove that $P \subseteq \posmp{\alpha}{1}{w}$. We fix a position $i \in P$ for the proof. By definition of $P$, there exists $j \leq n$ such that the position~$i$ is the one labeled by the highlighted letter $a_j$ in $w = w_0a_1w_1 \cdots a_nw_n$. We let $u = w_0a_1w_1 \cdots w_{j-1} \in H_0a_1H_1\cdots H_{j-1}$. Moreover, we let $v = w_j \cdots a_nw_n \in H_j \cdots a_nH_n$. Clearly, we have $w = ua_jv$. Since $H_0a_1H_1 \cdots a_nH_n$ is mixed deterministic, we know that the marked concatenation $(H_0a_1H_1\cdots H_{j-1}) a_j (H_j \cdots a_nH_n)$ is either left deterministic or right deterministic. By symmetry, we only treat the former case and prove that $i \in \poslp{\alpha}{1}{w}$ (in the latter case, one may prove that $i \in  \posrp{\alpha}{1}{w}$). Consequently, we assume that $(H_0a_1H_1\cdots H_{j-1}) a_j (H_j \cdots a_nH_n)$ is left deterministic. Recall that $i$ is the position carrying the highlighted letter $a_j$ in the decomposition $w = ua_jv$ of $w$. Hence, we have to prove that $\alpha(ua_j) \Rords \alpha(u)$. This will imply $i \in\poslp{\alpha}{1}{w}$ as desired. By contradiction, assume that $\alpha(ua_j) \Rrel \alpha(u)$. This yields $x \in A^*$ such that $\alpha(ua_jx) = \alpha(u)$. By definition of $u$, we have $u \in H_0a_1H_1\cdots H_{j-1}$. Moreover, since the whole product $H_0a_1H_1 \cdots a_nH_n$ is mixed deterministic, one may verify that $H_0a_1H_1\cdots H_{j-1}$ is unambiguous which means that it is recognized by $\alpha$ (it is a unambiguous product of $j \leq n \leq m$ languages in \Hb). Hence, as $\alpha(ua_jx) = \alpha(u)$, we get $ua_jx \in H_0a_1H_1 \cdots H_{j-1}$. Since it is clear that $ua_jx\in H_0a_1H_1\cdots H_{j-1}a_jA^*$, this contradicts the hypothesis that $(H_0a_1H_1\cdots H_{j-1}) a_j (H_j \cdots a_nH_n)$ is left deterministic. This concludes the proof for the left to right implication.

	We turn to the converse implication. We fix a \Cs-morphism $\eta: A^* \to N$ and $k \in \nat$. We prove that every \eqmek-class is defined by a mixed deterministic marked product of languages in \Cs. Since equivalence classes are pairwise disjoint and \eqmek has finite index, this implies that every union of \eqmek-classes belongs to \mdet{\Cs} as desired. We fix $w \in A^*$ and consider its \eqmek-class. We define $\sigma_\eta(w,\posmek{w}) =  (s_0,a_1,s_1,\dots,a_n,s_n)$. Let $L_h = \eta\inv(s_h)$ for every $h \leq n$. We have $L_h \in \Cs$ since $\eta$ is a \Cs-morphism. Let $L = L_0a_1L_1 \cdots a_n L_n$. We know from Lemma~\ref{lem:detprod} that $L_0a_1L_1 \cdots a_n L_n$ is mixed deterministic. Hence, $L \in \mdet{\Cs}$. We show that $L$ is the \eqmek-class of $w$, completing the proof. Let $w' \in A^*$. We prove that $w \eqmek w'$ if and only if $w' \in L$. If $w' \eqmek w$, then $\sigma_\eta(w',\posmek{w'}) = \sigma_\eta(w,\posmek{w})$. Hence, $\sigma_\eta(w',\posmek{w'})=(s_0,a_1,s_1,\dots,a_n,s_n)$ which yields $w' \in L$ by definition of $\eta$-snapshots. Assume now that $w' \in L$. By definition of $L$, we have $w' = w'_0a_1w'_1 \cdots a_nw'_n$ with \mbox{$\alpha(w'_h) = s_h$} for every $h \leq n$. Let $P' \subseteq \posc{w'}$ be the set containing all positions carrying the highlighted letters $a_1,\dots,a_n$. Clearly, $\sigma_\eta(w',P') = (s_0,a_1,s_1,\dots,a_n,s_n)$. Therefore, $\sigma_\eta(w,\posmek{w}) = \sigma_\eta(w',P')$ which yields $w \eqmek w'$ as desired by Corollary~\ref{cor:eqbij}.
\end{proof}

We complete Proposition~\ref{prop:opcar} with a useful technical corollary which strengthens the ``only if'' implication in the statement. %We use it to handle finitely many languages simultaneously.

\begin{corollary} \label{cor:opcar}
	Let \Cs be a \vari and $L_1,\dots,L_m$ finitely many languages in \ldet{\Cs} (resp. \rdet{\Cs}, \mdet{\Cs}). There exists a \Cs-morphism $\eta: A^* \to N$ and $k \in \nat$ such that  $L_1,\dots,L_m$ are unions of \eqlek-classes (resp. \eqrek-classes, \eqmek-classes).
\end{corollary}

\begin{proof}
	We consider \mdet{\Cs} (the others are left to the reader). Let $L_1,\dots,L_m \in \mdet{\Cs}$. For every $i \leq m$, Proposition~\ref{prop:opcar} yields a \Cs-morphism $\eta_i: A^* \to N_i$ and $k_i \in \nat$ such that $L_i$ is a union of $\eqmp{\eta_i,k_i}$-classes. Let $M = N_1 \times \cdots \times N_m$ be the monoid equipped with the componentwise multiplication and $\alpha: A^* \to M$ be the morphism  defined by $\alpha(w) = (\eta_1(w),\dots,\eta_m(w))$ for all $w \in A^*$. We let $\eta: A^* \to N$ as the surjection induced by $\alpha$. One may verify that $\eta$ is a \Cs-morphism since \Cs is a \vari and $\eta_i: A^* \to N_i$ was a \Cs-morphism for all $i \leq m$. Finally, let $k = max(k_1,\dots,k_m)$. One may verify that \eqmek is finer than \eqmp{\eta_i,k_i} for every $i \leq m$. Thus, $L_1,\dots,L_m$ are unions of \eqmek-classes as desired.
\end{proof}

We may now present a first application of this framework. We prove that the operators \ldeto, \rdeto and \mdeto preserve the property of being a \vari.

\begin{theorem}\label{thm:detclos}
	Let \Cs a be a \vari. Then, \ldet{\Cs}, \rdet{\Cs} and \mdet{\Cs} are \varis as well.
\end{theorem}

\begin{proof}
	We present a proof for \mdeto (the argument is symmetrical for \ldeto and \rdeto). Let $K,L \in \mdet{\Cs}$ and $w \in A^*$. We show that $K \cup L$, $A^* \setminus L$, $w\inv L$ and $Lw\inv$  belong to \mdet{\Cs}. By Corollary~\ref{cor:opcar}, there exist a \Cs-morphism $\eta: A^* \to N$ and $k \in \nat$ such that $K$ and $L$ are unions of \eqmek-classes. Hence, by Proposition~\ref{prop:opcar}, it suffices to prove that $K \cup L$, $A^* \setminus L$, $w\inv L$ and $Lw\inv$ are also unions of \eqmek-classes. This is immediate for $K \cup L$ and $A^* \setminus L$. Hence, we concentrate on $w\inv L$ and $Lw\inv$. By symmetry, we only treat the former. Let $u,v \in A^*$ such that $u \eqmek v$. We show that  $u \in w\inv L \Leftrightarrow v \in w\inv L$. Since \eqmek is a congruence by Lemma~\ref{lem:areparts}, we have $wu \eqmek wv$. Since $L$ is a union of \eqmek-classes, this yields $wu  \in L \Leftrightarrow wv \in L$. Therefore, $u \in w\inv L \Leftrightarrow v \in w\inv L$ as desired.
\end{proof}

\subsection{The special case of group languages}

As we explained in Section~\ref{sec:polc}, we are particularly interested in input classes of the form \bpol{\Gs} and \bpol{\Gs^+} where \Gs is an arbitrary group \vari.  Consequently, we shall apply the above framework in the special case when the morphism $\eta: A^* \to N$ is either a \bpol{\Gs}- or~a \bpol{\Gs^+}-morphism. We prove that when $\eta$ is such a morphism, the three equivalences \eqlek, \eqrek and \eqmek can be simplified: we may restrict ourselves to the special case when $k = 1$. This property will be crucial in Section~\ref{sec:logic} when we characterize quantifier alternation for two-variable first-order logic in terms of mixed polynomial closure.

\begin{proposition} \label{prop:conspoint}
	Let \Gs be a group \vari and $\Cs \in \{\Gs,\Gs^+\}$. If $\eta: A^* \to N$ is a \bpol{\Cs}-morphism  and $k \in \nat$, there exists a \bpol{\Cs}-morphism, $\gamma: A^* \to Q$ such that $\poslek{w} \subseteq \poslp{\gamma}{1}{w}$ and $\posrek{w} \subseteq \posrp{\gamma}{1}{w}$.
\end{proposition}

\begin{proof}
	We fix the group \vari \Gs and $\Cs \in \{\Gs,\Gs^+\}$ for the proof. Let us start with preliminary terminology and results. Let $\alpha: A^* \to M$ be a morphism. An \emph{$\alpha$-monomial} is a marked product of the form $\alpha\inv(s_0)a_1 \alpha\inv(s_1) \cdots a_d \alpha\inv(s_d)$ where $s_1, \dots s_d \in M$. The number $d$ is called the degree of this $\alpha$-monomial. Moreover, an \emph{$\alpha$-polynomial} is a finite union of $\alpha$-monomials. Its degree is the maximum among the degrees of all $\alpha$-monomials in the finite union. We have the following simple lemma.
	
	\begin{lemma} \label{lem:intmon}
		Let $\alpha$ be a morphism and $K,L$ which are defined by $\alpha$-polynomials of degrees $m,n \in \nat$. Then $K \cap L$ is defined by an $\alpha$-polynomial of degree at most $m + n$.
	\end{lemma}
	
	\begin{proof}
		Since intersection distributes over union, we may assume without loss of generality that $K,L$ are defined by $\alpha$-\emph{monomials} of degrees $m,n \in \nat$. Moreover, since there are finitely many $\alpha$-monomials of degree at most $m + n$, it suffices to prove that for every $w \in K \cap L$, there exists $H \subseteq A^*$ which is defined by an $\alpha$-monomial of degree at most $m + n$ and such that $w \in H \subseteq K \cap L$. The finite union of all these languages $H$ will then define $K \cap L$. We fix $w \in K \cap L$ . By hypothesis on $K$ and $L$, we have $K = \alpha\inv(s_0)a_1 \alpha\inv(s_1) \cdots a_m \alpha\inv(s_m)$ and $L = \alpha\inv(t_0)b_1 \alpha\inv(t_1) \cdots b_m \alpha\inv(t_m)$. Hence, since we have $w \in K \cap L$, there are $P,Q \subseteq \pos{w}$ such that $\sigma_\alpha(w,P) = (s_0,a_1,s_1,\dots,a_m,s_m)$ and $\sigma_\alpha(w,Q) = (t_0,b_1,t_1,\dots,b_n,t_n)$. We define $R = P \cup Q$. Clearly, $\ell = |R| \leq |P| + |Q| = m + n$. Let $(q_0,c_1,q_1,\dots,c_\ell,q_\ell) = \sigma_\alpha(w,R)$. We let $H$ as the language defined by $\alpha\inv(q_0)c_1 \alpha\inv(q_1) \cdots c_\ell \alpha\inv(q_\ell)$ of degree $\ell \leq m +n$. One may now verify that $w \in H \subseteq K \cap L$.
	\end{proof}
	
	We complete the definition with two lemmas for $\alpha$-polynomials. They consider the special case when $\alpha$ is a \Cs-morphism. There are actually two kinds of \Cs-morphisms since $\Cs \in \{\Gs, \Gs^+\}$. We start with the simplest kind.

	\begin{lemma} \label{lem:gloop}
		Let $\alpha: A^* \to G$ be a morphism into a finite group and $x,y,w \in A^*$ such that $\alpha(xw) = \alpha(w)$ and $\alpha(wy)= \alpha(w)$. For every $\alpha$-polynomial $H \subseteq A^*$, we have $w \in H \Rightarrow xwy \in H$. 
	\end{lemma}
	
	\begin{proof}
		Assume that $w \in H$. Since $G$ is a group, our hypotheses on $x$ and $y$ imply that $\alpha(x) = \alpha(y) = 1_G$. Moreover, if $w \in H$, there exists an $\alpha$-monomial $K$ in the union defining $H$ such that $w \in K$. One may now verify that $K = \alpha\inv(1_G) K\alpha\inv(1_G)$. Hence, $xwy \in K \subseteq H$ as desired.
	\end{proof}
	
	The second lemma considers arbitrary \Cs-morphisms.
	
	\begin{lemma} \label{lem:gploop}
		Let $\alpha: A^* \to M$ be a $\Cs$-morphism and $u,v \in A^*$ such that $|u| = |v|$. Let $x,y,w \in A^*$ such that $\alpha(xw) = \alpha(w)$, $\alpha(wy) = \alpha(w)$, $w \in uA^*v$ and $xwy \in uA^*v$. For every $\alpha$-polynomial $H \subseteq A^*$ of degree at most $|u|$, we have $w \in H \Rightarrow xwy \in H$. 
	\end{lemma}
	
	\begin{proof}
		We write $n = |u| = |v|$. When $n = 0$, the lemma is trivial. The $\alpha$-polynomials of degree $0$ are exactly the languages recognized by $\alpha$. Thus, since our hypotheses yields $\alpha(xwy) = \alpha(w)$, we get that $w \in H \Rightarrow xwy \in H$ for every $\alpha$-polynomial $H$ of degree $0$.
		
		Assume that $n \geq 1$ and $w \in H$. We get an $\alpha$-monomial $K$ in the union defining $H$ such that $w \in K$. We write $d \leq n$ for the degree of $K$. By definition, we know that $K$ is of the form $K =  \alpha\inv(s_0)a_1 \alpha\inv(s_1) \cdots a_d \alpha\inv(s_d)$. Consequently, we have $w = w_0a_1w_1\cdots a_dw_d$ where $\alpha(w_i) = s_i$ for every $i \leq d$. Since $w \in uA^*v$ and $|u| = |v| = n$, we know that $|w| \geq 2n$. Thus, since $d\leq n$, there exists $i \leq d$ such that $w_i \neq \veps$. We let $h \leq d$ and $\ell \leq d$ as the least and the greatest such $i$ respectively, $u' = w_0a_1 \cdots w_{h-1}a_h = a_1 \cdots a_h$ (if $h = 0$, then $u' = \veps$) and $v' = a_{\ell+1}w_{\ell+1} \cdots a_dw_d =a_{\ell+1} \cdots a_d$ (if $\ell = d$, then $v' = 0$). By definition, we have $y = u'w_ha_{h+1}w_{h+1} \cdots a_\ell w_\ell v'$ and $w_h,w_\ell \in A^+$.  By definition, $|u'| \leq d \leq n$ and $|v'| \leq d \leq n$. Thus, since $y \in uA^*v$ and $|u| = |v| = n$, it follows that $u'$ is a prefix of $u$ and $v'$ is a suffix of $v$. Since we also know that $xwz \in uA^*v$, this yields $z \in A^*$ such that $xwy = u'zv'$. By hypothesis on $w$, we also know that $xwy = xu'w_ha_{h+1}w_{h+1} \cdots a_\ell w_\ell v'y$. Thus, we get $x',y' \in A^*$ such that $u'x' = xu'$ and $y'v' = v'y$. Altogether, it follows that $xwy = u'x'w_ha_{h+1}w_{h+1} \cdots a_\ell w_\ell y'v'$. We now prove that $\alpha(x'w_h) = s_h$ and $\alpha(w_\ell y')= s_\ell$. By symmetry, we only detail the former. This is trivial if $x' = \veps$. Thus, we assume that $x' \in A^+$. Since $u'x' = xu'$, we have $x \in A^+$ as well. Let $G = \alpha(A^+)$. Since $\alpha$ is a \Cs-morphism, $\Cs \subseteq \Gs^+$ and \Gs is a group \vari, Lemma~\ref{lem:gmorph} yields that $G$ is a group. Hence, since $\alpha(xw) = \alpha(w)$ and $w \in A^+$, we get $\alpha(x) = 1_G$. Thus, since $u'x' = xu'$ and $u' \in A^+$, we get $\alpha(u'x') = \alpha(u')$. It follows that $\alpha(x') = 1_G$. Finally, since $w_h \in A^+$, we have  $\alpha(w_h) \in G$ and it follows that $\alpha(x'w_h) = \alpha(w_h) = s_h$. We may now complete the proof that $xwy \in H$. We obtain,
		\[
		x'w_ha_{h+1} \cdots a_\ell w_\ell y' \in \alpha\inv(s_h)a_{h+1} \alpha\inv(s_{h+1}) \cdots a_\ell \alpha\inv(s_\ell).
		\]
		By definition, we know that $u' \in \alpha\inv(s_0)a_1  \cdots \alpha\inv(s_{h-1})a_h$ and $v' \in a_\ell \alpha\inv(a_\ell)\cdots a_d \alpha\inv(s_d)$. Consequently, we obtain that $xwy =  u'x'w_ha_{h+1}w_{h+1} \cdots a_\ell w_\ell y'v' \in K \subseteq H$.
	\end{proof}
	
	We may now prove Proposition~\ref{prop:conspoint}. Let $\eta: A^* \to N$ be a \bpol{\Cs}-morphism and $k \in \nat$. We first define the \bpol{\Cs}-morphism $\gamma: A^* \to Q$ and then prove that $\poslek{w} \subseteq \poslp{\gamma}{1}{w}$ and $\posrek{w} \subseteq \posrp{\gamma}{1}{w}$.
	
	By hypothesis on $\eta$, there exists a finite set \Lb of languages in \Cs such that all languages recognized by $\eta$ are Boolean combinations of marked products of  languages in \Lb. Proposition~\ref{prop:genocm} yields a \Cs-morphism $\alpha: A^* \to M$ recognizing every $L \in \Lb$. Therefore, since union distributes over marked concatenation, every language recognized by $\eta$ is a Boolean combination of $\alpha$-monomials. These Boolean combinations can be put into disjunctive normal form. Moreover, intersection of $\alpha$-monomials are finite unions of \Cs-monomials by Lemma~\ref{lem:intmon}. Consequently, there exists a number $n \in \nat$ such that every language recognized by $\eta$ is a finite union of languages of the form $L \setminus H$ where $L$ is an $\alpha$-monomial of degree at most $n$ and $H$ is a finite union of $\alpha$-monomials of degree at most $n$ (\emph{i.e.}, an $\alpha$-polynomial of degree at most~$n$). Clearly, there are finitely many $\alpha$-polynomials of degree at most $(3n+1) \times k$ and since $\alpha$ is a \Cs-morphism, they all belong to $\pol{\Cs} \subseteq \bpol{\Cs}$. Hence, Proposition~\ref{prop:genocm} yields a \bpol{\Cs}-morphism $\gamma: A^* \to Q$ recognizing every $\alpha$-polynomial of degree at most $(3n+1) \times k$.
	
	It remains to prove the inclusions $\poslek{w} \subseteq \poslp{\gamma}{1}{w}$ and $\posrek{w} \subseteq \posrp{\gamma}{1}{w}$ for every $w \in A^*$. By symmetry, we only prove the former. We fix $w \in A^*$ for the proof. The hypothesis that $\Cs \in \{\Gs,\Gs^+\}$ implies the following lemma.
	
	\begin{lemma} \label{lem:conspoint}
		Let $h$ such that $1 \leq h \leq k$, $i \in \poslp{\eta}{h}{w}$ and $a \in A$ the label of $i$. There is an  $\alpha$-monomial $K$ of degree at most $(3n+1)h -1$ such that $\prefix{w}{i} \in K$ and $\prefix{w}{i} \not\in KaA^*$.
	\end{lemma}
	
	Let us first apply Lemma~\ref{lem:conspoint} to complete the main argument. Let $i \in \poslek{w}$. We show that $i \in \poslp{\gamma}{1}{w}$. Let $a$ be the label of $i$. By definition, we have to prove that $\gamma(\prefix{w}{i}a) \Rords \gamma(\prefix{w}{i})$. Since $\gamma$ is surjective (recall that it is a \bpol{\Cs}-morphism), this boils down to proving that $\gamma(\prefix{w}{i}) \neq \gamma(\prefix{w}{i}au)$ for every $u \in A^*$. We fix $u$ for the proof. Lemma~\ref{lem:conspoint} yields an $\alpha$-monomial $K$ of degree at most $(3n+1)k -1$ such that $\prefix{w}{i} \in K$ and $\prefix{w}{i} \not\in KaA^*$. Clearly, $KaA^*$ is defined by an $\alpha$-polynomial of degree at most $(3n+1)k$. Hence, $KaA^*$  is recognized by $\gamma$. Since we have $\prefix{w}{i}au \in KaA^*$ and $\prefix{w}{i} \not\in KaA^*$, we obtain $\gamma(\prefix{w}{i}) \neq \gamma(\prefix{w}{i}au)$ which completes the proof.

	\smallskip
	
	It remains to prove Lemma~\ref{lem:conspoint}. We consider a number $h$ such that $1 \leq h \leq k$, $i \in \poslp{\eta}{h}{w}$ and $a \in A$ the label of $i$. We have to construct an $\alpha$-monomial $K$ of degree at most $(3n+1)h -1$ such that $\prefix{w}{i} \in K$ and $\prefix{w}{i} \not\in KaA^*$. We proceed by induction on $h$. By definition, there exists $j \in \poslp{\eta}{h-1}{w} \cup \{0\}$ such that $\eta(\infix{w}{j}{i}a) \Rords \eta(\infix{w}{j}{i})$. We first prove an important result about the word \infix{w}{j}{i}. Let $E \subseteq A^*$ be the language of all words $u \in A^+$ such that $\alpha(u)$ is \emph{idempotent}.  We prove that there exists an $\alpha$-monomial $V$ of degree at most $3n$ which satisfies the following property: 
	\begin{equation} \label{eq:thev}
		\infix{w}{j}{i} \in V  \ \text{and} \ \infix{w}{j}{i} \not\in EVaA^*. %VaA^* \cup \alpha\inv(1_G)VaA^*
	\end{equation}
	Let $t = \eta(\infix{w}{i}{j})$. By construction, since $\infix{w}{i}{j} \in \eta\inv(t)$, there exist an $\alpha$-monomial $L$ and an $\alpha$-polynomial $H$, both of degree at most $n$ and such that $\infix{w}{i}{j} \in L \setminus H \subseteq \eta\inv(t)$. We now consider two cases depending on whether the monoid $M$ is a group or not.

	\smallskip
	\noindent
	{\it Construction of $V$, first case.} We assume that $M$ is a group. It follows that $1_M$ is the only idempotent in $M$ and therefore that $E = \alpha\inv(1_M)$. We let $V = L$ which is an $\alpha$-monomial of degree at most $n \leq 3n$. We already know that $\infix{w}{i}{j} \in L$. We show that $\infix{w}{j}{i} \not\in ELaA^*$. We proceed by contradiction. Assume that $\infix{w}{j}{i} = xyaz$ with $\alpha(x) = 1_M$, $y \in L$ and $z \in A^*$. We show that $\eta(xy) = \eta(\infix{w}{j}{i}) = t$. Since $\infix{w}{j}{i} = xyaz$, this yields $\eta(\infix{w}{j}{i}) = \eta(\infix{w}{j}{i}az)$, contradicting the hypothesis that $\eta(\infix{w}{j}{i}a) \Rords \eta(\infix{w}{j}{i})$. Since $L \setminus H \subseteq \eta\inv(t)$, it suffices to prove that $xy \in L\setminus H$. Since $\alpha(x) = 1_M$, we have $\alpha(xy) = \alpha(y)$. We also have $y \in L$ which is an $\alpha$-monomial. Thus, since $M$ is a group, Lemma~\ref{lem:gloop} yields $xy \in L$. It remains to prove $xy \not\in H$. By contradiction, we assume that $xy \in H$. Since $xy \in L$ and $\infix{w}{j}{i} \in L$, one may verify from the definition of $\alpha$-monomials that $\alpha(xy) = \alpha(\infix{w}{j}{i})$. Since $\infix{w}{j}{i} = xyaz$, we obtain $\alpha(xy) = \alpha(xyaz)$. Moreover, $H$ is an $\alpha$-polynomial by definition. Thus, since  $M$ is a group, Lemma~\ref{lem:gloop} yields $\infix{w}{j}{i} = xyaz \in H$. This is a contradiction since $\infix{w}{j}{i} \in L \setminus H$ by hypothesis.

	\medskip
	\noindent
	{\it Construction of $V$, second case.} We now assume that $M$ is not a group. We define $G = \alpha(A^+)$. Since $\Cs \subseteq \Gs^+$, we know that $\alpha$ is a $\Gs^+$-morphism.  Thus, Lemma~\ref{lem:gmorph} implies that $G$ is a group. Since $M = \{1_M\} \cup G$ by definition of $G$, it follows that $1_M \not\in G = \alpha(A^+)$ and we conclude that $\alpha\inv(1_M) = \{\veps\}$. We consider two sub-cases. First, assume that $|\infix{w}{j}{i}| \leq 3n$. In this case, we let $V = \{\infix{w}{j}{i}\}$. Since $\alpha\inv(1_M) = \{\veps\}$, this is an $\alpha$-monomial of degree $|\infix{w}{j}{i}| \leq 3n$. Since $\infix{w}{j}{i} \in V$ and $\infix{w}{j}{i} \not\in (\{\veps\} \cup \alpha\inv(1_G))VaA^*$, \eqref{eq:thev} is proved.
	
	We now consider the sub-case when $|\infix{w}{j}{i}| > 3n$. This hypothesis yields $u,v \in A^+$ such that $|u| = |v| = n$ and $\infix{w}{j}{i} \in uA^*v$. Since $\alpha\inv(1_M) = \{\veps\}$, it is immediate that $uA^*v$ is defined by an $\alpha$-polynomial of degree $2n$. Since $L$ is an $\alpha$-monomial of degree at most $n$, Lemma~\ref{lem:intmon} yields that $L \cap uA^*v$ is defined by an $\alpha$-polynomial of degree at most $3n$. Since $\infix{w}{j}{i} \in L \cap uA^*v$, we get an $\alpha$-monomial $V$ of degree at most $3n$ such that $\infix{w}{j}{i} \in V \subseteq L\cap uA^*v$. It remains to prove that $\infix{w}{j}{i} \not\in (\{\veps\} \cup \alpha\inv(1_G))VaA^*$. By contradiction, we assume that $\infix{w}{j}{i} = xyaz$ with $x = \veps$ or $\alpha(x) = 1_G$, $y \in V$ and $z \in A^*$. We prove that $\eta(xy) = \eta(\infix{w}{j}{i}) = t$. Since $\infix{w}{j}{i} = xyaz$, this implies that $\eta(\infix{w}{j}{i}) = \eta(\infix{w}{j}{i}az)$, contradicting the hypothesis that $\eta(\infix{w}{j}{i}a) \Rords \eta(\infix{w}{j}{i})$. Since $L \setminus H \subseteq \eta\inv(t)$, it suffices to prove that $xy \in L\setminus H$. By hypothesis on $V$, we have $y \in  L \cap uA^*v$. Thus, $xy \in A^*uA^*v$ and since $\infix{w}{j}{i} = xyaz \in uA^*v$, it follows that $xy \in uA^*v$.  Since $y \in A^+$ (which means that $\alpha(y) \in G$) and either $x = \veps$ or $\alpha(x)  =1_G$, we also have $\alpha(xy) = \alpha(y)$. Hence, since $L$ is an $\alpha$-monomial of degree at most $n$ and Lemma~\ref{lem:gploop} that $xy \in L$. It remains to show that $xy \not\in H$. By contradiction, we assume that $xy \in H$. Since $\infix{w}{j}{i} = xyaz$ and $xy$ both belong to $L$ which is an $\alpha$-monomial, we have $\alpha(xy) = \alpha(xyaz)$. Moreover, $xy \in uA^*v$ and $xyaz = \infix{w}{i}{j} \in uA^*v$. Hence, since $H$ is an $\alpha$-polynomial of degree at most $n$ by definition, Lemma~\ref{lem:gploop} yields $\infix{w}{j}{i} = xyaz \in H$. This is a contradiction since $\infix{w}{j}{i} \in L \setminus H$. This completes the construction of $V$.
	
	\medskip
	\noindent
	{\it Construction of $K$.} Using our $\alpha$-monomial $V$ of degree at most $3n$, we build $K$. There are two cases depending on whether $j = 0$ or $j \geq 1$. When $j = 0$, we choose $K = V$ which has degree $3n \leq (3n+1)h -1$. By~\eqref{eq:thev}, we have $\prefix{w}{i} \in K$ and $\prefix{w}{i} \not\in KaA^*$ as desired.
	
	Assume now that $1 \leq j < i$. Since $j \in \poslp{\eta}{h-1}{w}$, it follows that $h-1 \geq 1$. Let $b$ be the label of $j$. Induction on $h$ in Lemma~\ref{lem:conspoint} yields an $\alpha$-monomial $U$ with degree at most $(3n+1)(h-1) -1$  such that $\prefix{w}{j} \in U $ and $\prefix{w}{j} \not\in UbA^*$. We define $K = UbV$. By hypothesis on $U$ and $V$, we know that $K$ is an $\alpha$-monomial of degree at most $(3n+1)(h-1) -1 +1 + 3n = (3n+1)h -1$. Moreover, $\prefix{w}{i} = \prefix{w}{j}b\infix{w}{j}{i} \in UbV = K$. We now prove that $\prefix{w}{i} \not\in KaA^*$. By contradiction, assume that $\prefix{w}{i} \in KaA^* = UbVaA^*$. We get $x \in U$, $y \in V$ and $z \in A^*$ such that $\prefix{w}{i} = xbyaz$. Moreover, we know that $\prefix{w}{i} = \prefix{w}{j}b\infix{w}{j}{i}$ and since $\prefix{w}{j} \not\in UbA^*$, the word $xb \in Ub$ cannot be a prefix $\prefix{w}{j}$. Hence, we have $x' \in A^*$ such that $xb = \prefix{w}{j}bx'$ and $x'yaz = \infix{w}{j}{i}$. Since $U$ is an $\alpha$-monomial and $x,\prefix{w}{j} \in U$, we have $\alpha(x) = \alpha(\prefix{w}{j})$. Hence, $\alpha(xb) = \alpha(\prefix{w}{j}b)$ and since $xb = \prefix{w}{j}bx'$, it follows that either $x' = \veps$ or $\alpha(x') = 1_G$. We conclude that $\infix{w}{j}{i} = x'yaz \in (\{\veps\} \cup \alpha\inv(1_G))VaA^*$. This contradicts~\eqref{eq:thev} in the definition of $V$.
\end{proof}

%% file: carac.tex
We present generic algebraic characterizations of \ldet{\Cs}, \rdet{\Cs} and \mdet{\Cs} when \Cs is an arbitrary \vari. They imply that if \Cs has decidable membership, then so do \ldet{\Cs}, \rdet{\Cs} and \mdet{\Cs}. We organize the presentation in two parts. First, we consider the classes \ldet{\Cs} and \rdet{\Cs} which are handled symmetrically. Then, we turn to \mdet{\Cs}.

\subsection{Left/right polynomial closure}

We present the symmetrical algebraic characterizations of \ldeto and \rdeto. Given an arbitrary \vari \Cs, they characterize the \ldet{\Cs}- and \rdet{\Cs}-morphisms using the \Cs-pairs and the canonical equivalence \canec.

\begin{theorem} \label{thm:cldet}%
	Let \Cs be a \vari and $\alpha: A^* \to M$ a surjective morphism. The following properties are equivalent:
	\begin{enumerate}[label=\alph*)]
		\item\label{itm:el0} $\alpha$ is an \ldet{\Cs}-morphism.
		\item\label{itm:el1} $s^{\omega+1} = s^{\omega}t$ for all \Cs-pairs $(s,t) \in M^2$.
		\item\label{itm:el2} $s^{\omega+1} = s^{\omega}t$ for all $s,t \in M$ such that $s \canec t$. 
	\end{enumerate}		
\end{theorem}

\begin{theorem} \label{thm:crdet}
	Let \Cs be a \vari and $\alpha: A^* \to M$ a surjective morphism. The following properties are equivalent:
	\begin{enumerate}[label=\alph*)]
		\item $\alpha$ is an \rdet{\Cs}-morphism.
		\item $s^{\omega+1} = ts^{\omega}$ for all \Cs-pairs $(s,t) \in M^2$.
		\item $s^{\omega+1} = ts^{\omega}$ for all $s,t \in M$ such that $s \canec t$. 
	\end{enumerate}		
\end{theorem}

By Fact~\ref{fct:eqmemb}, computing the equivalence \canec boils down to \Cs-membership. Hence, by Proposition~\ref{prop:synmemb}, we get the following corollary of Theorem~\ref{thm:cldet} and Theorem~\ref{thm:crdet}.

\begin{corollary} \label{cor:lrmemb}
	Let \Cs be a \vari. If \Cs-membership is decidable, then so are \ldet{\Cs}- and \rdet{\Cs}-membership.
\end{corollary} 

We now concentrate on the proofs of Theorem~\ref{thm:cldet} and Theorem~\ref{thm:crdet}. Since the arguments are symmetrical, we only prove the former.

\begin{proof}[Proof of Theorem~\ref{thm:cldet}]
	We first prove that $\ref{itm:el0}  \Rightarrow \ref{itm:el1} $. We assume $\alpha$ is an \ldet{\Cs}-morphism and prove that~\ref{itm:el1} holds. Consider a \Cs-pair $(s,t) \in M^2$. We show that $s^{\omega+1} = s^\omega t$. Corollary~\ref{cor:opcar} yields a \Cs-morphism $\eta: A^* \to N$ and $k \in \nat$ such that every language recognized by $\alpha$ is a union of \eqlek-classes. Since $(s,t)$ is a \Cs-pair and $\eta$ is a \Cs-morphism, Lemma~\ref{lem:cmorph} yields $u,v \in A^*$ such that $\eta(u) = \eta(v)$, $\alpha(u) = s$ and $\alpha(v) = t$. Let $p = \omega(M) \times \omega(N)$, $w = u^{pk}u$ and $w' = u^{pk}v$. We have the following lemma.
	
	\begin{lemma} \label{lem:leq}
		For every $i \in \poslek{w}$, we have $i \leq |u^{pk}|$.
	\end{lemma}
	
	\begin{proof}
		We use induction on $h$ to show that for all $h \leq k$ and $i \in \poslp{\eta}{h}{w}$, we have $i \leq |u^{ph}|$. The case $h =k$ implies the lemma. We write $w = a_1 \cdots a_\ell$ for the proof. Let $h \leq k$. By contradiction, assume that there exists $i\in\poslp{\eta}{h}{w}$ such that $i>|u^{ph}|$. By definition, there exists $j \in \poslp{\eta}{h-1}{w} \cup \{0\}$ such that $j < i$ and the strict inequality  $\eta(\infix{w}{j}{i}a_i) \Rords \eta(\infix{w}{j}{i})$ holds. By induction, $j \leq |u^{p(h-1)}|$. Hence, since $i>|u^{ph}|$ and $w = u^{pk}u$, the infix  $\infix{w}{j}{i}$ must contain an infix $u^p$: we have $x,y \in A^*$ and $n \in \nat$ such that $\infix{w}{j}{i} = xu^py$ and $\suffix{w}{j} = xu^n$. Let $q \in \nat$ such that $n +q$ is a multiple of $p$. By definition, $\eta(u^{p}) \in E(N)$ is idempotent. Hence, $\eta(\suffix{w}{j}u^qy) = \eta(xu^{p}y) = \eta(\infix{w}{j}{i})$. Since $\infix{w}{j}{i}a_i$ is a prefix of $\suffix{w}{j}$, it follows that $\eta(\infix{w}{j}{i}) \Rord  \eta(\infix{w}{j}{i}a_i)$. This is a contradiction since $\eta(\infix{w}{j}{i}a_i) \Rords \eta(\infix{w}{j}{i})$ by hypothesis.
	\end{proof}
	
	We may now prove that $s^{\omega+1} = s^\omega t$. By Lemma~\ref{lem:leq}, every position in \poslek{w} belong to the prefix $u^{pk}$ of $w = u^{pk}u$. Therefore, since $u^{pk}$ is also a prefix of $w'=u^{pk}v$, $\poslek{w}\subseteq\posc{w'}$. Since $\eta(u) = \eta(v)$, we get $\sigma_\eta(w,\poslek{w}) = \sigma_\eta(w',\poslek{w})$. Hence, Corollary~\ref{cor:eqbij} yields $w \eqlek w'$ and it follows that $\alpha(w) = \alpha(w')$ since the languages recognized by $\alpha$ are unions of \eqlek-classes. By definition of $w,w'$ and since $p$ is a multiple of $\omega(M)$, this yields $s^{\omega+1} = s^\omega t$ as desired.
	
	\smallskip
	
	We turn to the implication $\ref{itm:el1} \Rightarrow \ref{itm:el2}$. We assume that~\ref{itm:el1} holds and consider $s,t \in M$ such that $s \canec t$. We show that $s^{\omega+1} = s^\omega t$. By Lemma~\ref{lem:transclos}, there exist $r_0,\dots,r_n \in M$ such that $r_0 = s$, $r_n = t$ and $(r_i,r_{i+1})$ is a \Cs-pair for all $i < n$. We use induction on $i$ to show that $s^{\omega+1} = s^\omega r_i$ for every $i \leq n$. The case $i = n$ yields the desired result as $t = r_n$. When $i = 0$, the result is immediate as $r_0  = s$. Assume now that $i \geq 1$. Since $(r_{i-1},r_{i})$ is a \Cs-pair,  $(s^\omega r_{i-1},s^\omega r_{i})$ is a \Cs-pair as well by Lemma~\ref{lem:mult}. Therefore, we get from~\ref{itm:el1} that $(s^\omega r_{i-1})^{\omega+1} = (s^\omega r_{i-1})^{\omega}s^\omega r_{i}$. Finally, induction yields $s^{\omega+1} = s^\omega r_{i-1}$. Combined with the previous equality, this yields $s^{\omega+1} = (s^{\omega+1})^{\omega+1} = (s^{\omega+1})^\omega s^\omega r_i = s^\omega r_i$ as desired.
	
	\smallskip
	
	It remains to prove $\ref{itm:el2} \Rightarrow \ref{itm:el0}$. We assume that~\ref{itm:el2} holds and show that $\alpha$ is an \ldet{\Cs}-morphism. Let $N = M/{\canec}$ and recall that $N$ is a monoid since \canec is a congruence by Lemma~\ref{lem:caquot}. We write $\eta = \ctype{\cdot} \circ \alpha: A^* \to N$ which is a \Cs-morphism by Lemma~\ref{lem:smult}. We let $k = |M|$ and consider the equivalence \eqlek on $A^*$. We prove the following property:
	\begin{equation} \label{eq:ldetcl}
		\text{for every $w,w' \in A^*$,} \quad w \eqlek w' \Rightarrow \alpha(w) = \alpha(w').
	\end{equation}
	This implies that every language recognized by $\alpha$ is a union of \eqlek-classes. Together with Proposition~\ref{prop:opcar} this yields that every language recognized by $\alpha$ belongs to \ldet{\Cs} since $\eta$ is a \Cs-morphism. We now concentrate on~\eqref{eq:ldetcl}. Let $w,w' \in A^*$ such that $w \eqlek w'$. We show that $\alpha(w)=\alpha(w')$. For the proof, we write $P= \poslp{\alpha}{1}{w}$. We use the hypothesis that $w \eqlek w'$ to prove the following lemma.
	
	\begin{lemma} \label{lem:lbij}
		There exists $P' \subseteq \posc{w'}$ such that $\sigma_\eta(w,P) = \sigma_\eta(w',P')$.
	\end{lemma}
	
	\begin{proof}
		Since~\ref{itm:el2} holds, we know that for all $s,t\in M$ such that $s\canec t$, we have $s^{\omega+1} = s^\omega t$. We may multiply by $s^\omega$ on the right to get $s^{\omega+1} = s^\omega ts^\omega$. Hence, it follows from Theorem~\ref{thm:cupol} that $\alpha$ is a \upol{\Cs}-morphism. Since $k = |M|$, Lemma~\ref{lem:isdet} yields $P = \poslp{\alpha}{1}{w}  \subseteq \poslek{w}$. Finally, since $w \eqlek w'$, we have $\sigma_\eta(w,\poslek{w}) = \sigma_\eta(w',\poslek{w'})$. Thus, Fact~\ref{fct:snap} yields a set $P' \subseteq \sigma_\eta(w',\poslek{w'})$ such that $\sigma_\eta(w,P) = \sigma_\eta(w',P')$ as desired.
	\end{proof}
	
	Let $(s_0,a_1,s_1,\dots,a_n,s_n) = \sigma_\alpha(w,P)$ and $(t_0,b_1, t_1,\dots,b_m,t_m) = \sigma_\alpha(w',P')$. It follows from Lemma~\ref{lem:lbij} that $\sigma_\eta(w,P) = \sigma_\eta(w',P')$. We obtain $n = m$, $a_i = b_i$ for $1 \leq i \leq n$ and $s_i \canec t_i$ for $0 \leq i \leq n$ by definition of $\eta$. Therefore, we have $\alpha(w) = s_0a_1s_1 \cdots a_ns_n$ and $\alpha(w') = t_0a_1t_1 \cdots a_nt_n$ by definition of $\alpha$-snapshots. It now remains to prove that $s_0a_1s_1 \cdots a_hs_h = t_0a_1t_1 \cdots a_ht_h$. We let $q_h = s_0a_1s_1 \cdots a_h$ and $r_h = t_0a_1t_1 \cdots a_h$ for every $h$ such that $0 \leq h \leq n$ (in particular, $q_0 = r_0 = 1_M$). We use induction on $h$ to show that $q_hs_h = r_ht_h$ for $0 \leq h \leq n$. Clearly, the case $h = n$ yields the desired result.
	
	We fix $h \leq n$ and show that $q_hs_h = r_ht_h$. Since $P=\poslp{\alpha}{1}{w}$, one may verify from the definitions that $q_hs_h \Rrel q_h$. We get $x \in M$ such that $q_h = q_hs_hx$. Since $s_h \canec t_h$ and \canec is a congruence, we have $xs_h \canec xt_h$. Hence, it follows from~\ref{itm:el2} that $(xs_h)^{\omega+1} = (xs_h)^{\omega}xt_h$. We may now multiply on the left by $s_h$ to obtain  $(s_hx)^{\omega+1} s_h = (s_hx)^{\omega+1}t_h$. We combine this with $q_h = q_hs_hx$ to obtain  $q_hs_h = q_ht_h$. This concludes the proof when $h = 0$ since $q_0 = r_0 = 1_M$, we get $q_0s_0 = r_0t_0$ as desired. Finally, if \mbox{$h \geq 1$}, induction yields $q_{h-1}s_{h-1} =r_{h-1}t_{h-1}$.  Since $q_h = q_{h-1}a_{h}$ and $r_h = r_{h-1}a_h$ by definition, it follows that $q_h = r_h$. Altogether, we get $q_hs_h = r_ht_h$ which completes the proof.
\end{proof}

We conclude the presentation with an important corollary of Theorems~\ref{thm:cldet} and~\ref{thm:crdet}. We shall use it later when considering the determinsitic hierarchies built uniformly from a single input class \Cs be applying \ldeto and \rdeto alternately (we define them properly in Section~\ref{sec:deth}). Intuitively, the class \Ds in the statement is meant to be a level in  such a hierarchy.

\begin{lemma} \label{lem:lradv}
	Let $\Cs,\Ds$ be \varis such that $\Cs \subseteq \Ds \subseteq \upol{\Cs}$ and $\alpha: A^* \to M$ a surjective morphism. The following properties hold:
	\begin{itemize}
		\item if $\alpha$ is an \ldet{\Ds}-morphism, then for every $e \in E(M)$ and $q,r \in M$ such that $q \caned r$ and $\ctype{e} \Rord \ctype{q}$, we have $e q = e r$.
		\item if $\alpha$ is a \rdet{\Ds}-morphism, then for every $e \in E(M)$ and $q,r,s \in M$ such that $q \caned r$ and $\ctype{e} \Lord \ctype{q}$, we have $qe = re$.
	\end{itemize}
\end{lemma}

\begin{proof}
	By symmetry, we only prove the first assertion. Assume that $\alpha$ is an \ldet{\Ds}-morphism. Given $e \in E(M)$ and $q,r \in M$ such that $q \caned r$ and $\ctype{e} \Rord \ctype{q}$, we show that $e q = e r$. Note that since $\ctype{e} \Rord \ctype{q}$, there exists $s \in M$ such that $\ctype{e} = \ctype{qs}$ which exactly says that $e \canec qs$. Since $\Ds \subseteq \upol{\Cs}$, we have $\ldet{\Ds} \subseteq \upol{\Cs}$. Thus, $\alpha$ is a \upol{\Cs}-morphism and since $e \canec qs$, Theorem~\ref{thm:cupol} yields $e = eqse$. Hence, $eq = eqseq$. Moreover, since $q \caned r$ and \caned is a congruence we have $seq \caned ser$. Since $\alpha$ is an \ldet{\Ds}-morphism, Theorem~\ref{thm:cldet} yields $(seq)^{\omega+1} = (seq)^{\omega} ser$. We now combine this with $eq = eqseq$ to get $eq = eqser$. Finally, since $e = eqse$, we obtain $eq = er$ as desired.
\end{proof}

\subsection{Mixed polynomial closure}

We now consider the operator $\Cs \mapsto \mdet{\Cs}$. In this case, the characterization is more involved.

\begin{theorem} \label{thm:cmdet}
	Let \Cs be a \vari and $\alpha: A^* \to M$ a surjective morphism. The following properties are equivalent:
	\begin{enumerate}[label=\alph*)]
		\item\label{itm:em0} $\alpha$ is an \mdet{\Cs}-morphism.
		\item\label{itm:em1} $(sq)^{\omega}s(rs)^\omega = (sq)^{\omega}t(rs)^\omega$ for all \Cs-pairs $(s,t) \in M^2$ and all $q,r \in M$.
		\item\label{itm:em2} $(sq)^{\omega}s(rs)^\omega = (sq)^{\omega}t(rs)^\omega$ for all $q,r,s,t \in M$ such that $s \canec t$. 
	\end{enumerate}		
\end{theorem}

By Fact~\ref{fct:eqmemb}, one may compute the equivalence \canec associated to a morphism provided that \Cs-membership is decidable. Hence, in view of Proposition~\ref{prop:synmemb}, we obtain the following corollary of Theorems~\ref{thm:cldet}, \ref{thm:crdet} and~\ref{thm:cmdet}.

\begin{corollary} \label{cor:pmemb}
	Let \Cs be a \vari. If \Cs-membership is decidable, then so is \mdet{\Cs}-membership.
\end{corollary}

 \begin{proof}[Proof of Theorem~\ref{thm:cmdet}]
	We fix a \vari \Cs and a surjective morphism $\alpha: A^* \to M$. We start with $\ref{itm:em0} \Rightarrow \ref{itm:em1}$. Assume that $\alpha$ is an \mdet{\Cs}-morphism. Let $q,r,s,t \in M$ such that $(s,t)$ is a \Cs-pair. We show that $(sq)^{\omega}s(rs)^\omega = (sq)^{\omega}t(rs)^\omega$. Corollary~\ref{cor:opcar} yields a \Cs-morphism $\eta: A^* \to N$ and $k \in \nat$ such that every language recognized by $\alpha$ is a union of \eqmek-classes. Since $(s,t)$ is a \Cs-pair and $\eta$ is a \Cs-morphism, Lemma~\ref{lem:cmorph} yields $u,v \in A^*$ such that $\eta(u) = \eta(v)$, $\alpha(u) = s$ and \mbox{$\alpha(v) = t$}. Let $x,y \in A^*$ such that $\alpha(x) = q$ and \mbox{$\alpha(y) = r$}. We define $p = \omega(M) \cdot \omega(N)$. Let $w = (ux)^{pk}u(yu)^{pk}$ and $w' = (ux)^{pk}v(yu)^{pk}$. %We have the following lemma.
	
	\begin{lemma} \label{lem:posm}
		For every $i \in \posmek{w}$, either $i \leq |(ux)^{pk}|$ or $i > |(ux)^{pk}u|$.
	\end{lemma}

	\begin{proof}
		Since $\posmek{w} = \poslek{w} \cup \posrek{w}$, there are two cases depending on whether $i\in  \poslek{w}$ or $i \in \posrek{w}$. lBy symmetry, we only treat the former case. Given a position $i \in \poslek{w}$, we show that either $i\leq |(ux)^{pk}|$ or $i>|(ux)^{pk}u|$. We write $w = a_1 \cdots a_\ell$ for the proof. We consider a slightly stronger property. Let $h \leq k$. Using induction on $h$, we show that for every $i \in \poslp{\eta}{h}{w}$, either $i \leq |(ux)^{ph}|$ or $i > |(ux)^{pk}u|$. By contradiction, assume that there exists $i \in \poslp{\eta}{h}{w}$ such that $|(ux)^{ph}| < i \leq |(ux)^{pk}u|$. This yields $j \in \poslp{\eta}{h-1}{w} \cup \{0\}$ such that $j < i$ and $\eta(\infix{w}{j}{i}a_i) \Rords \eta(\infix{w}{j}{i})$. By induction, we have $j \leq |(ux)^{p(h-1)}|$. Therefore, since we have \mbox{$|(ux)^{ph}| < i \leq |(ux)^{pk}u|$} and $w = (ux)^{pk}u(yu)^{pk}$, the infix  $\infix{w}{j}{i}$ must contain an infix $(ux)^p$: we have $z,z' \in A^*$ and $n \in \nat$ such that $\infix{w}{j}{i} = z(ux)^pz'$ and $\infix{w}{j}{|(ux)^{pk}u|+1} = z(ux)^n u$. Let $m \in \nat$ be a number such that $n +1 + m$ is a multiple of $p$. By definition of $p$, $\eta(u^p)$ is an idempotent of $N$. Hence, $\eta(\infix{w}{j}{|(ux)^{pk}u|+1}x(ux)^hz') = \eta(z(ux)^pz') = \eta(\infix{w}{j}{i})$. By definition, $\infix{w}{j}{i}a_{i}$ is a prefix of $\eta(\infix{w}{j}{|(ux)^{pk}u|+1}$. Consequently, it follows that $\eta(\infix{w}{j}{i}) \Rord  \eta(\infix{w}{j}{i}a_i)$. This is a contradiction since $\eta(\infix{w}{j}{i}a_i) \Rords \eta(\infix{w}{j}{i})$ by hypothesis. 
	\end{proof}
	
	Lemma~\ref{lem:posm} states that all positions in \posmek{w} belong either to the prefix $(ux)^{pk}$ or to the suffix $(yu)^{pk}$. We consider the set $P'$ made of the corresponding positions in \posc{w'}:
	\[
	 \begin{array}{ll}
		 P'  =  & \{i \mid \text{$i \in \posmek{w}$ and $i \leq |(ux)^{pk}|$}\} \quad \cup\\
		  & \{i - |u|+|v| \mid \text{$i \in \posmek{w}$ and $i > |(ux)^{pk}u|$}\}.
	\end{array}
	\]
	Since $\eta(u) = \eta(v)$, one may verify from the definition that $\sigma_\eta(w,\poslek{w}) = \sigma_\eta(w',P')$. Thus, Corollary~\ref{cor:eqbij} yields $w \eqmek w'$. Since the languages recognized by $\alpha$ are unions of \eqmek-classes, we get \mbox{$\alpha(w)\! =\! \alpha(w')$}. By definition, this yields $(sq)^{\omega}s(rs)^\omega = (sq)^{\omega}t(rs)^\omega$.
	
	\smallskip
	
	We turn to the implication $\ref{itm:em1} \Rightarrow \ref{itm:em2}$. Assume that~\ref{itm:em1} holds and consider $q,r,s,t \in M$ such that $s \canec t$. We show that $(sq)^{\omega}s(rs)^\omega = (sq)^{\omega}t(rs)^\omega$. We start with a preliminary remark. By hypothesis, the second assertion in Theorem~\ref{thm:cupol} holds (this is the special case of~\ref{itm:em1} when $q = r = 1_M$). Thus, Theorem~\ref{thm:cupol} yields the following property:
	\begin{equation} \label{eq:mdet23}
		x^{\omega+1} = x^\omega y x^\omega \quad  \text{for all $x,y \in M$ such that $x \canec y$}.
	\end{equation}
	Since $s \canec t$, Lemma~\ref{lem:transclos} yields $s_0,\dots,s_n \in M$ such that $s_0 =s$, $s_n = t$ and $(s_i,s_{i+1})$ is a \Cs-pair for all $i < n$. We now prove that $(sq)^{\omega}s_i(rt)^\omega = (sq)^{\omega}s_{i+1}(rt)^\omega$ for every $i < n$. Since $s = s_0$ and $t = s_n$, this yields the desired result by transitivity. We fix $i < n$. By definition, $s \canec t \canec s_{i}$. Hence, since \canec is a congruence, we get $sq \canec s_iq$ and $rt \canec rs_i$. It then follows from~\eqref{eq:mdet23} that $(sq)^{\omega+1} = (sq)^{\omega}s_iq(sq)^{\omega}$ and $(rs)^\omega = (rs)^\omega rs_i(rs)^\omega$. Thus,
	\[
	\begin{array}{lllll}
		(sq)^{\omega} & = & ((sq)^{\omega}s_iq(sq)^{\omega})^\omega &=& (sq)^{\omega}(s_iq(sq)^{\omega})^\omega. \\
		(rs)^\omega & = & ((rs)^\omega rs_i(rs)^\omega)^\omega &=& ((rs)^\omega rs_i)^\omega (rs)^\omega.
	\end{array}
	\]
	Moreover, we have $(s_iq(sq)^{\omega})^\omega s_i ((rs)^\omega rs_i)^\omega =(s_iq(sq)^{\omega})^\omega s_{i+1}((rs)^\omega rs_i)^\omega$ since $(s_{i},s_{i+1})$ is a \Cs-pair and~\ref{itm:em1} holds. Hence,
	\[
	\begin{array}{lll}
		(sq)^{\omega}s_i(rs)^\omega & = & (sq)^{\omega} (s_iq(sq)^{\omega})^\omega s_i ((rs)^\omega rs_i)^\omega (rs)^\omega \\
		&= & (sq)^{\omega} (s_iq(sq)^{\omega})^\omega s_{i+1} ((rs)^\omega rs_i)^\omega (rs)^\omega \\
		&= &(sq)^{\omega}s_{i+1}(rs)^\omega.
	\end{array}
	\]
	This concludes the proof for the implication $\ref{itm:em1} \Rightarrow \ref{itm:em2}$.
	
	\smallskip
	
	It remains to prove $\ref{itm:em2} \Rightarrow \ref{itm:em0}$. We assume that~\ref{itm:em2} holds and show that $\alpha$ is an \mdet{\Cs}-morphism. Let $N = M/{\canec}$ and recall that $N$ is a monoid since \canec is a congruence by Lemma~\ref{lem:caquot}. We write $\eta = \ctype{\cdot} \circ \alpha: A^* \to N$ which is a \Cs-morphism by Lemma~\ref{lem:smult}. We let $k = |M|$ and consider the equivalence \eqmek on $A^*$. We prove the following property:
	\begin{equation} \label{eq:mdetcl}
		\text{for every $w,w' \in A^*$,} \quad w \eqmek w' \Rightarrow \alpha(w) = \alpha(w').
	\end{equation}
	This implies that every language recognized by $\alpha$ is a union of \eqmek-classes. Together with Proposition~\ref{prop:opcar} this yields that every language recognized by $\alpha$ belongs to \mdet{\Cs} since $\eta$ is a \Cs-morphism. We now concentrate on~\eqref{eq:mdetcl}. Let $w,w' \in A^*$ such that $w \eqmek w'$. We show that $\alpha(w)=\alpha(w')$. We first use our hypothesis to prove the following lemma.

	\begin{lemma} \label{lem:mbij}
		There exist $P \subseteq \posc{w}$ and $P' \subseteq \posc{w'}$ which satisfy $\poslp{\alpha}{1}{w} \subseteq P$, $\posrp{\alpha}{1}{w'} \subseteq P'$ and $\sigma_\eta(w,P) = \sigma_\eta(w',P')$.
	\end{lemma}

	\begin{proof}
		We write $Q = \posmek{w}$ and $Q' =\posmek{w'}$. Since $w \eqmek w'$, we have $\sigma_\eta(w,Q) = \sigma_\eta(w',Q')$. In particular, we have $|Q| = |Q'|$ and there is a unique increasing bijection \mbox{$f: Q \to Q'$}. Since $\alpha$ satisfies~\ref{itm:em2}, one may verify from Theorem~\ref{thm:cupol} that it is a \upol{\Cs}-morphism. Thus, since $k = |M|$, Lemma~\ref{lem:isdet} yields $\poslp{\alpha}{1}{w}\!  \subseteq\! \poslek{w}\! \subseteq\! Q$ and $\posrp{\alpha}{1}{w'}\! \subseteq\! \posrek{w'}\! \subseteq\! Q'$. Therefore, the set $f(\poslp{\alpha}{1}{w}) \subseteq Q'$ is well-defined. We define $P' = f(\poslp{\alpha}{1}{w}) \cup \posrp{\alpha}{1}{w'} \subseteq Q'$ and $P = f\inv(P')$. It is clear from the definition that $\poslp{\alpha}{1}{w} \subseteq P$ and $\posrp{\alpha}{1}{w'} \subseteq P'$. Moreover, since $\sigma_\eta(w,Q) = \sigma_\eta(w',Q')$, it is immediate from the definition that $\sigma_\eta(w,P) = \sigma_\eta(w',P')$ as well.
\end{proof}
	
	Let $(s_0,a_1,s_1,\dots,a_n,s_n) = \sigma_\alpha(w,P)$ and $(t_0,b_1,t_1,\dots,b_m,t_m) = \sigma_\alpha(w',P')$. Since $\sigma_\eta(w,P) = \sigma_\eta(w',P')$, we get $n = m$, $a_i = b_i$ for $1 \leq i \leq n$ and $s_i \canec t_i$ for $0 \leq i \leq n$ by definition of $\eta$. Therefore, we have $\alpha(w) = s_0a_1s_1 \cdots a_ns_n$ and $\alpha(w') = t_0a_1t_1 \cdots a_nt_n$ by definition of $\alpha$-snapshots (for the sake of avoiding clutter, we abuse terminology and write $a_i$ for $\alpha(a_i)$). We now prove that $s_0a_1s_1 \cdots a_ns_n = t_0a_1t_1 \cdots a_nt_n$. For all $h$ such that $0 \leq h \leq n$, we write $q_h = s_0a_1 \cdots s_{h-1}a_h$ and $r_h = a_{h+1}t_{h+1} \cdots a_nt_n$ ($q_0 = 1_M$ and $r_n = 1_M$). Since $\poslp{\alpha}{1}{w} \subseteq P$ and $\posrp{\alpha}{1}{w'} \subseteq P'$, one may verify from the definitions that $q_hs_h \Rrel q_h$ and $t_hr_h \Lrel r_h$ for $0 \leq h \leq n$. We prove that $q_hs_hr_h = q_ht_hr_h$ for $0 \leq h \leq n$.
	
	Let us first explain why this implies $\alpha(w) = \alpha(w')$. One may verify from the definition that  $q_hs_hr_h = q_{h+1}t_{h+1}r_{h+1}$ for $0 \leq h < n$. Together with $q_hs_hr_h = q_ht_hr_h$, this yields $q_ht_hr_h = q_{h+1}t_{h+1}r_{h+1}$. By transitivity, we get $q_0t_0r_0 = q_{n}t_nr_n$. Together with the equality $q_0s_0r_0 = q_0t_0r_0$, this yields $q_0s_0r_0 = q_{n}t_nr_n$. Hence, we get $s_0a_1s_1 \cdots a_ns_n = t_0a_1t_1 \cdots a_nt_n$, \emph{i.e.}  $\alpha(w) = \alpha(w')$ as desired.
	
	We now fix an index $h$ such that $0 \leq h \leq n$ and show that $q_hs_hr_h = q_ht_hr_h$. Recall that $q_hs_h \Rrel q_h$ and $t_hr_h \Lrel r_h$. Hence, we get $x,y \in M$ such that $q_h = q_hs_hx = q_h(s_hx)^\omega$ and $r_h = yt_hr_h = (yt_h)^\omega r_h$. Since $s_h \canec t_h$ and $\canec$ is a congruence, we get $ys_h \canec yt_h$ which yields $(yt_h)^{\omega+1} = (yt_h)^{\omega} ys_h (yt_h)^{\omega}$ by~\ref{itm:em2}.  Thus,   $(yt_h)^{\omega} = ((yt_h)^{\omega} ys_h (yt_h)^{\omega})^\omega = ((yt_h)^{\omega} ys_h)^\omega  (yt_h)^{\omega}$. Moreover, since $s_h \canec t_h$ and $\alpha$ satisfies~\ref{itm:em2}, we have,
	\[
	(s_hx)^\omega s_h ((yt_h)^{\omega} ys_h)^\omega = (s_hx)^\omega t_h ((yt_h)^{\omega} ys_h)^\omega.
	\]
	We now multiply by $(yt_h)^{\omega}$ on the right. This yields $(s_hx)^\omega s_h (yt_h)^{\omega} = (s_hx)^\omega t_h (yt_h)^{\omega}$. Hence, since we have $q_h = q_h(s_hx)^\omega$ and $r_h = (yt_h)^\omega r_h$, it  follows that $q_hs_hr_h = q_h t_h r_h$ as desired which completes the proof.
\end{proof}

%% file: deth.tex
We present a construction process which take a single input class \Cs and uses \ldeto and \rdeto to build a hierarchy which classifies the languages in \upol{\Cs}. Then, we prove that mixed polynomial closure is a key ingredient for investigating these hierarchies.

\subsection{Definition}

The definition is motivated by a result of~\cite{pzupol,pzupol2}. Let \Cs be a \vari. We define the alternating polynomial closure of \Cs (\adet{\Cs}) as the least class containing \Cs and closed under both left deterministic and right deterministic marked products and under disjoint union. The following theorem is proved in~\cite{pzupol,pzupol2}.

\begin{theorem} \label{thm:apol}
	If \Cs is a \vari, then $\upol{\Cs} = \adet{\Cs}$.
\end{theorem}

In view of Theorem~\ref{thm:apol}, given a \vari \Cs, alternately applying \ldeto and \rdeto builds a classification of \upol{\Cs}. For all $n \in \nat$, there are two levels \ldetn{\Cs} and \rdetn{\Cs}. We let $\ldetp{0}{\Cs} = \rdetp{0}{\Cs} = \Cs$. Then, for every $n \geq 1$, we define $\ldetn{\Cs} = \ldet{\rdetp{n-1}{\Cs}}$ and $\rdetn{\Cs} = \rdet{\ldetp{n-1}{\Cs}}$. Clearly, the union of all levels \ldetn{\Cs} (or \rdetn{\Cs}) is exactly the class \adet{\Cs}, \emph{i.e.} \upol{\Cs} by Theorem~\ref{thm:apol}. In general these are strict hierarchies (we discuss a well-known example below) and the levels \ldetn{\Cs} and \rdetn{\Cs} are incomparable for every $n \geq 1$. This motivates the introduction of intermediary levels ``combining'' the two.

Consider two classes $\Ds_1$ and $\Ds_2$. We write $\Ds_1 \cap \Ds_2$ for the class made of all languages which belong simultaneously to $\Ds_1$ and $\Ds_2$. Moreover, we write $\Ds_1 \vee \Ds_2$ for the least Boolean algebra containing both $\Ds_1$ and $\Ds_2$. We consider the additional levels \idetn{\Cs} and \jdetn{\Cs}. The following statement can be verified from Theorem~\ref{thm:detclos}.

\begin{corollary} \label{cor:detclos}
	Let \Cs be a \vari. For every $n \in \nat$, \ldetn{\Cs}, \rdetn{\Cs}, \idetn{\Cs} and \jdetn{\Cs} are \varis.
\end{corollary}

A specific hierarchy of this kind is well-known. Its input \Cs is the class \pt of piecewise testable languages: the class \bpol{\stzer} with $\stzer = \{\emptyset,A^*\}$ as the trivial \vari. It is known that this hierarchy is strict. It admits many distinct characterizations based on algebra~\cite{twhiera,subDA} or logic~\cite{kwfo2alt2,kwfo2alt3} (we come back to the second point in Section~\ref{sec:logic}). Moreover, it is known~\cite{subDA} that membership is decidable for \ldetn{\pt}, \rdetn{\pt} and \idetn{\pt} for every $n \in \nat$. This can be reproved using Corollary~\ref{cor:lrmemb} and the decidability of \pt-membership~\cite{simonthm}. It is also know~\cite{almjoin,kljoin1,kljoin2} that for every $n \in \nat$, membership is decidable for \jdetn{\pt}. We explain below that part of these results can also be reproved using Corollary~\ref{cor:pmemb}.

\smallskip

We complete the definition of determinsitic hierarchies with a useful result. We prove that when applying \ldeto, \rdeto or \mdeto to some level in a deterministic hierarchy, one may strengthen the requirements on marked products. Let \Cs be a \vari. We say that a marked product $L_0a_1L_1 \cdots a_nL_n$ is \emph{left (resp. right, mixed) \Cs-deterministic} when there exist $H_0,\dots,H_n \in \Cs$ such that $L_i \subseteq H_i$ for each $i \leq n$ and $H_0a_1H_1 \cdots a_nH_n$ is left (resp. right, mixed) deterministic. In other words, $L_0a_1L_1 \cdots a_nL_n$ can be ``over-approximated''  by a left (resp. right, mixed) deterministic marked product of languages in \Cs. We use Lemma~\ref{lem:isdet} and Proposition~\ref{prop:opcar} to prove the following result.

\begin{proposition} \label{prop:cguard}
	Let $\Cs,\Ds$ be two \varis such that $\Cs \subseteq \Ds$ and $\Ds \subseteq \upol{\Cs}$. Moreover, consider a language $L$ in \ldet{\Ds} (resp. \rdet{\Ds}, \mdet{\Ds}). Then, $L$ is a finite union of left (resp. right, mixed) \Cs-deterministic marked products of languages in \Ds.
\end{proposition}

\begin{proof}
	We treat the case when $L \in \mdet{\Ds}$ (the other cases are symmetrical). Proposition~\ref{prop:opcar} yields a \Ds-morphism $\alpha: A^* \to M$ and $k \in \nat$ such that $L$ is a union of \eqmak-classes. Thus, it suffices to prove that each \eqmak-class is a finite union of mixed \Cs-deterministic marked products of languages in~\Ds. Let $w \in A^*$ and $K \subseteq A^*$ its \eqmak-class. For every $u \in A^*$ such that $u \eqmak w$, we build a language $H_u \subseteq A^*$ defined by a mixed \Cs-deterministic marked product of languages in \Ds and such that $u \in H_u \subseteq L$. Moreover, we show that while there might be infinitely many words $u \in A^*$ such that $u \eqmak w$, there are only finitely many distinct languages $H_u$. Altogether, it will follow that $K$ is equal to the \emph{finite} union of all languages $H_u$ for $u \in A^*$ such that $u \eqmak w$ which completes the proof. For the construction, we consider the canonical equivalence \canec on $M$ and write $N = {M}/{\canec}$. We also define $\eta$ as the morphism $\eta = \ctype{\cdot} \circ \alpha: A^* \to N$.  By Lemma~\ref{lem:smult}, $\eta$ is a \Cs-morphism. 
	
	We now consider $u \in A^*$ such that $u \eqmak w$ and build $H_u$. We write $P_u =  \posmp{\eta}{k|M|}{u}$. One may verify from the definition that $|P_u| \leq 2|N|^{k|M|}$ (the key point is that this bound is independent from $u$). We let $(s_0,a_1,s_1,\dots,a_n,s_n) = \sigma_\alpha(u,P_u)$ and define \mbox{$H_u = \alpha\inv(s_0)a_1\alpha\inv(s_1) \cdots a_n\alpha\inv(s_n)$}. Since \mbox{$|P_u| \leq 2|N|^{k|M|}$}, we know that $H_u$ is the marked product of at most \mbox{$2|N|^{k|M|}+1$} languages recognized by $\alpha$. Hence, there are only finitely many languages $H_u$ for $u \in A^*$ such that $u \eqmak w$. Moreover, the languages in the product defining $H_u$ belong to \Ds by hypothesis on $\alpha$. We now prove that this marked product is mixed \Cs-deterministic. Let $(t_0,a_1,t_1,\dots,a_n,t_n) = \sigma_\eta(u,P_u)$. Since we have $P_u =  \posmp{\eta}{k|M|}{u}$ and $\eta$ is a \Cs-morphism, Lemma~\ref{lem:isdet} implies that $\eta\inv(t_0)a_1\eta\inv(t_1) \cdots a_n\eta\inv(t_n)$ is a mixed deterministic marked product of languages in \Cs. Moreover, since $\eta = \ctype{\cdot} \circ \alpha$, we have $\alpha\inv(s_i) \subseteq \eta\inv(t_i)$ for every $i \leq n$. Thus, the product $\alpha\inv(s_0)a_1\alpha\inv(s_1) \cdots a_n\alpha\inv(s_n)$ which defines $H_u$ is mixed \Cs-deterministic as desired. 
	
	It remains to prove that  $u \in H_u \subseteq L$. That $u \in H_u$ is immediate by definition since $(s_0,a_1,s_1,\dots,a_n,s_n) = \sigma_\alpha(u,P_u)$. Hence, we let $v \in H_u$ and prove that $v \in L$, \emph{i.e.} $v \eqmak u$. By definition of $H_u$, we know that there exists a set $Q \subseteq \pos{w}$ such that $\sigma_\alpha(v,Q) = (s_0,a_1,s_1,\dots,a_n,s_n) = \sigma_\alpha(u,P_u)$. Moreover, since $\Ds \subseteq \upol{\Cs}$ by hypothesis, we know $\alpha$ is a \upol{\Cs}-morphism. Therefore, $\posmak{w} \subseteq \posmp{\eta}{k|M|}{u} =P_u$ by Lemma~\ref{lem:isdet}. Hence, since $\sigma_\alpha(v,Q) = \sigma_\alpha(u,P_u)$, one may verify that there exists $Q' \subseteq Q$ such that $\sigma_\alpha(v,Q') = \sigma_\alpha(u,\posmak{u})$ and Corollary~\ref{cor:eqbij} yields $v \eqmak u$ as desired.
\end{proof}

\subsection{Connection with mixed polynomial closure}

We associated \emph{four} closely related hierarchies to every \vari \Cs. Their construction processes can be unified using \mdeto. As seen in Section~\ref{sec:polc}, \mdeto is not idempotent: given a \vari \Ds, it may happen that \mdet{\Ds} is \emph{strictly} included in \mdet{\mdet{\Ds}}. Hence, a hierarchy is built by applying \mdeto iteratively to \Ds. It turns out that deterministic hierarchies can be built in this way. First, the levels \ldetn{\Cs} and \rdetn{\Cs} are built from \ldet{\Cs} and \rdet{\Cs} using only \mdeto.

\begin{lemma} \label{lem:halt}
	Let \Cs be a \vari. Then, we have $\ldetp{n+1}{\Cs} = \mdet{\rdetp{n}{\Cs}}$ and $\rdetp{n+1}{\Cs} = \mdet{\ldetp{n}{\Cs}}$ for every $n \geq 1$.
\end{lemma}

\begin{proof}
	We prove that $\ldetp{n+1}{\Cs} = \mdet{\rdetp{n}{\Cs}}$ (the other property is symmetrical). Since $\ldetp{n+1}{\Cs} = \ldet{\rdetn{\Cs}}$ by definition, the left to right inclusion is immediate.  We concentrate on the converse one. We write $\Ds =  \ldetp{n-1}{\Cs}$ for the proof. By definition, we need to prove that	$\mdet{\rdet{\Ds}} \subseteq \ldet{\rdet{\Ds}}$.
	
	Every language in $\mdet{\rdet{\Ds}}$ is a finite disjoint union of mixed deterministic marked products of languages in \rdet{\Ds}. Hence, since \ldet{\rdet{\Ds}} is closed under union, it suffices to prove that if $L = L_0a_1L_1\cdots a_kL_k$ is a mixed deterministic marked product such that $L_1,\dots,L_k \in \rdet{\Ds}$, then $L \in \ldet{\rdet{\Ds}}$. We proceed by induction on $k$. If $k = 0$, then $L = L_0 \in \rdet{\Ds} \subseteq \ldet{\rdet{\Ds}}$ and we are finished. Assume now that $k \geq 1$. Since $L_0a_1L_1\cdots a_kL_k$ is mixed deterministic, we know that the marked concatenation $(L_0a_1L_1\cdots L_{k-1})a_k(L_k)$ is either left deterministic or right deterministic. We handle these two cases separately. Assume first that $(L_0a_1L_1\cdots a_{k-1}L_{k-1})a_k(L_k)$ is left deterministic. One may verify that the product of $k-1$ languages $L_0a_1L_1\cdots a_{k-1}L_{k-1}$ remains a mixed deterministic product. Hence, $L_0a_1L_1\cdots a_{k-1}L_{k-1} \in \ldet{\rdet{\Ds}}$ by induction. Moreover, since $L_0 \in \rdet{\Ds} \subseteq \ldet{\rdet{\Ds}}$ and the marked concatenation $(L_0a_1L_1\cdots a_{k-1}L_{k-1})a_k(L_k)$ is left deterministic, we get $L_0a_1L_1\cdots a_kL_k \in \ldet{\rdet{\Ds}}$ from Lemma~\ref{lem:dleast}. Assume now that $(L_0a_1L_1\cdots a_{k-1}L_{k-1})a_k(L_k)$ is right deterministic. Hence, $L_{k-1}a_kL_k$ is right deterministic. Thus, since $L_{k-1},L_k \in \rdet{\Ds}$, we obtain from Lemma~\ref{lem:dleast} that $L_{k-1}a_kL_k \in \rdet{\Ds}$. One may now verify that the product of $k-1$ languages $L_0a_1L_1\cdots a_{k-1}(L_{k-1}a_kL_k)$ is mixed deterministic. Thus, we obtain from induction on $k$ that $L = L_0a_1L_1\cdots a_kL_k \in \ldet{\rdet{\Ds}}$ This completes the proof.
\end{proof}

Moreover, the levels \idetn{\Cs} can all be built from $\ldet{\Cs} \cap \rdet{\Cs}$ using only \mdeto (the proof is based on the algebraic characterizations of \ldeto, \rdeto and \mdeto).

\begin{theorem} \label{thm:intcarn}
	If $\Cs$ is a \vari, then $\idetp{n+1}{\Cs} = \mdet{\idetp{n}{\Cs}}$ for every $n \geq 1$.
\end{theorem}

\begin{proof}
	We first present a preliminary lemma which applies to all classes of the form $\Ds_1 \cap \Ds_2$. 
	
	\begin{lemma} \label{lem:interweak}
	Let $\Ds_1,\Ds_2$ be \varis and $\Ds = \Ds_1 \cap \Ds_2$. Let $\alpha: A^* \to M$ be a surjective  morphism. The equivalence \caned on $M$ is the least one containing both \canedo and \canedt.
	\end{lemma}

\begin{proof}
	We write $\equiv$ for the least equivalence of $M$ containing $\canedo$ and $\canedt$. We prove that $\equiv = \caned$. Clearly, $\equiv \subseteq \caned$ since $\caned$ contains $\canedo$ and $\canedt$ (this is immediate since $\Ds_1$ and $\Ds_2$ both contain \Ds). Conversely, let $s,t \in M$ such that $s \caned t$. We show that $s \equiv t$. Let $F \subseteq M$ be the $\equiv$-class of $s$. We show that $t \in F$. By definition of $\equiv$, $F$ is a union of $\canedo$-classes and a union of $\canedt$-classes. Thus, Lemma~\ref{lem:smult} yields that $\alpha\inv(F)$ belongs to $\Ds_1 \cap \Ds_2 = \Ds$. Since $s \in F$ and $s \caned t$, we get $t \in F$ by definition of $\caned$.
\end{proof}

	We may now prove Theorem~\ref{thm:intcarn}. We fix a \vari \Cs and $n \geq 1$. We have to prove that $\idetp{n+1}{\Cs} = \mdet{\idetp{n}{\Cs}}$. We start with right to left inclusion. It is immediate that $\mdet{\idetp{n}{\Cs}}$ is included in both \mdet{\ldetn{\Cs}} and \mdet{\rdetn{\Cs}}. Since these classes are equal to \rdetp{n+1}{\Cs} and \ldetp{n+1}{\Cs} respectively by Lemma~\ref{lem:halt}, we get $\mdet{\idetp{n}{\Cs}} \subseteq \idetp{n+1}{\Cs}$.
	
	We turn to the converse inclusion. For the sake of avoiding clutter, we write \Ds for the class \idetn{\Cs}. Let $L\in\idetp{n+1}{\Cs}$. We show that $L \in \mdet{\Ds}$. By Theorem~\ref{thm:detclos}, \Ds and \mdet{\Ds} are \varis. Hence, by Proposition~\ref{prop:synmemb}, it suffices to verify that the syntactic morphism $\alpha: A^* \to M$ of $L$ satisfies the characterization of \mdet{\Ds} given in Theorem~\ref{thm:cmdet}. Let $q,r,s,t \in M$ such that $s \caned  t$. We prove that $(sq)^{\omega}s(rs)^\omega = (sq)^{\omega}t(rs)^\omega$. Since $\Ds = \idetn{\Cs}$, Lemma~\ref{lem:interweak} yields $p_0,\dots,p_\ell \in M$ such that $p_0= s$, $p_\ell = t$ and for $i < \ell$, either $p_i \sim_{\ldetn{\Cs}} p_{i+1}$ or $p_i \sim_{\rdetn{\Cs}} p_{i+1}$. We prove that for all $i < \ell$, we have $(sq)^\omega p_i(rs)^\omega = (sq)^\omega p_{i-1}(rs)^\omega$. By transitivity, this implies that $(sq)^{\omega}s(rs)^\omega = (sq)^{\omega}t(rs)^\omega$ as desired. We fix $i < \ell$ for the proof. We only treat the case when $p_{i-1} \sim_{\ldetn{\Cs}} p_{i}$ (the case $p_{i-1} \sim_{\rdetn{\Cs}} p_{i}$ is symmetrical and left to the reader). With this hypothesis in hand, we prove that $p_i(rs)^\omega = p_{i-1}(rs)^\omega$ which implies the desired result.
	
	We have $L \in \rdet{\ldetn{\Cs}}$ by hypothesis. Consequently, its syntactic morphism $\alpha$ is a \rdet{\ldetn{\Cs}}-morphism by Proposition~\ref{prop:synmemb}. It is also clear that $\Cs \subseteq \ldetn{\Cs} \subseteq \upol{\Cs}$. Moreover, by hypothesis, we have $p_{i-1} \sim_{\ldetn{\Cs}} p_{i}$ and $(rs)^\omega$ is an idempotent. Finally, since \Cs is included in both \ldetn{\Cs} and \rdetn{\Cs}, the equivalences $\sim_{\ldetn{\Cs}}$ and $\sim_{\rdetn{\Cs}}$ are included in \canec. Hence, we have $s \canec p_i$ by definition which implies that  $\ctype{(rs)^\omega} \Lord \ctype{p_i}$. Altogether, it follows from Lemma~\ref{lem:lradv} that $p_i(rs)^\omega = p_{i-1}(rs)^\omega$ as desired. 
\end{proof}

A similar result holds for the levels \jdetn{\Cs}: they can all be built from $\ldet{\Cs} \vee \rdet{\Cs}$ using only \mdeto.

\begin{theorem} \label{thm:joincarn}
	If $\Cs$ is a \vari, then $\jdetp{n+1}{\Cs} = \mdet{\jdetp{n}{\Cs}}$ for every $n \geq 1$.
\end{theorem}

Theorem~\ref{thm:joincarn} has an interesting application. Since \mdeto preserves the decidability of membership by Corollary~\ref{cor:pmemb}, we get that for all \varis \Cs, if membership is decidable for \jdet{\Cs}, then this is also the case for \emph{all} levels \jdetn{\Cs}. This can be applied for $\Cs = \pt$. It is known that \jdet{\pt}~\cite{almjoin,kljoin1}. Thus, we lift this result to every level \jdetn{\pt} ``for free''. This reproves a result of~\cite{kljoin2}.

\begin{proof}[Proof of Theorem~\ref{thm:joincarn}]
	We fix a \vari \Cs and $n\geq 1$. Let us start with the inclusion $\jdetp{n+1}{\Cs}\subseteq\mdet{\jdetp{n}{\Cs}}$. By Theorem~\ref{thm:detclos}, \mdet{\jdetp{n}{\Cs}} is a \vari. Hence, it suffices to prove that \ldetp{n+1}{\Cs} and \rdetp{n+1}{\Cs} are included in \mdet{\jdetp{n}{\Cs}}. By symmetry, we only prove the former. By definition,  $\ldetp{n+1}{\Cs} = \ldet{\rdetn{\Cs}}$ which yields $\ldetp{n+1}{\Cs} \subseteq \mdet{\rdetn{\Cs}}$. Finally, since it is immediate by definition that $\rdetn{\Cs} \subseteq \jdetn{\Cs}$, we obtain the inclusion $\ldetp{n+1}{\Cs} \subseteq \mdet{\jdetp{n}{\Cs}}$ as desired which completes the proof for the left to right inclusion.
	
	We now prove that $\mdet{\jdetp{n}{\Cs}}$ is included in \jdetp{n+1}{\Cs}. We write $\Ds = \jdetp{n}{\Cs}$. Corollary~\ref{cor:detclos} implies that \Ds is a \vari. Moreover, it is immediate that $\Cs \subseteq \Ds \subseteq \upol{\Cs}$ (\upol{\Cs} is a \vari by Theorem~\ref{thm:pupol} and it contains both \ldetn{\Cs} and \rdetn{\Cs}). Hence, Proposition~\ref{prop:cguard} implies that every language in \mdet{\Ds} is a disjoint union of mixed \Cs-deterministic marked products of languages in \Ds. It now remains to prove that for every mixed \Cs-deterministic marked product $L = L_0a_1L_1 \cdots a_nL_n$ such that $L_0,\dots,L_n \in \Ds$, we have $L \in \jdetp{n+1}{\Cs}$. The definition yields $H_i \in \Cs$ for each $i \leq n$ such that $L_i \subseteq H_i$ and $H_0a_1H_1 \cdots a_nH_n$ is mixed deterministic.
	
	Consider $i \leq n$. We have $L_i \in \Ds$ and $\Ds = \jdetn{\Cs}$. Hence, by definition $L_i$ is a Boolean combination of languages in \ldetn{\Cs} and \rdetn{\Cs}. We can put the Boolean combination in disjunctive normal form. Moreover, since \ldetn{\Cs} and \rdetn{\Cs} are \varis by Corollary~\ref{cor:detclos}, each disjunct is the intersection of a single language in \ldetn{\Cs} with a single language in \rdetn{\Cs}. Altogether, it follows that  $L_i$ is a finite union of languages $P_i \cap Q_i$ with $P_i \in \ldetn{\Cs}$ and $Q_i \in\rdetn{\Cs}$. Moreover, since $L_i \subseteq H_i \in \Cs$, we may assume without loss of generality that all languages $P_i$ and $Q_i$ are included in $H_i$ as well (otherwise we may replace them by $P_i\cap H_i$ and $Q_i\cap H_i$).  Consequently, since marked concatenation distributes over union, we obtain that $L = L_0a_1L_1 \cdots a_nL_n$ is a finite union of products $(P_0 \cap Q_0)a_1(P_1 \cap Q_1) \cdots a_n(P_n \cap Q_n)$ such that $P_i \in \ldetn{\Cs}$ and $Q_i \in\rdetn{\Cs}$ are included in $H_i$ for every $i \leq n$. It now suffices to prove that every such marked product belongs to \jdetp{n+1}{\Cs}. Since $H_0a_1H_1 \cdots a_nH_n$ is mixed deterministic, it is also unambiguous. Hence, since $P_i$ and $Q_i$ are included in $H_i$ for every $i \leq n$, one may verify that the language $(P_0 \cap Q_0)a_1(P_1 \cap Q_1) \cdots a_n(P_n \cap Q_n)$ is equal to the intersection,
	\[
	\left(P_0a_1P_1 \cdots a_nP_n\right) \cap \left(Q_0a_1Q_1 \cdots a_nQ_n\right).
	\]
	Finally, it is clear that $P_0a_1P_1 \cdots a_nP_n$ and $Q_0a_1Q_1 \cdots a_nQ_n$ are mixed deterministic marked products since this is the case for $H_0a_1H_1 \cdots a_nH_n$. By definition, it follows that they both belong to \mdet{\ldetn{\Cs}} and \mdet{\rdetn{\Cs}} respectively. Thus, we obtain $P_0a_1P_1 \cdots a_nP_n \in \rdetp{n+1}{\Cs}$ and $Q_0a_1Q_1 \cdots a_nQ_n \in \ldetp{n+1}{\Cs}$ by Lemma~\ref{lem:halt}. Hence, the intersection of these two languages belongs to \jdetp{n+1}{\Cs} as desired.
\end{proof}

%% file: logic.tex
We now look at quantifier alternation hierarchies for two-variable first-order logic over words (\fod). We characterize several hierarchies of this kind with  mixed polynomial closure.

\subsection{Definitions}

We first recall the definition of first-order logic over words. We view a word $w \in A^*$ as a logical structure. Its domain is the set $\pos{w} = \{0,\dots,|w|+1\}$ of positions in $w$. A position $i$ such that $1 \leq i \leq |w|$ carries a label in $A$. On the other hand, $0$ and \mbox{$|w|+1$} are artificial \emph{unlabeled} positions. We use first-order logic (\fo) to express properties of words~$w$: a formula can quantify over the positions in $w$ and use a predetermined set of predicates to test properties of these positions. We also allow two constants ``$min$'' and ``$max$''  interpreted as the artificial unlabeled positions $0$ and $|w|+1$. Given a formula $\varphi(x_1,\dots,x_n)$ with free variables $x_1,\dots,x_n$, $w \in A^*$ and $i_1,\dots,i_n \in \pos{w}$, we write $w \models \varphi(i_1,\dots,i_n)$ to indicate that $w$ satisfies $\varphi$ when $x_1,\dots,x_n$ are interpreted as the positions $i_1,\dots,i_n$. As usual, a sentence $\varphi$ is a formula without free variables. It defines the language $L(\varphi) = \{w \in A^* \mid w \models \varphi\}$. We use standard predicates. For each $a \in A$, we use a unary predicate (also denoted by~$a$)  selecting all positions labeled by ``$a$''. We also use three binary predicates: equality ``$=$'', the (strict) linear order ``$<$'' and the successor ``$+1$''.

\begin{example}
	The language $A^*aA^*bA^*c$ is defined by the following sentence fo first-order logic:
 $(\exists x \exists y\ (x < y) \wedge a(x) \wedge b(y)) \wedge (\exists x\ c(x) \wedge (x+1 = max))$. 
\end{example}

A \emph{fragment} of first-order logic consists in the specification of a (possibly finite) set $V$ of variables and a set \Fs of \fo formulas using only the variables in $V$ which contains all quantifier-free formulas and is closed under disjunction, conjunction and  quantifier-free substitution (if $\varphi \in  \Fs$, replacing a quantifier-free sub-formula of $\varphi$ with another quantifier-free formula in \Fs yields a new formula in \Fs). If \frS is a set of predicates and \Fs is a fragment, we let $\Fs(\frS)$ be the class containing all languages $L(\varphi)$ where $\varphi$ is a sentence of \Fs using only the predicates in \frS, equality \emph{and} the label predicates.

In this paper, we use generic sets of predicates which are built from an arbitrary input class \Cs. There are two of them. The first one, written $\frI_{\Cs}$, contains a binary ``infix'' predicate $I_L(x,y)$ for every $L \in \Cs$. Given $w \in A^*$ and two positions $i,j \in \pos{w}$, we have $w \models I_L(i,j)$ if and only if $i < j$ \emph{and} $\infix{w}{i}{j} \in L$. The second set, written $\frP_{\Cs}$,  contains a unary ``prefix'' predicate $P_L(x)$ for every $L \in \Cs$. Given $w \in A^*$ and a position $i \in \pos{w}$, we have $w \models P_L(i)$ if and only if $0 < i$ \emph{and} $\infix{w}{0}{i} \in L$. The predicates in $\frP_{\Cs}$ can be expressed by those in $\frI_{\Cs}$: $P_L(x)$ is equivalent to $I_L(min,x)$. In practice, we consider the sets $\frP_{\Cs}$ when \Cs is either a \emph{group \vari} \Gs or its \wsuit extension $\Gs^+$.  This is motivated by the following lemma.

\begin{lemma} \label{lem:gensig}
	If \Gs is a group \vari and \Fs is a fragment of \fo, then $\Fs(\frI_{\Gs}) = \Fs(<,\frP_{\Gs})$ and $\Fs(\frI_{\Gs^+}) = \Fs(<,+1,\frP_{\Gs})$.
\end{lemma}

\begin{proof}
	We first prove the inclusions $\Fs(<,\prefsigg) \subseteq \Fs(\infsigg)$ and $\Fs(<,+1,\prefsigg) \subseteq \Fs(\infsiggp)$. The formula $x < y$ is equivalent to $I_{A^*}(x,y)$ ($I_{A^*}$ belongs to \infsigg and \infsiggp since \Gs is a \vari which yields $A^* \in \Gs$). Moreover, for all $L \in \Gs$, the formula $P_L(x)$ is equivalent to $I_L(min,x)$ (again, $I_L$ belongs to both \infsigg and \infsiggp).  It follows that $\Fs(<,\prefsigg) \subseteq\Fs(\infsigg)$. Finally, the formula $x+1 = y$ is equivalent to $I_{\{\veps\}}(x,y)$ (which is available in \infsiggp as $\{\veps\} \in \Gs^+$ but not necessarily in \infsigg). Thus, we~get $\Fs(<,+1,\prefsigg) \subseteq \Fs(\infsiggp)$. We turn to the converse inclusions.
	
	Let us start with $\Fs(\infsigg) \subseteq \Fs(<,\prefsigg)$. By definition of fragments, it suffices to prove that for each $L \in \Gs$, the atomic formula $I_L(x,y)$ is equivalent to a quantifier-free formula of $\Fs(<,\prefsigg)$. Proposition~\ref{prop:genocm} yields a \Gs-morphism $\eta: A^* \to G$ recognizing $L$. We have $L = \alpha\inv(F)$ for some $F \subseteq N$. Since \Gs is a group \vari, $G$ is a group by Lemma~\ref{lem:gmorph}. Let $T = \{(g,a,h) \in G \times A \times G \mid (g\alpha(a))\inv h \in F\}$. Since $\alpha\inv(g) \in \Gs$, we know that $P_{\alpha\inv(g)}$ is a predicate in \prefsigg for all $g \in \Gs$. Hence, the following is a quantifier-free formula of $\Fs(<,\prefsigg)$:
	\[
	\varphi(x,y) := (x < y) \wedge \Big(\bigvee_{(g,a,h) \in T} \big(P_{\alpha\inv(g)}(x) \wedge a(x) \wedge P_{\alpha\inv(h)}(y)\big)\Big).
	\]
	One may now verify that $I_L(x,y)$ is equivalent to $(x = min \wedge P_L(y)) \vee  \varphi(x,y)$ which is a quantifier-free formula of $\Fs(<,\prefsigg)$. This concludes the proof for $\Fs(\infsigg) \subseteq \Fs(<,\prefsigg)$.
	
	Finally, we prove that $\Fs(\infsiggp) \subseteq \Fs(<,+1,\prefsigg)$. By definition, it suffices to show that for every language $K \in \Gs^+$, the atomic formula $I_K(x,y)$ is equivalent to a quantifier-free formula of $\Fs(<,+1,\prefsigg)$. By definition of $\Gs^+$, there exists $L \in \Gs$ such that either $L = \{\veps\} \cup K$ or $L = A^+ \cap K$. Consequently, $I_K(x,y)$ is equivalent to either $I_{\{\veps\}}(x,y) \vee I_L(x,y)$ or $I_{A^+}(x,y) \wedge I_L(x,y)$. Since, $L \in \Gs$, we already proved above that $I_L(x,y)$ is equivalent to a quantifier-free formula of $\Fs(<,\prefsigg) \subseteq \Fs(<,+1,\prefsigg)$. Moreover, $I_{\{\veps\}}(x,y)$ is equivalent to $x+1 = y$ and $I_{A^+}$ is equivalent to $x < y \wedge \neg (x+1 = y)$. This concludes the proof.
\end{proof}

Lemma~\ref{lem:gensig} covers many important sets of predicates.  If \Gs is the trivial \vari $\stzer = \{\emptyset,A^*\}$, all predicates in $\frP_{\stzer}$ are trivial. Hence, we get the classes $\Fs(<)$ and $\Fs(<,+1)$. We also look at the class \md of \emph{modulo languages}: the Boolean combinations of languages $\{w \in A^* \mid |w| \equiv k \bmod m\}$ with $k,m \in \nat$ such that $k < m$. One may verify that in this case, we obtain $\Fs(<,MOD)$ and $\Fs(<,+1,MOD)$ where ``$MOD$'' is the set of \emph{modular predicates} (for all $k,m \in \nat$ such that $k < m$, it contains a unary predicate $M_{k,m}$ selecting the positions $i$ such that $i \equiv k \bmod m$). Finally, consider the class \abg of \emph{alphabet modulo testable languages}. If $w\in A^*$ and $a \in A$, we let $\#_a(w) \in \nat$ be the number of occurrences of ``$a$'' in $w$. \abg contains the Boolean combinations of languages $\{w \in A^* \mid \#_a(w) \equiv k \bmod m\}$ where $a \in A$ and $k,m \in \nat$ such that $k < m$ (these are the languages recognized by commutative groups). In this case, we get \mbox{$\Fs(<,AMOD)$} and $\Fs(<,+1,AMOD)$ where ``$AMOD$'' is the set of \emph{alphabetic modular predicates} (for all $a \in A$ and $k,m \in \nat$ such that $k < m$, it contains a unary predicate $M^a_{k,m}$ selecting the positions $i$ such $\#_a(\prefix{w}{i}) \equiv k \bmod m$).

\smallskip
\noindent
{\bf Quantifier alternation in \fod.} We now present the particular fragments that we consider. First, we write \fod for the fragment consisting of all first-order formulas which use at most \emph{two} distinct variables (which can be reused). In the formal definition, this boils down to picking a set $V$ of variables which has size two. We do not look at \fod itself. Instead, we consider its quantifier-alternation hierarchy. We first present the one of full first-order logic.

For every $n \in \nat$, we associate two fragments \sic{n} and \bsc{n} of \fo. We present the definition by induction on $n \in \nat$. When $n = 0$, we let $\sic{0} = \bsc{0}$ as the fragment containing exactly the quantifier-free formulas of \fo. Assume now that $n\geq 1$. We let \sic{n} as the least set of expressions which contains the \bsc{n-1} formulas and is closed under disjunction ($\vee$), conjunction ($\wedge$) and existential quantification ($\exists$). Moreover, we let \bsc{n} as the set of all Boolean combinations of \sic{n} formulas, \emph{i.e.} the least one containing \sic{n} and closed under  disjunction ($\vee$), conjunction ($\wedge$) and negation ($\neg$).

For every $n \in \nat$, we define \fodsn (resp. \fodbn) as the fragment containing all formulas which belong simultaneously to \fod and \sic{n} (resp. \bsc{n}). In this paper, we look at classes of the form $\fodbn(\infsigc)$ where \Cs is a \vari. Our results only apply in the case when \Cs is either a group \vari \Gs or its \wsuit extension $\Gs^+$ (in which case Lemma~\ref{lem:gensig} applies). Yet, we shall use the following general result which is specific to the first non-trivial level.

\begin{theorem} \label{thm:level1}
	Let \Cs be a \vari. Then, $\fodb{1}(\infsigc) = \bsc{1}(\infsigc) = \bpol{\Cs}$.
\end{theorem}

\begin{proof}
	That $\bsc{1}(\infsigc) = \bpol{\Cs}$ is proved in~\cite{PZ:generic18}. This is a specific case of the generic correspondence between the quantifier alternation hierarchies of \fo and concatenation hierarchies (which are built with \poln and \booln). The inclusion $\fodb{1}(\infsigc) \subseteq \bsc{1}(\infsigc)$ is trivial. Hence, it suffices to show that $\bpol{\Cs} \subseteq \fodb{1}(\infsigc)$. By definition, \bpol{\Cs} contains all Boolean combinations of marked products $L_0a_1L_1 \cdots a_nL_n$ with $L_0,\dots,L_n \in \Cs$. Since $\fodb{1}(\infsigc)$ is closed under Boolean operations, it suffices to prove that all marked products of this kind belong to $\fodsi{1}(\infsigc)$. We use induction to build a formula $\varphi_k(x)$ of $\fodsi{1}(\infsigc)$ for each $k \leq n$ which has one free variable $x$ and such that for all $w \in A^*$ and $i \in \pos{w}$, we have $w \models \varphi_k(i)$ if and only if $0 < i$ and $\prefix{w}{i} \in L_0a_1L_1 \cdots a_kL_k$. It will then follow that $L_0a_1L_1 \cdots a_nL_n$ is defined by the sentence $\varphi_n(max)$ of $\fodsi{1}(\infsigc)$, completing the proof. If $k = 0$, it suffices to define $\varphi_0(x) := I_{L_0}(min,x)$. Assume now that $k \geq 1$. It suffices to define $\varphi_k(x) := \exists y\ (\varphi_{k-1}(y) \wedge a_{k}(y) \wedge I_{L_k}(y,x))$ (the definition involves implicit renaming of the variables in $\varphi_{k-1}$, this is standard in \fod). Clearly $\varphi_{k}(x)$ is a formula of $\fodsi{1}(\infsigc)$.
\end{proof}

\subsection{\texorpdfstring{Properties of the quantifier alternation hierarchy of \fod}{Properties of the quantifier alternation hierarchy of FO2}}

We present results that we shall need to prove the language theoretic characterization of the quantifier alternation hierarchy of \fod by mixed polynomial closure.   First, we recall standard notions from finite model theory (yet, our terminology is tailored to the generic signatures \infsigc). For a morphism $\eta: A^* \to N$ and $k,n  \in  \nat$, we associate an equivalence \eqtlkn on $A^*$. Given a \vari \Cs and $n \in \nat$, we use the equivalences \eqtlkn where $\eta$ is a \Cs-morphism to characterize $\fodbn(\infsigc)$. Then, we present properties of these preorders which are specific to the paper. %We shall later use them to prove that $\fodbn(\infsigc)$ is included in another class.

\smallskip
\noindent
{\bf Definitions.} We start with two preliminary notions. The first one is standard. Given a \fod formula $\varphi$, the \emph{quantifier rank} of $\varphi$ is defined as the maximal nesting depth of quantifiers in $\varphi$. Moreover, for each morphism $\eta: A^* \to N$, we associate a set $\infsig{\eta}$ of predicates. For each language $L \subseteq A^*$ which \emph{is recognized} by $\eta$, the set $\infsig{\eta}$ contains the binary predicate $I_L$. Recall that $w \models I_L(i,j)$ if and only if $i < j$ and $\infix{w}{i}{j} \in L$. Note that \infsig{\eta} is a \emph{finite} set.

Let $\eta: A^* \to N$ be a morphism, $k \in \nat$ and $n \geq 1$. We associate a preorder \qtlkn which compares pairs $(w,i)$ where $w \in A^*$ and $i \in \pos{w}$. Consider $w,w' \in A^*$, $i \in \pos{w}$ and $i' \in \pos{w'}$. We let $w,i \qtlkn w',i'$ if and only if for every formula $\varphi(x)$ of $\fodsn(\infsig{\eta})$ with quantifier rank at most $k$ and at most one free variable ``$x$'' the following  implication holds:
\[
w \models \varphi(i) \Rightarrow w' \models \varphi(i').
\]
By definition, \qtlkn is a preorder and has \emph{finitely many upper sets}. This is standard: one may verify that there are finitely many non-equivalent formulas of $\fodsn(\infsig{\eta})$ with quantifier-rank at most $k$ (here, it is important that \infsig{\eta} is finite). Moreover, one may verify the following fact.

\begin{fct} \label{fct:dclass}
	Let $\eta: A^* \to N$ be a morphism, $k \in \nat$, $n \geq 1$, $w \in A^*$ and $i \in \pos{w}$. There exists a formula $\varphi(x)$ of $\fodsn(\infsig{\eta})$ with quantifier rank at most $k$ such that for all $w' \in A^*$ and $i' \in \pos{w'}$, we have $w' \models \varphi(i')$ if and only if $w,i \qtlkn w',i'$.
\end{fct}

We restrict the preorders $\qtlkn$ to single words in $A^*$. Let $w,w' \in A^*$. We let $w \qtlkn w'$ if and only if $w,0 \qtlkn w',0$. This is a preorder on $A^*$. Finally, we write \eqtlkn for the equivalence associated to \qtlkn:  $w \eqtlkn w'$ if and only if $w \qtlkn w'$ and $w'\qtlkn w$. Clearly, this equivalence has finite index. We use it to characterize the classes $\fodbn(\infsigc)$.

\begin{lemma}\label{lem:uset}
	Let \Cs be a \vari, $n \geq 1$ and $L \subseteq A^*$. Then, $L \in \fodbn(\infsigc)$ if and only if there exists a \Cs-morphism $\eta: A^* \to N$ and $k \in \nat$ such that $L$ is a union of $\eqtlkn$-classes.
\end{lemma}

\begin{proof}
	For the ``only if'' direction, assume that $L \in \fodbn(\infsigc)$ and let $\varphi$ be the sentence of $\fodbn(\infsigc)$ which defines $L$. Let $k \in \nat$ be the rank of $\varphi$. Proposition~\ref{prop:genocm} yields a \Cs-morphism $\eta: A^* \to N$ such that $\varphi$ is a formula of $\fodbn(\infsig{\eta})$. One may now verify that $L$ is a union of \eqtlkn-classes. For the ``if'' direction, consider a \Cs-morphism $\eta: A^* \to N$ and $k \in \nat$. We prove that every union of \eqtlkn-classes belongs to $\fodbn(\infsigc)$. As \eqtlkn has finite index, it suffices to show that all \eqtlkn-classes belong to $\fodbn(\infsigc)$. For every $u \in A^*$, Fact~\ref{fct:dclass} yields a formula $\psi_u(x)$ of $\fodsn(\infsig{\eta})$ with rank at most $k$ such that for every $v \in A^*$ and $j \in \pos{v}$, we have $v \models \psi_u(j)$ if and only if $u,0 \qtlkn v,j$. Let $w \in A^*$. We define, 
	\[
	\varphi_{w} = \psi_w(min) \wedge \left(\bigwedge_{w \qtlkn u \text{ and } u \not\eqtlkn w} \neg \psi_u(min)\right). 
	\]
	Note that the conjunction boils down to a finite one since there are finitely many non-equivalent $\fodsn(\infsig{\eta})$ of rank at most $k$. One may now verify that $\varphi_w$ defines the \eqtlkn-class of $w$ which concludes the proof: this is a $\fodbn(\infsigc)$ sentence since $\eta$ is a \Cs-morphism.
\end{proof}

We complete the definitions with an alternate inductive definition of the preorders \qtlkn. Roughly, it is inspired from \efgame games. Yet, formulating it as an inductive definition rather than a game is more convenient. We start with a preliminary notion. Let $\eta: A^* \to N$ be a morphism, $w,w' \in A^*$, $i \in \pos{w}$ and $i' \in \pos{w}$. We say that $(w,i)$ and $(w',i')$ are $\eta$-equivalent if and only if one of the three following conditions holds:
\begin{itemize}
	\item $i = i' = 0$, and $\eta(w) = \eta(w')$ or,
	\item $i = |w|+1$, $i' = |w'|+1$ and $\eta(w) = \eta(w')$ or,
	\item $i \in \posc{w}$, $i' \in \posc{w}$, the positions $i$ and $i'$ have the same label, $\eta(\prefix{w}{i}) = \eta(\prefix{w'}{i'})$ and $\eta(\suffix{w}{i}) = \eta(\suffix{w'}{i'})$
\end{itemize}
\begin{proposition}\label{prop:efgame}
	Let $\eta: A^* \to N$ be a morphism, $k \in \nat$, $n \geq 1$, $w,w' \in A^*$, $i \in \pos{w}$ and $i' \in \pos{w'}$. Then, we have $w,i \qtlkn w',i'$ if and only if the four following properties hold:
	\begin{enumerate}
		\item\label{ef:1} $(w,i)$ and $(w',i')$ are $\eta$-equivalent.
		\item\label{ef:2} If $n \geq 2$, then $w',i' \qtlp{k,n-1} w,i$.
		\item\label{ef:3} If $k \geq 1$, then for all $j \in \pos{w}$ such that $i < j$, there exists $j' \in \pos{w'}$ such that $i' < j'$, $\eta(\infix{w}{i}{j}) = \eta(\infix{w'}{i'}{j'})$ and $w,j \qtlp{k-1,n} w',j'$.
		\item\label{ef:4} If $k \geq 1$, then for all $j \in \pos{w}$ such that $j < i$, there exists $j' \in \pos{w'}$ such that $j' < i'$, $\eta(\infix{w}{j}{i}) = \eta(\infix{w'}{j'}{i'})$ and $w,j \qtlp{k-1,n} w',j'$.
	\end{enumerate}
\end{proposition}

\begin{proof}
	We start with the ``only if'' implication. Assume that $w,i \qtlkn w',i'$. We show that the four conditions in the lemma are satisfied. The first one is immediate as one may check $\eta$-equivalence using quantifier-free formulas in $\fodsi{n}(\infsig{\eta})$. We turn to Condition~\ref{ef:2}. Assume that $n \geq 2$. We prove $w',i' \qtlp{k,n-1} w,i$. Given a formula $\varphi(x)$ of $\fodsi{n-1}(\infsig{\eta})$ with rank at most $k$, we show that $w' \models \varphi(i') \Rightarrow w \models \varphi(i)$. By definition, $\neg \varphi(x) \in \fodsn(\infsig{\eta})$ and it has  rank at most $k$. Hence, since $w,i \qtlkn w',i'$, we have $w \models \neg \varphi(i) \Rightarrow w' \models \neg \varphi(i')$. The contrapositive is exactly the desired implication. It remains to handle Conditions~\ref{ef:3} and~\ref{ef:4}. By symmetry, we only detail the former. Assume that $k \geq 1$ and let $j \in \pos{w}$ such that $i < j$. We have to exhibit $j' \in \pos{w'}$ such that $i' < j'$, $\eta(\infix{w}{i}{j}) = \eta(\infix{w'}{i'}{j'})$ and $w,j \qtlp{k-1,n} w',j'$. Fact~\ref{fct:dclass} yields a formula $\varphi(x)$ of  $\fodsn(\infsig{\eta})$ with rank at most $k-1$ such that for all $u \in A^*$ and $h \in \pos{u}$, $u \models \varphi(h)$ if and only if $w,j \qtlp{k-1,n} u,h$. Moreover, we let $s = \eta(\infix{w}{i}{j}) \in N$ (recall that $i < j$) and $L = \eta\inv(s)$. Let $\psi(x)$ be the formula $\exists y\ (I_L(x,y) \wedge \varphi(y))$ of $\fodsn(\infsig{\eta})$. Clearly, $\psi(x)$ has rank at most $k$. Moreover, $w \models\psi(i)$ (one~may~use $j$ as the position quantified by $y$). Since $w,i \qtlkn w',i'$, we get $w' \models \psi(i')$. This yields $j' \in \pos{w'}$ such that $i' < j'$, $\infix{w'}{i'}{j'} \in L$ and $w' \models \varphi(j')$. Since $L = \eta\inv(s)$, we get $\eta(\infix{w}{i'}{j'}) =s = \eta(\infix{w}{i}{j})$ . Finally, since $w' \models \varphi(j')$, we obtain $w,j \qtlp{k-1,n} w',j'$ by definition of $\varphi$. %This concludes the proof for the ``only if'' direction.
	
	\smallskip
	
	We turn to the ``if'' implication. Assume that the four conditions are satisfied. We show that $w,i \qtlkn w',i'$. We have to prove that given a $\fodsn(\infsig{\eta})$ formula $\varphi(x)$ with rank at most~$k$, the implication $w \models \varphi(i) \Rightarrow w'  \models \varphi(i')$ holds. First, we put $\varphi(x)$ into normal form. The following lemma can be verified from the definition of \fodsn and DeMorgan's laws.
	
	\begin{lemma} \label{lem:nform}
		The formula $\varphi(x)$ is equivalent to a formula of rank at most $k$ belonging to the least set closed under disjunction, conjunction and existential quantification, and containing atomic formulas, their negations and, if $n \geq 2$, the negations of $\fodsi{n-1}(\infsig{\eta})$ formulas.
	\end{lemma}
	
	We assume that $\varphi(x)$ is of the form described in Lemma~\ref{lem:nform} and use structural induction on $\varphi$ to prove that $w \models \varphi(i) \Rightarrow w'  \models \varphi(i')$. If $\varphi(x)$ is an atomic formula of its negation, the implication can be verified from Condition~\ref{ef:1}. We turn to the case when  $\varphi(x) := \neg \psi(x)$ where $\psi(x)$ is a $\fodsi{n-1}(\infsig{\eta})$ formula (this may only happen when $n \geq 2$). Clearly, $\psi(x)$ has rank at most $k$ by hypothesis on $\varphi(x)$. Since $w',i'\qtlp{k,n-1} w,i$ by Condition~\ref{ef:2},  \mbox{$w' \models \psi(i') \Rightarrow w \models \psi(i)$}. The contrapositive yields $w \models \varphi(i) \Rightarrow w' \models \varphi(i')$. We turn to conjunction and disjunction. If $\varphi = \psi_1~X~\psi_2$ for $X \in \{\vee,\wedge\}$, we get $w \models \psi_h(i) \Rightarrow w' \models \psi_h(i)$ for $h = 1,2$ by structural induction. Hence, $w \models \varphi(i) \Rightarrow w' \models \varphi(i')$ as desired.
	
	It remains to handle existential quantification. Assume that $\varphi(x) = \exists y\ \psi(x,y)$ (since variables can be renamed, we may assume that $y \neq x$). By hypothesis on $\varphi$, we know that $\psi$ has rank at most $k-1$. Assume that $w \models \varphi(i)$. We show that $w \models \varphi(i')$. By hypothesis on $\varphi$, we get $j \in \pos{w}$ such that $w \models \psi(i,j)$. We use it define $j' \in \pos{w'}$. There are several cases depending on whether $j = i$, $i < j$ or $j < i$. By symmetry, we only treat the case when $i < j$. In this case, Condition~\ref{ef:3} yields $j' \in \pos{w'}$ such that $i' < j'$, $w,j \qtlp{k-1,n} w',j'$ and $\eta(\infix{w}{i}{j}) = \eta(\infix{w}{i'}{j'})$. We use a sub-induction on the structure of $\psi(x,y)$ to show that $w' \models \psi(i',j')$ which implies that $w',i' \models \varphi(i')$ as desired. If $x$ is the \emph{only} free variable in $\psi$, then our hypothesis states that $w \models \psi(i)$ and the main induction yields  $w' \models \psi(i')$ as desired. If $y$ is the \emph{only} free variable in $\psi$, then our hypothesis states that $w \models \psi(j)$. Hence, since $w,j \qtlp{k-1,n} w',j'$ and $\psi$ has rank at most $k-1$, we obtain $w' \models \psi(j')$ has desired. If $\psi(x,y)$ is an atomic formula or its negation involving both $x$ and $y$ (\emph{i.e.} $x = y$, $\neg (x= y)$, $I_L(x,y)$ or $\neg I_{L}(x,y)$ with $L$ recognized by $\eta$), since $w \models \psi(i,j)$, $i< j$, $i' < j'$ and $\eta(\infix{w}{i}{j}) = \eta(\infix{w}{i'}{j'})$, one may verify that $w \models \psi(i',j')$. Finally, disjunction and conjunction are handled by sub-induction as in the main induction.
\end{proof}

\smallskip
\noindent
{\bf Properties.} We now present important properties of these relations. First, we have the following simple property of the preorders \qtlkn which can be verified using Proposition~\ref{prop:efgame}.

\begin{lemma}\label{lem:compa}
	Let $\eta: A^* \to N$ be a morphism, $k \in \nat$ and $n \geq 1$. Let $x_1,x_2,y_1,y_2 \in A^*$ and $a \in A$ such that $x_1 \qtlkn y_1$ and $x_2 \qtlkn y_2$. Moreover, let $i = |x_1|+1$ and $j = |y_1|+1$. Then, $x_1x_2 \qtlkn y_1y_2$ and  $x_1ax_2,i \qtlkn y_1ay_2,i'$.
\end{lemma}

We turn to properties that are specific to morphisms $\eta: A^* \to N$ such that the set $\eta(A^+)$ is a \emph{finite group}. This reflects the fact our characterization of the quantifier-alternation hierarchy of \fod is restricted to the sets of predicates \infsigg and \infsiggp when \Gs is a \emph{group} \vari. We first present two preliminary results for the preorders \qtlp{k,1}. The first one considers the case when $\eta$ is a morphism into a group.

\begin{lemma} \label{lem:alt2:efg1}
	Consider a morphism $\eta: A^* \to G$ into a group and $p$ a multiple of $\omega(G)$. Let $u,v,x,y \in A^*$ and $\ell \in \nat$ such that $\eta(u) = \eta(v)$. Then, $v \qtlp{\ell,1} u(yv)^{p}$ and $v \qtlp{\ell,1} (vx)^{p}u$.
\end{lemma}

\begin{proof}
	By symmetry, we only prove $v \qtlp{\ell,1} u(yv)^{p}$. Since $G$ is a group, $\eta((vy)^p) = 1_G$. Since $\eta(u) = \eta(v)$, this yields $\eta(uy(vy)^{p-1}) = 1_G$. Thus, one may verify from Proposition~\ref{prop:efgame} that $\veps \qtlp{\ell,1} uy(vy)^{p-1}$. Hence, Lemma~\ref{lem:compa} yields $v \qtlp{\ell,1} u(yv)^{p}$ as desired.
\end{proof}

We now consider the case of morphisms $\eta: A^* \to N$ such that $\eta(A^+)$ is a group. We prove a slightly weaker result. 

\begin{lemma} \label{lem:alt2:efg2}
	Consider a morphism $\eta: A^* \to N$ such that $G = \alpha(A^+)$ is group, $\ell \in \nat$ and $p$ a multiple of $\omega(G)$. We consider $u,v,w,x \in A^*$ such that $|w| \geq \ell$ and $\eta(u) = \eta(v)$. We have $wv \qtlp{\ell,1} wu(xwv)^{p}$ and $vw \qtlp{\ell,1} (vwx)^{p}uw$.
\end{lemma}

\begin{proof}
	By symmetry, we only prove that $wv \qtlp{\ell,1} wu(xwv)^{p}$. We consider a slightly more general property that we prove by induction. We let $z = wv$ and $z' = wu(xwv)^{p}$. Let $m = |wu(xwv)^{p}x|$. Clearly, if $i \in \pos{z}$, then $m+i$ is the corresponding position in the suffix $z = wv$ of $z' =wu(xwv)^{p}$. We prove the two following properties for every $h \leq \ell$:
	\begin{itemize}
		\item if $i \leq \ell-h$, then $z,i \qtlp{h,1} z',i$.
		\item if $i > \ell-h$, then $z,i \qtlp{h,1} z',m+i$.
	\end{itemize}
	In the case when $h = \ell$ and $i = 0$, the first assertion yields $wv \qtlp{\ell,1} wu(xwv)^{p}$ as desired.
	
	We now prove that the two above properties hold for every $i \in \pos{wv}$ and $h \leq \ell$. We proceed by induction on $h$. By symmetry, we only consider the first property and leave the other to the reader. Thus, we assume that $i \leq \ell-h$ and show that $z,i \qtlp{h,1} z',i$. We use Proposition~\ref{prop:efgame}. There are only three conditions to verify: Condition~\ref{ef:2} is trivial since we are in the case $n= 1$. Moreover, it is straightforward to verify Condition~\ref{ef:1} from our hypotheses. We turn to Conditions~\ref{ef:3} and~\ref{ef:4}. By symmetry, we only detail the former. Assume that $h \geq 1$ and let $j \in \pos{v}$ such that $i < j$, we show that there exists $j' \in \pos{w}$ such that $i < j'$, $\eta(\infix{z}{i}{j}) = \eta(\infix{z'}{i}{j'})$ and $z,j \qtlp{h-1,1} z,j'$. There are two sub-cases depending on $j$. First, assume that $j \leq \ell - (h-1)$. In this case, we let $j' = j$. Clearly, we have $\eta(\infix{z}{i}{j}) = \eta(\infix{z'}{i}{j})$ since $\infix{z}{i}{j} = \infix{z'}{i}{j}$ (this is because $w$ is a common prefix of $z$ and $z'$, and $|w| \geq \ell$). Since $j \leq \ell - (h-1)$, we get $v,j \qtlp{h-1,1} w,j$ by induction on $h$. We turn to the second sub-case. Assume that $\ell - (h-1) < j$.  We define $j' = m+j$. Clearly, $i < j'$  since we have $i < j$. Moreover, since $j > \ell - (h-1)$ and $j' = m +j$, induction on $h$ yields $z,j \qtlp{h-1,1} z',j'$. We show that $\eta(\infix{z}{i}{j}) = \eta(\infix{z'}{i}{j'})$. By definition $j'$ is the position corresponding to $j \in \pos{z}$ in the suffix $z = wv$ of $z'$. Hence, there exists $y \in A^*$ such that $\suffix{z}{i} = \infix{z}{i}{j}y$ and  $\suffix{z'}{i} = \infix{z'}{i}{j'}y$. Moreover, by definition of $z'$, we have $\suffix{z'}{i} = \suffix{z}{i} (xwv)^{p}$. Since $p$ is a multiple of $\omega(G)$ and $xwv \in A^+$ (we have $|w| \geq \ell$), we get $\eta(xwv) = 1_G$. Moreover, $\suffix{z}{i} \in A^+$ since we have $i \leq \ell -h$ and $h \geq 1$. Altogether, it follows that $\eta(\infix{z}{i}{j}y) = \eta(\infix{z'}{i}{j'}y)$. If $y = \veps$, this concludes the proof. Otherwise, $y \in A^+$ and since $i \leq \ell = h$ and $\ell - (h-1) < j$, we also have $\infix{z}{i}{j},\infix{z'}{i}{j'} \in A^+$. Since $G = \alpha(A^+)$ is a group, we get  $\eta(\infix{z}{i}{j}) = \eta(\infix{z'}{i}{j'})$ as desired.
\end{proof}

We are ready to present the main property. We state it in the following proposition.

\begin{proposition} \label{prop:mainefg}
	Consider a morphism $\eta: A^* \to N$ such that $G = \alpha(A^+)$ is a group. For all \mbox{$k\in \nat$}, we have $p \geq 1$ such that if $n \geq 1$ and $u,v,x,y,z \in A^*$ satisfy $u \qtlp{k,n} v \qtlp{k,1} z$, 
	\[ 
	(zx)^{p}u(yz)^{p} \qtlp{k,n+1} (zx)^{p}v(yz)^{p}.
	\]
\end{proposition}

\begin{proof}
	We fix $k \in \nat$. Let us first define $p \geq 1$. By Lemma~\ref{lem:compa} the equivalence \eqtlp{k,1} is a congruence of finite index. Hence, the quotient set ${A^*}/{\eqtlp{k,1}}$ is a \emph{finite} monoid. We now define $p = \omega(G) \times \omega({A^*}/{\eqtlp{k,1}})$. By definition, we have the following key property of $p$:
	\begin{equation} \label{eq:pchoice}
		\text{for every $\ell \leq k$ and $w \in A^*$, $w^{2p} \eqtlp{\ell,1} w^p$.}
	\end{equation}
	Let $n \geq 1$ and $x,y,z \in A^*$. Moreover we write $w_1 = (zx)^{p}$ and $w_2 = (yz)^{p}$. We prove a more general property.
	
	\begin{lemma} \label{lem:efind}
		Let $\ell \leq k$, $1 \leq m \leq n$ and $u,v \in A^*$ such that $u \qtlp{\ell,1} z$ and $v \qtlp{\ell,1} z$. Let $w = w_1uw_2$ and $w' = w_1vw_2$. The three following properties hold:
		\begin{enumerate}
			\item if $0 \leq i \leq |w_1|$ and $u \qtlp{\ell,m} v$, then $w,i \qtlp{\ell,m+1} w',i$.
			\item if $1 \leq i \leq |w_2|+1$ and $u \qtlp{\ell,m} v$, then\\ $w,|w_1u|+i \qtlp{\ell,m+1} w',|w_1v|+i$.
			\item if $i \in \posc{u}$ and $i' \in \posc{v}$ satisfy $u,i \qtlp{\ell,m} v,i'$, then $w,|w_1|+i \qtlp{\ell,m+1} w',|w_1|+i'$.
		\end{enumerate}
	\end{lemma}
	
	Let us first apply the lemma to compete the main argument. Consider $u,v \in A^*$ such that $u \qtlp{k,n} v \qtlp{k,1} z$. The first assertion in Lemma~\ref{lem:efind} yields $w_1uw_2,0 \qtlp{k,n+1} w_1vw_2,0$. This exactly says that $w_1uw_2 \qtlp{k,n+1} w_1vw_2$ by definition and Proposition~\ref{prop:mainefg} is proved. It remains to prove Lemma~\ref{lem:efind}.
	
	\smallskip
	
	We fix $\ell \leq k$, $1 \leq m \leq n$ and $u,v \in A^*$ such that $u \qtlp{\ell,1} z$ and $v \qtlp{\ell,1} z$. We write $w = w_1uw_2$ and $w' = w_1vw_2$. We use induction on $\ell$ and $m$ (in any order) to prove that the three properties in the lemma hold. Since the three of them are handled using similar arguments, we only detail the third one and leave the other two to the reader. Hence, we consider $i \in \posc{u}$ and $i' \in \posc{v}$ such that $u,i \qtlp{\ell,m} v,i'$. We show that $w,|w_1|+i \qtlp{\ell,m+1} w',|w_1|+i'$. The argument is based on Proposition~\ref{prop:efgame}. There are four conditions to verify. For Condition~\ref{ef:1}, that $(w,|w_1|+i)$ and $(w',|w_1|+i')$ are $\eta$-equivalent can be verified from $u,i \qtlp{\ell,m} v,i'$ which implies that $(u,i)$ and $(v,i')$ are $\eta$-equivalent. We turn to Condition~\ref{ef:2}. we have to prove that $w',|w_1|+i' \qtlp{\ell,m} w,|w_1|+i$. There are two sub-cases depending on $m$. First, assume that $m \geq 2$. Since $u,i \qtlp{\ell,m} v,i'$, Proposition~\ref{prop:efgame} implies that $v,i' \qtlp{\ell,m-1} u,i$. Hence, by induction on $m$, the third assertion in Lemma~\ref{lem:efind} yields \mbox{$w',|w_1|+i \qtlp{\ell,m} w,|w_1|+i$} as desired. We now assume that $m = 1$: we prove that \mbox{$w',|w_1|+i' \qtlp{\ell,1} w,|w_1|+i$}. Consider the decompositions $u = u_1au_2$ and $v = v_1av_2$ where the positions carrying the highlighted letters ``$a$'' are $i$ and $i'$. We prove that $w_1v_1 \qtlp{\ell,1} w_1u_1$ and $v_2w_2 \qtlp{\ell,1} u_2w_2$, Since $w = w_1u_1au_2w_2$ and $w' = w_1v_1av_2w_2$, it will then follow from Lemma~\ref{lem:compa} that $w',|w_1|+i' \qtlp{\ell,1} w,|w_1|+i$ as desired. By symmetry, we only prove that $v_2w_2 \qtlp{\ell,1} u_2w_2$. If $u_2 = v_2$, this is trivial. Hence, we assume that $u_2 \neq v_2$. Since $u,i \qtlp{\ell,1} v,i'$, one may verify from Proposition~\ref{prop:efgame} that $\eta(u_2) = \eta(v_2)$. We prove that $v_2 \qtlp{\ell,1} u_2(yv)^{p}$. Let us first explain why this implies the desired result. By~\eqref{eq:pchoice}, we have $(yv)^{2p} \qtlp{\ell,1} (yv)^p$. Together, with $v_2 \qtlp{\ell,1} u_2(yv)^{p}$ and Lemma~\ref{lem:compa}, this implies $v_2(yv)^{p} \qtlp{\ell,1} u_2(yv)^{2p}\qtlp{\ell,1} u_2(yv)^{p}$ as desired. It remains to prove that $v_2 \qtlp{\ell,1} u_2(yv)^{p}$. Let $y' = yv_1a$. Clearly, we have $yv = y'v_2$. Thus, we have to show that $v_2 \qtlp{\ell,1} u_2(y'v_2)^{p}$. There are two cases depending on $\eta$. If $\eta(A^*) = G$, the result is immediate from Lemma~\ref{lem:alt2:efg1} since $\eta(u_2) = \eta(v_2)$ and $p$ is a multiple of $\omega(G)$. Assume now that $\eta(A^*) \neq G$. Since $\eta(A^+) = G$, it follows that $\eta\inv(1_N) =  \{\veps\}$. Hence, since $u,i \qtlp{\ell,1} v,i'$ and $u_2 \neq v_2$, one may verify from Proposition~\ref{prop:efgame} that $|u_2| \geq \ell$, $|v_2| \geq \ell$ and $\prefix{u_2}{\ell+1} = \prefix{v_2}{\ell+1}$. Hence, we may apply Lemma~\ref{lem:alt2:efg2} to obtain $v_2 \qtlp{\ell,1} u_2(y'v_2)^{p}$ since $p$ is a multiple of $\omega(G)$. This completes the proof for Condition~\ref{ef:2}.

	It remains to handle Conditions~\ref{ef:3} and~\ref{ef:4}. Since those are symmetrical, we only present an argument for the former. Let $j \in\pos{w}$ such that $|w_1|+i < j$. We have to exhibit $j' \in \pos{w'}$ such that $|w_1|+i' < j'$, $\eta(\infix{w'}{|w_1|+i'}{j'}) = \eta(\infix{w}{|w_1|+i}{j})$ and $w,j \qtlp{\ell-1,m+1} w',j'$. We distinguish two sub-cases depending on $j$. First, assume that $|w_1|+i < j \leq |w_1u|$. In this case, there exists a position $h \in \posc{u}$ such that $j = |w_1| + h$. In particular, we have $i \leq h$. Hence, since $u,i \qtlp{\ell,m} v,i'$, Proposition~\ref{prop:efgame} yields $h' \in \posc{v}$ such that $\eta(\infix{u}{i}{h}) = \eta(\infix{v}{i'}{h'})$ and $u,h \qtlp{\ell-1,m} v,h'$. We now define $j' = |w_1|+h'$. Clearly, $\infix{w'}{|w_1|+i'}{j'} = \infix{v}{i'}{h'}$ and $\infix{w}{|w_1|+i}{j} = \infix{u}{i}{h}$. Hence, it is immediate that $\eta(\infix{w'}{|w_1|+i'}{j'}) = \eta(\infix{w}{|w_1|+i}{j})$. Moreover, since $u,h \qtlp{\ell-1,m} v,h'$, it follows from induction on $\ell$ that we may apply the third assertion in Lemma~\ref{lem:efind} to get $w,j \qtlp{\ell-1,m+1} w',j'$. We turn to the second sub-case: $j > |w_1u|$. In this case, there exists a position $1 \leq h \leq |w_2|+1$ of $w_2$ such that $j = |w_1u| + h$. We let $j' = |w_1v|+h$. Clearly, we have $|w_1|+i' < j'$. It is also immediate that $\infix{w'}{|w_1|+i'}{j'} = \infix{v}{i'}{|v|+1}\infix{w_2}{0}{h}$ and $\infix{w}{|w_1|+i}{j} = \infix{u}{i}{|u|+1}\infix{w_2}{0}{h}$. Additionally, since  $u,i \qtlp{\ell,m} v,i'$, one may verify from Proposition~\ref{prop:efgame} that $\eta(\infix{u}{i}{|u|+1}) = \eta(\infix{v}{i'}{|v|+1})$. Hence, we get $\eta(\infix{w'}{|w_1|+i'}{j'}) = \eta(\infix{w}{|w_1|+i}{j})$. Finally, it follows from induction on $\ell$ that we may apply the second assertion in Lemma~\ref{lem:efind} to get $w,j \qtlp{\ell-1,m+1} w',j'$. %This completes the proof of Lemma~\ref{lem:efind}.
\end{proof}

Finally, we complete Proposition~\ref{prop:mainefg} with a useful corollary. %In fact, this is the result that we shall actually use. 

\begin{corollary} \label{cor:efgmain}
	Consider a morphism $\eta: A^* \to N$ such that $G = \alpha(A^+)$ is a group. For all $k \in \nat$, we have $p \geq 1$ such that for $n \geq 1$ and $u,v,x,y \in A^*$ satisfying $u \eqtlp{k,n} v$, we have $(vx)^{p}u(yv)^{p} \eqtlp{k,n+1} (vx)^{p}v(yv)^{p}$.
\end{corollary}

\begin{proof}
	We fix $k \in \nat$ and define $p \geq 1$ as the number given by Proposition~\ref{prop:mainefg}. Since $u \qtlkn v \qtlp{k,1} v$, the case $z = v$ in the proposition yields $(vx)^{p}u(yv)^{p} \qtlp{k,n+1} (vx)^{p}v(yv)^{p}$. Moreover, $v \qtlkn u \qtlp{k,1} v$. Thus, we may apply Proposition~\ref{prop:mainefg} in the case when $u$ and $v$ have been swapped and $z = v$. This yields $(vx)^{p}v(yv)^{p} \qtlp{k,n+1} (vx)^{p}u(yv)^{p}$ as desired.
\end{proof}

\subsection{Characterization}
We show that for a set of predicates built from a group \vari, one may ``climb'' the quantifier alternation hierarchy of \fod with mixed polynomial closure.

\begin{theorem} \label{thm:mhiera}
	If \Gs is a group \vari,  then we have $\fodb{n+1}(\infsigg)= \mdet{\fodb{n}(\infsigg)}$ and $\fodb{n+1}(\infsiggp)= \mdet{\fodb{n}(\infsiggp)}$  for every $n \geq 1$.
\end{theorem}

Theorem~\ref{thm:level1} and Theorem~\ref{thm:mhiera} imply that for every group \vari  \Gs, if $\Cs \in \{\Gs,\Gs^+\}$, then all levels \fodb{n}(\infsigc) are built iteratively from \bpol{\Cs} by applying \mdeto. By Proposition~\ref{prop:bpvar}, \bpol{\Cs} is a \vari. Moreover, Theorem~\ref{thm:detclos} and Corollary~\ref{cor:pmemb} imply that when \mdeto is applied to a \vari, it outputs a \vari and preserves the decidability of membership. It follows that when membership is decidable for \bpol{\Cs}, this is also the case for \emph{all} levels $\fodb{n}(\infsigc)$. Since $\Cs \in \{\Gs,\Gs^+\}$, it follows from Theorem~\ref{thm:bpgm} that membership is decidable for \bpol{\Cs} provided that separation is decidable for \Gs. Finally, we have $\fodb{n}(\frI_{\Gs}) = \fodb{n}(<,\frP_{\Gs})$ and $\fodb{n}(\frI_{\Gs^+}) = \fodb{n}(<,+1,\frP_{\Gs})$ by Lemma~\ref{lem:gensig}. Altogether, we obtain the following corollary.

\begin{corollary} \label{cor:mhiera}
	Let \Gs be a group \vari with decidable separation. For every $n \geq 1$, membership is decidable for $\fodb{n}(<,\prefsigg)$ and $\fodb{n}(<,+1,\prefsigg)$.
\end{corollary}

Corollary~\ref{cor:mhiera} reproves earlier results. Separation is clearly decidable for $\stzer = \{\emptyset,A^*\}$. Hence, $\fodb{n}(<)$ and $\fodb{n}(<,+1)$ have decidable membership for all $n \geq 1$. For \mbox{$\fodb{n}(<)$}, this was first proved independently by Kufleitner and Weil~\cite{kwfo2alt3} and Krebs and Straubing~\cite{ksfo2alt}. For $\fodb{n}(<,+1)$, this was first proved by Kufleitner and Lauser~\cite{klfo2alts}.

\begin{remark}
	In~\cite{kwfo2alt3}, it is also shown that $\fodb{n}(<) = \idetp{n}{\pt}$ for every $n \geq 1$  (with $\pt =\bpol{\stzer}$). This can be reproved using Theorem~\ref{thm:intcarn}, Theorem~\ref{thm:mhiera} and the fact that $\pt = \idet{\pt}$. This is specific to $\fodb{n}(<)$: this fails in general. This is because the equality $\pt = \idet{\pt}$ is specific to $\pt =\bpol{\stzer}$.
\end{remark}

Additionally, it is known that separation is decidable for the group \varis \md and \abg. This is straightforward for \md and proved in~\cite{abelianp} for \abg (see also~\cite{pzgroups} for recent proofs). Hence, we also obtain the decidability of membership for all levels $\fodb{n}(<,MOD)$, $\fodb{n}(<,+1,MOD)$, $\fodb{n}(<,AMOD)$ and $\fodb{n}(<,+1,AMOD)$. Note that this was already known for the levels $\fodb{n}(<,+1,MOD)$. This was proved in~\cite{dpfo2} using a reduction to the levels $\fodb{n}(<,+1)$ which is based on independent techniques

Theorem~\ref{thm:mhiera} also yields characterizations of \fod. Indeed, one may verify from Theorem~\ref{thm:apol} that given a \vari \Ds, the union of all classes built from \Ds by iteratively applying \mdeto is \upol{\Ds}. Hence, we obtain the following corollary.

\begin{corollary} \label{cor:fod}
	If \Gs is a group \vari, then $\fod(<,\prefsigg) = \upol{\bpol{\Gs}}$ and $\fod(<,+1,\prefsigg) = \upol{\bpol{\Gs^+}}$.
\end{corollary}

Since \upolo preserves the decidability of membership by Theorem~\ref{thm:cupol}, the above argument also implies that for all group \varis \Gs with decidable separation, $\fod(<,\prefsigg)$ and $\fod(<,+1,\prefsigg)$ have decidable membership. This  yields known results~\cite{twfo2,DartoisP13} in the cases $\Gs = \stzer$ and $\Gs = \md$.

\begin{remark}
	Another proof of Corollary~\ref{cor:fod} is available in~\cite{pzupol2}. It is more direct (and simpler) since it considers the classes $\fod(<,\prefsigg)$ and $\fod(<,+1,\prefsigg)$ directly without looking at their quantifier-alternation hierarchies. In fact, specialized characterizations of $\fod(<,\prefsigg)$ and $\fod(<,+1,\prefsigg)$ are also presented in~\cite{pzupol2}.
\end{remark}

%The remainder of the section is now devoted to the proof of Theorem~\ref{thm:mhiera}. 

\begin{proof}[Proof of Theorem~\ref{thm:mhiera}]
	We fix a group \vari \Gs and let $\Cs \in \{\Gs,\Gs^+\}$. We use induction on $n$ to show that $\fodb{n}(\infsigc)$ is a \vari and $\fodb{n+1}(\infsigc)  = \mdet{\fodb{n}(\infsigc)}$ for all $n \geq 1$. We fix $n\geq 1$ for the proof. We first show that $\fodb{n}(\infsigc)$ is a \vari. If $n=1$, then $\fodb{1}(\infsigc) = \bpol{\Cs}$ by Theorem~\ref{thm:level1} and \bpol{\Cs} is a \vari Proposition~\ref{prop:bpvar}. Otherwise, induction on $n$ yields that $\fodb{n}(\infsigc)  = \mdet{\fodb{n-1}(\infsigc)}$ and $\fodb{n-1}(\infsigc)$ is a \vari.  Hence, we obtain from Theorem~\ref{thm:detclos} that $\fodb{n}(\infsigc)$ is a \vari. It remains to prove the equality $\fodb{n+1}(\infsigc)= \mdet{\fodb{n}(\infsigc)}$.  We start with the left to right inclusion.
	
	\medskip
	\noindent
	{\bf Inclusion $\fodb{n+1}(\infsigc) \subseteq \mdet{\fodb{n}(\infsigc)}$.}  The argument is based on the algebraic characterization of \mdeto. Let $L\in\fodb{n+1}(\infsigc)$. Since $\fodb{n}(\infsigc)$ is a \vari, Proposition~\ref{prop:synmemb} yields that it suffices to prove that the syntactic morphism $\alpha: A^* \to M$ of $L$ is an \mdet{\fodb{n}(\infsigc)}-morphism. By Theorem~\ref{thm:cmdet}, this boils down to proving that for every $q,r,s,t \in M$ such that $(s,t) \in M^2$ is a $\fodb{n}(\infsigc)$-pair, we have $(sq)^\omega s (rs)^\omega = (sq)^\omega t (rs)^\omega$. 
	
	Since $L \in \fodb{n+1}(\infsigc)$, Lemma~\ref{lem:uset} yields a \Cs-morphism $\eta: A^* \to N$ and \mbox{$k \in \nat$} such that $L$ is a union of \eqtlp{k,n+1}-classes. Let $K$ be the union of all \eqtlp{k,n}-classes which intersect $\alpha\inv(s)$. By Lemma~\ref{lem:uset}, we have $K \in \fodb{n}(\infsigc)$. Moreover, $\alpha\inv(s) \subseteq K$ by hypothesis. Thus, since $(s,t) \in M^2$ is a $\fodb{n}(\infsigc)$-pair, we have $K \cap \alpha\inv(t) \neq \emptyset$. We get $u,v \in A^*$ such that $\alpha(v) = s$, $\alpha(u) = t$ and $u \eqtlp{k,n} v$. We also let $x,y \in A^*$ such that $\alpha(x) = q$ and $\alpha(y) = r$. Since $\Cs \in \{\Gs,\Gs^+\}$ and $\eta: A^* \to N$ is a \Cs-morphism, Lemma~\ref{lem:gmorph} implies that $G = \eta(A^+)$ is a group. Hence, since $u \eqtlp{k,n} v$, Corollary~\ref{cor:efgmain} and Lemma~\ref{lem:compa} yield $p \geq 1$ such that,
	\[
	w(vx)^{p}u (yv)^{p}w' \eqtlp{k,n+1} w(vx)^{p}v (yv)^{p}w' \text{ for all $w,w' \in A^*$.}
	\]
	Since $L$ is union of \eqtlp{k,n+1}-classes, it follows that $(vx)^{p}v (yv)^{p}$ and $(vx)^{p}u (yv)^{p}$ have the same image under the syntactic morphism $\alpha$ of $L$. Hence, $(sq)^{p}s (rs)^{p} =  (sq)^{p} t (rs)^{p}$. It now suffices to multiply by the right amount of copies of $sq$ on the left and of $rs$ on the right to obtain $(sq)^\omega s (rs)^\omega = (sq)^\omega t (rs)^\omega$.  This completes the proof that  $\fodb{n+1}(\infsigc) \subseteq \mdet{\fodb{n}(\infsigc)}$.
	
	\medskip
	\noindent
	{\bf Inclusion $\mdet{\fodb{n}(\infsigc)} \subseteq \fodb{n+1}(\infsigc)$.} This part of the proof is based on a key property of \mdeto that we present first. We say that a marked product $L_0a_1L_1\cdots a_mL_m$ is \Cs-\emph{pointed} if for all $1\leq i \leq m$, there are $K_i,K'_i \in \bpol{\Cs}$ such that $K_ia_iK'_i$ is \emph{unambiguous}, $L_0a_1L_1\cdots a_{i-1}L_{i-1} \subseteq K_i$ and $L_ia_{i+1}L_{i+1} \cdots a_mL_m \subseteq K'_i$. We now use the hypothesis that $\Cs \in \{\Gs,\Gs^+\}$ to apply Proposition~\ref{prop:conspoint} and prove the following lemma.
	
	\begin{lemma} \label{lem:mprod}
		Every language in \mdet{\fodb{n}(\infsigc)} is a finite union of \Cs-pointed marked products of languages in $\fodb{n}(\infsigc)$ 
	\end{lemma}
	
	\begin{proof}
	We fix $L \in \mdet{\fodb{n}(\infsigc)}$. Since $\fodb{n}(\infsigc)$ is a \vari,  Proposition~\ref{prop:opcar} yields a $\fodb{n}(\infsigc)$-morphism $\alpha: A^* \to M$ and $k \in \nat$ such that $L$ is a finite union of $\eqmak$-classes. Hence, it suffices to prove that every $\eqmak$-class is a finite union of \Cs-pointed marked products of languages in $\fodb{n}(\infsigc)$. First, we associate a language $U_w$ to every word $w \in A^*$.
	
	Let $\eta$ be the morphism $\eta: \ctype{\cdot} \circ \alpha: A^* \to {M}/{\sim_{\bpol{\Cs}}}$. We know that $\eta$ is a \bpol{\Cs}-morphism by Lemma~\ref{lem:smult}.    Moreover, observe that $\bpol{\Cs} \subseteq \fodb{n}(\infsigc) \subseteq \upol{\bpol{\Cs}}$. Indeed, we know that $\Ds_1 = \bpol{\Cs}$ by Theorem~\ref{thm:level1} and induction in Theorem~\ref{thm:mhiera} implies that $\fodb{n}(\infsigc)$ is built from $\Ds_1$ by applying \mdeto iteratively. Therefore, Lemma~\ref{lem:isdet} implies that $\posmak{w} \subseteq \posmp{\eta}{k|M|}{w}$. Finally, since $\Cs \in \{\Gs,\Gs^+\}$ and \Gs is a group \vari, it follows from Proposition~\ref{prop:conspoint} that there exists another \bpol{\Cs}-morphism, $\gamma: A^* \to Q$ such that $\posmp{\eta}{k|M|}{w} \subseteq \posmp{\gamma}{1}{w}$. We define,
	\[
	\begin{array}{lll}
		(s_0,a_1,s_1,\dots,a_h,s_h)& =& \sigma_\alpha(w,\posmak{w}). \\
		(q_0,a_1,q_1,\dots,a_h,q_h)& =& \sigma_\gamma(w,\posmak{w}). 
	\end{array}
	\]
	For all $i \leq h$, we let $V_i = \alpha\inv(s_i) \cap \gamma\inv(q_i)$. Finally, we define $U_w = V_0a_1V_1 \cdots a_hV_h$. By definition, $h = |\posmak{w}| \leq 2|M|^{k}$. Thus, there are finitely many languages $U_w$ even though there infinitely many $w \in A^*$. Moreover, it is clear that $w \in U_w$. We now prove that $U_w$ is included in the \eqmak-class of $w$ and that $V_0a_1V_1 \cdots a_hV_h$ is a \Cs-pointed marked product of languages in $\fodb{n}(\infsigc)$. It will then follow that each \eqmak-class is the \emph{finite} union of all languages $U_w$ for the words $w$ in the \eqmak-class, \emph{i.e.} a finite union of \Cs-pointed marked product of languages in $\fodb{n}(\infsigc)$ as desired. We first show that if $u \in U_w$, then $u \eqmak w$. By definition of $U_w$, there exists $P \subseteq \pos{u}$ such that $\sigma_\alpha(u,P) = (s_0,a_1,s_1,\dots,a_h,s_h) =  \sigma_\alpha(w,\posmak{w})$ and Corollary~\ref{cor:eqbij} yields $u \eqmak w$.
	
	It remains to show that $V_0a_1V_1 \cdots a_hV_h$ is a \Cs-pointed marked product of languages in $\fodb{n}(\infsigc)$. As $\alpha$ is a $\fodb{n}(\infsigc)$-morphism, $\gamma$ is a \bpol{\Cs}-morphism and $\bpol{\Cs} \subseteq \fodb{n}(\infsigc)$, it is immediate by definition that $V_i \in \fodb{n}(\infsigc)$ for all $i \leq h$. We prove that $V_0a_1V_1 \cdots a_hV_h$ is \Cs-pointed. We fix $i \leq h$ for the proof. Let $r_i = q_0\gamma(a_1)q_1 \cdots \gamma(a_{i-1})q_{i-1}$ and $K_i = \gamma\inv(r_i)$. Moreover, we let  $r'_i = q_i\gamma(a_{i+1})q_{i+1} \cdots \gamma(a_h)q_h$ and $K'_i = \gamma\inv(r'_i)$. One may verify that $V_0a_1V_1\cdots a_{i-1}V_{i-1} \subseteq K_i$ and $V_ia_{i+1}V_{i+1} \cdots a_hV_h \subseteq K'_i$. Hence, we have to prove that $K_ia_iK'_i$ is unambiguous. We have $\posmak{w} \subseteq \posmp{\gamma}{1}{w}$ by construction of $\gamma$. Therefore, all letters in the $\gamma$-snapshot $\sigma_\gamma(w,\posmak{w}) = (q_0,a_1,q_1,\dots,a_h,q_h)$ correspond to positions in \posmp{\gamma}{1}{w}. By definition, this implies that either $r_i\gamma(a_i) \Rords r_i$ or $\gamma(a_i)r'_i \Lords r'_i$. By symmetry, we assume that the former holds and prove that $K_ia_iK'_i$ is left deterministic. By contradiction, assume that there exists $x \in K_i \cap K_ia_iA^*$. Since $K_i = \gamma\inv(r_i)$, this yields $y \in A^*$ such that $r_i = r_i\gamma(a_i)\gamma(u)$, contradicting the hypothesis that $r\gamma(a_i) \Rords r$.
\end{proof}
	
	We now prove that $\mdet{\fodb{n}(\infsigc)} \subseteq \fodb{n+1}(\infsigc)$. In view of Lemma~\ref{lem:mprod}, it suffices to show that if $L_0,\dots,L_m \in \fodb{n}(\infsigc)$ and $L_0a_1L_1\cdots a_mL_m$ is a \Cs-pointed marked product, then \mbox{$L_0a_1L_1\cdots a_mL_m \in \fodb{n+1}(\infsigc)$}. We do so by building a $\fodb{n+1}(\infsigc)$ sentence defining $L_0a_1L_1\cdots a_mL_m$. We have $K_h,K'_h \in \bpol{\Cs}$ for every $h \leq m$ such that $K_ha_hK'_h$ is \emph{unambiguous}, $L_0a_1L_1\cdots a_{h-1}L_{h-1} \subseteq K_h$ and $L_ha_{h+1}L_{h+1} \cdots a_mL_m \subseteq K'_h$. Hence, for all $w \in A^*$, we have $w \in L_0a_1L_1\cdots a_mL_m$, if and only if the two following properties hold:
	\begin{enumerate}[label=\alph*)]
		\item\label{itm:p1} There are $i_0,i_1,\dots,i_m,i_{m+1} \in \pos{w}$ such that $0 = i_0 < i_1 < \cdots < i_m < i_{m+1} = |w|+1$ and for all $h$ such that $1 \leq h \leq m$, $i_h$ has label $a_h$, $\prefix{w}{i_h} \in K_h$ and $\suffix{w}{i_h} \in K'_h$. Observe that these positions must be \emph{unique} since $K_ha_hK'_h$ is unambiguous.
		\item\label{itm:p2} For $0 \leq h \leq m$, we have $\infix{w}{i_h}{i_{h+1}} \in L_h$.
	\end{enumerate}
	We show that both properties can be expressed in $\fodb{n+1}(\infsigc)$. First, we build $\fodb{1}(\infsigc)$ formulas that we shall use to pinpoint the positions $i_0,i_1,\dots,i_m,i_{m+1}$.
	
	\begin{lemma} \label{lem:pointing}
		For $1 \leq h \leq m$, there exists a formula $\psi_h(x)$ of $\fodb{1}(\infsigc)$ with one free variable $x$ such that for every $w \in A^*$ and $i \in \pos{w}$, we have $w \models \psi_h(i)$ if and only if $i$ has label $a_h$, $\prefix{w}{i} \in K_h$ and $\suffix{w}{i} \in K'_h$.
	\end{lemma}

	Lemma~\ref{lem:pointing} holds since $K_h,K'_h \in \bpol{\Cs}$ (the argument is identical to the one used in Theorem~\ref{thm:level1} to prove that $\bpol{\Cs} \subseteq \fodb{1}(\infsigc)$). We fix the $\fodb{1}(\infsigc)$ formulas $\psi_1,\dots,\psi_m$ for the proof. We use them to define new formulas $\Gamma_h(x)$ for $1 \leq h \leq m$. We let $\Gamma_1(x) := \psi_1(x)$. Additionally, for $h > 1$, we define $\Gamma_h(x) := \psi_h(x) \wedge \exists y\ ( y < x \wedge \Gamma_{h-1}(y))$ (the definition involves implicit variable renaming, this is standard in \fod). Finally, we let $\Gamma := \exists x\ \Gamma_m(x)$. By definition, $\Gamma$ is a sentence of $\fodb{2}(\infsigc) \subseteq \fodb{n+1}(\infsigc)$ and it expresses Condition~\ref{itm:p1} above. 
	
	We turn to Condition~\ref{itm:p2}. We define $\psi_0(x) := (x = min)$ and $\psi_{m+1}(x) := (x = max)$ for the construction. For every $h$ such that $0 \leq h \leq m$, we construct a  $\fodb{n+1}(\infsigc)$ sentence $\varphi_h$ which satisfies the following property: for every word $w \in A^*$ such that $w \models \Gamma$ (which yields \emph{unique} positions $i_h,i_{h+1} \in \pos{w}$ such that $w \models \psi_h(i_h)$ and $w \models \psi_{h+1}(i_{h+1})$), we have $w \models \varphi_h$ if and only $\infix{w}{i_h}{i_{h+1}} \in L_h$. It will then be immediate that $L_0a_1L_1\cdots a_mL_m$ is defined by the sentence $\varphi:= \Gamma \wedge \bigwedge_{0 \leq h \leq m} \varphi_h$ of $\fodb{n+1}(\infsigc)$, completing the proof.
	
	We now fix $h$ such that $0 \leq h \leq m$ and construct $\varphi_h$. By hypothesis, we have $L_h \in \fodb{n}(\infsigc) = \fodb{n}(\infsigc)$. Hence, we get a sentence $\delta_h$ of $\fodb{n}(\infsigc)$ defining $L_h$. We build $\varphi_h$ from $\delta_h$ by applying two kinds of modifications. First, we restrict the quantifications in $\delta_h$ to the positions that are in-between the two unique ones satisfying $\psi_h$ and $\psi_{h+1}$. We recursively replace each sub-formula of the form $\exists x\ \zeta$ by the following (we write ``$x \leq y$'' for the formula ``$x < y \vee x = y$''):
	\[
		\exists x\ \left(\zeta \wedge (\exists y\ (\psi_h(y) \wedge y \leq x))  \wedge (\exists y\ (\psi_{h+1}(y) \wedge x \leq y))\right).
	\]
	Intuitively, we are using the unique positions satisfying $\psi_h$ and $\psi_{h+1}$ as substitutes for the two artificial unlabeled positions. Hence, we also need to tweak the atomic sub-formulas in $\delta_h$. First, we replace all atomic sub-formulas $b(x)$ with $b \in A$ by,
	\[
	b(x) \wedge (\exists y\ (\psi_h(y) \wedge y < x))  \wedge (\exists y\ (\psi_{h+1}(y) \wedge x < y).
	\]
	We also need to modify the atomic sub-formulas  involving the constants $min$ and $max$. All sub-formulas $\xi(min,x)$ with $\xi(min,x) := (min = x)$ or $\xi(min,x) := I_L(min,x)$ where $L \in \Cs$ are replaced by $\exists y (\psi_h(y) \wedge \xi(y,x))$. Symmetrically, all sub-formulas $\xi(x,max)$ with $\xi(x,max) := (x = max)$ or $\xi(x,max) := I_L(x,max)$ where $L \in \Cs$ are replaced by $\exists y (\psi_{h+1}(y) \wedge \xi(x,y))$. Finally, all sub-formulas $I_L(min,max)$ for $L \in \Cs$ are replaced by the formula $\exists x \exists y (\psi_h(x) \wedge \psi_{h+1}(y) \wedge I_L(x,y))$. There can be other atomic sub-formulas involving $min$ and $max$ such as $b(min)$, ($min = max$) or $I_L(max,x)$. We do not modify them since they are equivalent to $\bot$ (\emph{i.e.}, false).
	
	By definition, $\varphi_h$ is built by nesting the $\fodb{1}(\infsigc)$ formulas $\psi_h$ and $\psi_{h+1}$ under the sentence $\delta_h$ of $\fodb{n}(\infsigc)$. Thus, one may verify that $\varphi_h$ is a sentence of  $\fodb{n+1}(\infsigc)$ as desired. One may also verify that $\varphi_h$ satisfies the desired property:  for every word $w \in A^*$ such that $w \models \Gamma$ (we get \emph{unique} positions $i_h,i_{h+1} \in \pos{w}$ such that $w \models \psi_h(i_h)$ and $w \models \psi_{h+1}(i_{h+1})$), $w \models \varphi_h$ if and only if $\infix{w}{i_h}{i_{h+1}} \models \delta_h$ (\emph{i.e.}, $\infix{w}{i_h}{i_{h+1}} \in L_h$). This concludes the proof.
\end{proof}

%% file: ratms.tex
We now consider separation and covering. In the paper, we mostly work with covering (it is more general by Lemma~\ref{lem:covsep}).  In particular, all results that we present are formulated and proved within a tailored framework that was introduced in~\cite{pzcovering}. The purpose of this preliminary section is to recall this framework. It is based on~algebraic objects called ``\ratms'' that we first define. Then, we connect them to covering. At the~end of the section, we present additional terminology designed to handle the particular classes that we shall consider. Namely, those built with \ldeto, \rdeto and \mdeto from a single \emph{finite} \vari.

\subsection{\Ratms} A \emph{semiring} is a tuple $(R,+,\cdot)$ where $R$ is a set and ``$+$'' and ``$\cdot$'' are two binary operations called addition and multiplication, which satisfy the following axioms:
\begin{itemize}
  \item $(R,{+})$ is a commutative monoid, whose identity element is denoted by $0_R$.
  \item $(R,{\cdot})$ is a monoid, whose identity element is denoted by $1_R$.
  \item Multiplication distributes over addition: for $r,s,t \in R$, $r \cdot (s + t) = (r \cdot s) + (r \cdot t)$  and $(r + s) \cdot t = (r \cdot t) + (s \cdot t)$.
  \item $0_R$ is a zero for $(R,{\cdot})$: $0_R \cdot r = r \cdot 0_R = 0_R$ for every $r \in R$.
\end{itemize}
A semiring $R$ is \emph{idempotent} when $r + r = r$ for every $r \in R$, \emph{i.e.}, when the additive monoid $(R,+)$ is idempotent (there is no additional constraint on the multiplicative monoid $(R,\cdot)$). Given an idempotent semiring $(R,+,\cdot)$, one may define a canonical ordering $\leq$ over $R$:
\[\text{For all }
  r, s\in R,\quad r\leq s \text{ when } r+s=s.
\]
One may verify that $\leq$ is a partial order which is compatible with both addition and multiplication. Moreover, every morphism between two such commutative and idempotent monoids is increasing for this ordering.

\begin{example}\label{exa:semiring}
  A key example of idempotent semiring is the set of all languages $2^{A^*}$. Union is the addition and language concatenation is the multiplication (with $\{\varepsilon\}$ as the identity element). Observe that in this case, the canonical ordering is inclusion. More generally, if $M$ is a monoid, then $2^M$ is an idempotent semiring whose addition is union, and whose multiplication is obtained by lifting the one of $M$ to subsets.
\end{example}

When dealing with subsets of an idempotent semiring $R$, we shall often apply a \emph{downset operation}. Given $S \subseteq R$, we write $\dclosr S = \{r \in R \mid r \leq s \text{ for some $s \in S$}\}$. We extend this notation to Cartesian products of arbitrary sets with $R$. Given some set $X$ and $S \subseteq X \times R$, we write $\dclosr S = \{(x,r) \in X \times R \mid \text{$\exists s \in R$ such that $r \leq s$ and $(x,s) \in S$}\}$.

\smallskip
\noindent
{\bf \Mratms.} We define a \emph{\mratm} as a semiring morphism $\rho: (2^{A^*},\cup,\cdot) \to (R,+,\cdot)$ where $(R,+,\cdot)$ is a \emph{finite} idempotent semiring, called the \emph{rating set of $\rho$}. That is, $\rho$ is a map from $2^{A^{*}}$ to $R$ satisfying the following properties:
\begin{enumerate}
  \item\label{itm:bgen:ford} $\rho(\emptyset) = 0_R$ and $\rho(K_1\cup K_2)=\rho(K_1)+\rho(K_2)$ for every $K_1,K_2 \subseteq A^*$.
  \item\label{itm:bgen:fmult}  $\rho(\{\varepsilon\}) = 1_R$ and $\rho(K_1K_2) = \rho(K_1) \cdot \rho(K_2)$ for every $K_1,K_2 \subseteq A^*$.
\end{enumerate}

For the sake of improved readability, when applying a \mratm $\rho$ to a singleton set $\{w\}$, we shall write $\rho(w)$ for $\rho(\{w\})$. Additionally, we write $\rho_*: A^* \to R$ for the restriction of $\rho$ to $A^*$: for every $w \in A^*$, we have $\rho_*(w) = \rho(w)$ (this notation is useful when referring to the language $\rho_*\inv(r) \subseteq A^*$, which consists of all words $w \in A^*$ such that $\rho(w) = r$). Note that $\rho_*$ is a morphism into the finite monoid $(R,\cdot)$.

\begin{remark}
  As the adjective ``\tame'' suggests, a more general notion, the ``\ratms'', is defined in~\cite{pzcovering}. These are morphisms of idempotent and commutative monoids ($R$ needs not be equipped with a multiplication). We do not use this notion in the paper.
\end{remark}

Most of the theory makes sense for arbitrary \mratms. Yet, in the paper, we work with special \mratms satisfying an additional property.

\medskip
\noindent
{\bf \Nice \mratms.} A \mratm $\rho: 2^{A^*} \to R$ is \emph{\nice} when, for every language $K \subseteq A^*$, there exist finitely many words $w_1,\dots,w_n \in K$ such that $\rho(K) = \rho(w_1) + \cdots + \rho(w_k)$.

A \nice \mratm $\rho: 2^{A^*} \to R$ is characterized by the canonical monoid morphism $\rho_*: A^* \to R$. Indeed, for $K \subseteq A^*$, we may consider the sum of all elements $\rho(w)$ for $w \in K$: while it may be infinite, this sum boils down to a finite one since $R$ is commutative and idempotent for addition. The hypothesis that $\rho$ is \nice implies that $\rho(K)$ is equal to this sum. The key point here is that \nice \mratms are finitely representable: clearly, a \nice \mratm $\rho$ is characterized by the morphism $\rho_*: A^* \to R$, which is finitely representable since it is a morphism into a finite monoid. Hence, we may speak about algorithms taking \nice \mratms as input.

\medskip
\noindent
{\bf Canonical \mratm associated to a monoid morphism.} Finally, one may associate a particular \nice \mratm $\rho_\alpha$ to every monoid morphism $\alpha: A^* \to M$ into a finite monoid. Its rating set is the idempotent semiring $(2^M,\cup,\cdot)$, whose multiplication is obtained by lifting the one of $M$ to subsets of~$M$. Moreover, for every language $K\subseteq A^*$, we let $\rho_\alpha(K)$ be the direct image $\alpha(K) \subseteq A^*$. In other words, we~define:
\[
  \begin{array}{llll}
    \rho_\alpha: & 2^{A^*} & \to     & 2^M                          \\
                 & K       & \mapsto & \{\alpha(w) \mid w \in A^*\}.
  \end{array}
\]
Clearly, $\rho_\alpha$ is a \nice \mratm.

\subsection{\Imprints, optimality and application to covering.} We may now define \imprints. Let $\rho: 2^{A^*} \to R$ be a \mratm. For every finite set of languages~\Kb, we define the $\rho$-\imprint of~\Kb. Intuitively, when \Kb is a cover of some language $L$, this object measures the ``quality'' of \Kb. The  \emph{$\rho$-\imprint of \Kb} is the subset of~$R$ defined by:
\[
  \prin{\rho}{\Kb} = \dclosr \big\{\rho(K) \mid K \in\Kb\big\}.
\]
We now define optimality. Consider an arbitrary \mratm $\rho: 2^{A^*} \to R$ and a lattice~\Ds. Given a language $L$, an \emph{optimal} \Ds-cover of $L$ for $\rho$ is a \Ds-cover \Kb of $L$ having the smallest possible imprint among all \Ds-covers, \emph{i.e.}, which satisfies the following~property:
\[
  \prin{\rho}{\Kb} \subseteq \prin{\rho}{\Kb'} \quad \text{for every \Ds-cover $\Kb'$ of $L$}.
\]
In general, there can be infinitely many optimal \Ds-covers for a given \mratm $\rho$. Yet, there always exists at least one, if \Ds is a lattice (see~\cite[Lemma~4.15]{pzcovering}).

\begin{lemma}\label{lem:opt}
  Let \Ds be a lattice. For every language $L$ and every \mratm~$\rho$, there exists an optimal \Ds-cover of $L$ for $\rho$.
\end{lemma}

Clearly, given a lattice \Ds, a language $L$ and a \mratm $\rho$, all optimal \Ds-covers of $L$ for $\rho$ have the same $\rho$-\imprint. Hence, this unique $\rho$-\imprint is a \emph{canonical} object for \Ds, $L$ and $\rho$. We call it the \emph{\Ds-optimal $\rho$-\imprint on $L$} and we write it $\opti{\Ds}{L,\rho}$:
\[
  \opti{\Ds}{L,\rho} = \prin{\rho}{\Kb} \quad \text{for any optimal \Ds-cover \Kb of $L$  for $\rho$}.
\]
An important special case is when $L = A^*$. In this case, we write \opti{\Ds}{\rho} for \opti{\Ds}{A^*,\rho}.

\smallskip
\noindent
{\bf Connection with covering.} We may now connect these definitions to the covering problem. The key idea is that solving \Ds-covering for a fixed class \Ds boils down to finding an algorithm that computes \Ds-optimal \imprints from \nice \mratms given as inputs. In~\cite{pzcovering}, two statements are presented. The first is simpler but it only applies Boolean algebras, while the second is more involved and applies to all lattices. Since all classes investigated in the paper are Boolean algebras, we only present the first statement.

\begin{proposition}\label{prop:breduc}
  Let \Ds be a Boolean algebra. There exists an effective reduction from \Ds-covering to the following problem:

  \begin{tabular}{ll}
    {\bf Input:} & A \nice \mratm $\rho: 2^{A^*} \to R$ and $F \subseteq R$. \\
    {\bf Question:} & Is it true that $\opti{\Ds}{\rho} \cap F = \emptyset$?
  \end{tabular}
\end{proposition}

\begin{proof}[Proof sketch]
  We briefly describe the reduction (we refer the reader to~\cite{pzcovering} for details). Consider an input pair $(L_0,\{L_1,\dots,L_n\})$ for \Ds-covering. Since the languages $L_i$ are regular, for every $i \leq n$, one may compute a morphism $\alpha_i: A^* \to M_i$ into a finite monoid recognizing~$L_i$ together with the set $F_i \subseteq M_i$ such that $L_i = \alpha_i\inv(F_i)$. Consider the associated \nice \mratms $\rho_{\alpha_i} : 2^{A^*} \to 2^{M_i}$. Moreover, let $R$ be the idempotent semiring $2^{M_0} \times \cdots \times 2^{M_n}$ equipped with the componentwise addition and multiplication. We define a \nice \mratm $\rho: 2^{A^*} \to R$ by letting $\rho(K) = (\rho_{\alpha_0}(K),\dots,\rho_{\alpha_n}(K))$ for every $K \subseteq A^*$. Finally, let $F \subseteq R$ be the set of all tuples $(X_0,\dots,X_n) \in R$ such that $X_i \cap F_i \neq \emptyset$ for every $i \leq n$. One may now verify that $(L_0,\{L_1,\dots,L_n\})$ is \Ds-coverable if and only if $\opti{\Ds}{\rho} \cap F = \emptyset$. Let us point out that this equivalence is only true when \Ds is a Boolean algebra. When \Ds is only a lattice, one has to handle the language $L_0$ separately.
\end{proof}

We complete Proposition~\ref{prop:breduc} with a second statement which handles the converse direction. We prove that if \Ds-covering is decidable, then one may compute the set \dopti{L,\rho} associated to a regular language $L$ and a \nice \mratm $\rho$.

\begin{proposition} \label{prop:covreduc}
	Let \Ds be a Boolean algebra, $L \subseteq A^*$ and $\rho: 2^{A^*} \to R$ a \nice \mratm. Then,
	\[
	\dopti{L,\rho} = \dclosr \left\{\sum_{q \in Q} q \mid \text{$Q \subseteq R$ such that } \left(L,\left\{\rho_*\inv(q) \mid q \in Q\right\}\right) \text{ is not \Ds-coverable}\right\}.
	\]
\end{proposition}

\begin{proof}
	We first prove the left to right inclusion. Let $r \in \dopti{L,\rho}$. We exhibit $Q \subseteq R$ such that $r \leq \sum_{q \in Q} q$ and $(L,\{\rho_*\inv(q) \mid q \in Q\})$ is not \Ds-coverable. Let $\tau: 2^{A^*} \to 2^R$ be the map defined by $\tau(K) = \{\rho(w)  \mid w \in K\}$. One may verify that $\tau$ is a \nice \mratm. Let $\Kb_\tau$ be an optimal \Ds-cover of $L$ for $\tau$. Since $r \in \dopti{L,\rho}$, we have $r \in \prin{\rho}{\Kb_\tau}$ and we get $K \in \Kb_\tau$ such that $r \leq \rho(K)$. Let $Q = \tau(K) \subseteq R$. Since $\rho$ is \emph{\nice}, one may verify that $\rho(K) = \sum_{q \in Q} q$. Thus, $r\leq \sum_{q \in Q} q$ and it remains to prove that $(L,\{\rho_*\inv(q) \mid q \in Q\})$ is not \Ds-coverable. We proceed by contradiction. Assume that there exists a \Ds-cover \Hb of $L$ which is separating for $\{\rho_*\inv(q) \mid q \in Q\}$. For every $H \in \Hb$, we know that there exists $q \in Q$ such that $H \cap \rho_*\inv(q) = \emptyset$. By definition of $\tau$, this implies that $Q \not\subseteq \tau(H)$ for every $H \in \Hb$. Consequently, $Q \not\in \prin{\tau}{\Hb}$ which yields $Q \not\in \dopti{L,\tau}$. This is a contradiction since $Q = \tau(K)$ by definition and $K \in \Kb_\tau$ which is an \emph{optimal} \Ds-cover of $L$ for $\tau$.

	It remains to prove the right to left inclusion. Since \dopti{L,\rho} is an \imprint, we have $\dclosr \dopti{L,\rho}= \dopti{L,\rho}$ by definition. Hence, it suffices to prove that for every $Q \subseteq R$ such that $(L,\{\rho_*\inv(q) \mid q \in Q\})$ is not \Ds-coverable, we have $\sum_{q \in Q} q \in \dopti{L,\rho}$. Let $\Kb_\rho$ be an optimal \Ds-cover of $L$ for $\rho$. Since $(L,\{\rho_*\inv(q) \mid q \in Q\})$ is not \Ds-coverable, $\Kb_\rho$ cannot be separating for $\{\rho_*\inv(q) \mid q \in Q\}$ and we get $K \in \Kb_\rho$ such that $K \cap \rho_*\inv(q) \neq \emptyset$ for every $q \in Q$. It follows that $\sum_{q \in Q} q \leq \rho(K)$. We get $\sum_{q \in Q} q \in \prin{\rho}{\Kb_\rho} = \dopti{L,\rho}$ as desired.
\end{proof}

\subsection{Application to the classes considered in the paper} Proposition~\ref{prop:breduc} implies that given a Boolean algebra \Ds, deciding \Ds-covering boils down to computing \opti{\Ds}{\rho} from a \nice \mratm $\rho$. We use this approach for several classes \Ds. Roughly, all of them are levels in the deterministic hierarchy built from an arbitrary \emph{finite} \vari \Cs. Hence, an algorithm computing \opti{\Ds}{\rho} should be parameterized by \Cs in some way. Let us explain how. We first  present a key property of the finite \varis.

\smallskip
\noindent
{\bf Canonical morphism of a finite \vari.} Consider a \emph{finite} \vari \Cs (\emph{i.e.}, \Cs contains finitely many languages). Proposition~\ref{prop:genocm} implies that there exists a \Cs-morphism recognizing \emph{all} languages in \Cs. The next lemma implies that it is unique (up to renaming).

\begin{lemma} \label{lem:cmdiv}
	Let \Cs be a finite \vari and let $\alpha: A^* \to M$ and $\eta: A^* \to N$ be two \Cs-morphisms. If $\alpha$ recognizes all languages in \Cs, then there exists a morphism $\gamma: M \to N$ such that $\eta = \gamma \circ \alpha$.
\end{lemma}

\begin{proof}
	For each $s \in M$, we fix a word $w_s \in \alpha\inv(s)$ (recall that \Cs-morphisms are surjective) and define $\gamma(s) = \eta(w_s)$. It remains to prove that $\gamma$ is a morphism and that $\eta = \gamma \circ \alpha$. It suffices to prove the latter: since $\alpha$ is surjective, the former is an immediate consequence. Let $v \in A^*$. We show that $\eta(v) = \gamma(\alpha(v))$. Let $s = \alpha(v)$. By definition, $\gamma(s) = \eta(w_s)$. Hence, we need to prove that $\eta(v) = \eta(w_s)$. Since $\eta$ is a \Cs-morphism, $\eta\inv(\eta(w_{s})) \in \Cs$. Hence, our hypothesis implies that $\eta\inv(\eta(w_{s}))$ is recognized by $\alpha$. Since it is clear that $w_s \in \eta\inv(\eta(w_{s}))$ and $\alpha(v) = \alpha(w_s) = s$, we get $v \in \eta\inv(\eta(w_{s}))$ which exactly says that $\eta(v) = \eta(w_s)$.
\end{proof}

By Lemma~\ref{lem:cmdiv}, if \Cs is a finite \vari and $\alpha: A^* \to M$ and $\eta: A^* \to N$ are two \Cs-morphisms which both recognize \emph{all} languages in \Cs, there are morphisms $\gamma: M \to N$ and $\beta: N \to M$ such that $\eta=\gamma \circ \alpha$ and $\alpha = \beta \circ \eta$. Since $\alpha$ and $\eta$ are surjective, $\beta \circ \gamma: M \to M$ is the identity morphism. Hence, $\beta$ and $\gamma$ are both isomorphisms which means that $\alpha$ and $\eta$ are the same object up to renaming. We call it the \emph{canonical \Cs-morphism} and denote it by $\etac: A^* \to \canc$. Let us emphasize that this object is only defined when \Cs is a \emph{finite \vari}.

\smallskip
\noindent
{\bf Pointed optimal \imprints.} We now come back to covering and optimal \imprints. The key idea is that when dealing with a Boolean algebra \Ds built from some finite \vari \Cs, an algorithm which computes $\opti{\Ds}{\rho} \subseteq R$ from a \nice \mratm $\rho: 2^{A^*} \to R$ does not consider this set directly. Instead, it looks at a more general object that \emph{records more information} (the idea being that this extra information is required in the computation). More precisely, we shall use an algorithm which computes all sets $\opti{\Ds}{\etac\inv(s),\rho}$ for $s \in \canc$ where $\etac: A^*\to \canc$ is the canonical \Cs-morphism (as seen in Lemma~\ref{lem:pointgen} below, their union is the desired set \dopti{\rho}). Yet, it will be more convenient to represent this family of sets by a single set of pairs. Here, we introduce terminology for this purpose.

Let \Ds be a Boolean algebra, $\eta: A^* \to N$ a morphism and $\rho: 2^{A^*} \to R$ a \mratm. The \emph{$\eta$-pointed \Ds-optimal $\rho$-\imprint} is the following set $\popti{\Ds}{\eta}{\rho} \subseteq N \times R$:
\[
  \popti{\Ds}{\eta}{\rho} = \bigl\{(s,r) \in N \times R \mid r \in \opti{\Ds}{\eta\inv(s),\rho}\bigl\}.
\]
Clearly, $\popti{\Ds}{\eta}{\rho} \subseteq N \times R$ encodes all sets \opti{\Ds}{\eta\inv(s),\rho} for $s \in N$. The following statement implies that this suffices in order to compute \dopti{\rho} (see~\cite[Lemma~4.15]{pzcovering} for the proof).

\begin{lemma}\label{lem:pointgen}
  Let \Ds be a Boolean algebra, $\eta: A^* \to N$ be a morphism into a finite monoid and $\rho: 2^{A^*} \to R$ be a \mratm. Then,
  \[
    \opti{\Ds}{\rho} = \bigcup_{s\in N}  \opti{\Ds}{\eta\inv(s),\rho} = \{r \in R \mid \text{there exists $s \in N$ such that $(s,r) \in \popti{\Ds}{\eta}{\rho}$}\}.
  \]
\end{lemma}

In the sequel, we shall present algorithms which compute the sets $\popti{\Ds}{\etac}{\rho} \subseteq \canc \times R$ from a \nice \mratm $\rho$ where \Cs is a \emph{finite} \vari and \Ds is a class built from \Cs using \ldeto, \rdeto and \mdeto.

\smallskip
\noindent
{\bf Properties.} We present a few useful generic properties of these sets. Let $\eta: A^* \to N$ be a morphism and $\rho: 2^{A^*} \to R$ a \mratm. We say that a set $S\subseteq N \times R$ is \emph{saturated} for $\eta$ and $\rho$ to indicate that it satisfies the three~following properties:
\begin{enumerate}
	\item {\bf Trivial elements.} For every $w \in A^*$, we have $(\eta(w) ,\rho(w)) \in S$.
	\item {\bf Downset.} We have $\dclosr S = S$.
	\item {\bf Multiplication.} For every $(s,q),(t,r) \in S$, we have $(st,qr) \in S$.
\end{enumerate}

We have the following lemma (see~\cite[Lemma 7.7]{pzcovering} for the proof).

\begin{lemma} \label{lem:genprin}
	Let \Ds be a \vari, $\eta: A^* \to N$ a morphism and $\rho: 2^{A^*} \to R$ a \mratm. Then, the set $\popti{\Ds}{\eta}{\rho} \subseteq N \times R$ is saturated for $\eta$ and $\rho$.
\end{lemma}

We now present two technical lemmas. When put together, they characterize the sets \popti{\Ds}{\eta}{\rho} in terms of \Ds-morphisms. This will be useful in proof arguments.

\begin{lemma} \label{lem:optmo}
	Let \Ds be a \vari, $\eta: A^* \to N$ a morphism and $\rho: 2^{A^*} \to R$ a \mratm. Moreover, let $\alpha: A^* \to M$ be a \Ds-morphism. For every $(s,r) \in \popti{\Ds}{\eta}{\rho}$, there exists $w \in A^*$ such that  $\eta(w) = s$ and $r \leq \rho(\atyp{w})$.
\end{lemma}

\begin{proof}
	We fix $(s,r) \in \popti{\Ds}{\eta}{\rho}$ for the proof. By definition $r \in \dopti{\eta\inv(s),r}$. Since $\alpha$ is a \Ds-morphism, the set $\Kb = \{\atyp{w} \mid w \in \eta\inv(s)\}$ is a \Ds-cover of $\eta\inv(s)$. Hence, $r \in \prin{\rho}{\Kb}$ by hypothesis. By definition of \Kb, this yields $w \in A^*$ such that  $\eta(w) = s$ and $r \leq \rho(\atyp{w})$.
\end{proof}

For the second lemma, we need a preliminary definition. Let \Cs be a \emph{finite} \vari and $\alpha: A^* \to M$ a morphism. We say that $\alpha$ is \emph{\Cs-compatible} to indicate that the morphism $\ctype{\cdot} \circ \alpha: A^* \to {A^*}/{\canec}$ (which is a \Cs-morphism by Lemma~\ref{lem:smult}) is exactly the canonical \Cs-morphism $\etac: A^* \to \canc$ (up to renaming).

\begin{lemma} \label{lem:moopt}
	Let \Cs be a finite \vari and \Ds a \vari such that $\Cs \subseteq \Ds$. Let $\eta: A^* \to N$ be a morphism and $\rho: 2^{A^*} \to R$ a \mratm. There exists a \Cs-compatible \Ds-morphism $\alpha: A^* \to M$ such that for every $w \in A^*$ and $r \leq \rho(\atyp{w})$, we have $(\eta(w),r) \in \popti{\Ds}{\eta}{\rho}$.
\end{lemma}

\begin{proof}
	For every $s \in N$, we let $\Kb_s$ as an optimal \Ds-cover of $\eta\inv(s)$. Since \Ds is a \vari and \Cs is a finite \vari such that $\Cs \subseteq \Ds$, Proposition~\ref{prop:genocm} yields a \Ds-morphism $\alpha$ recognizing all languages in \Cs and all languages $K \in \Kb_s$ for $s \in N$. It follows from Lemma~\ref{lem:smult} that $\ctype{\cdot} \circ \alpha$ is a \Cs-morphism which recognizes \emph{all} languages in \Cs. Hence, it is the canonical \Cs-morphism by Lemma~\ref{lem:cmdiv} and we conclude that $\alpha$ is \Cs-compatible. It remains to prove that for $w \in A^*$ and $r \leq \rho(\atyp{w})$, we have $(\eta(w),r) \in \popti{\Ds}{\eta}{\rho}$. Let $s = \eta(w)$. Since $w \in \eta\inv(s)$, there exists $K \in \Kb_s$ such that $w \in K$. Moreover, since $K$ is recognized by $\alpha$, we have $\atyp{w} \subseteq K$. Hence, $r \leq \rho(\atyp{w}) \leq \rho(K)$. Since $\Kb_S$ is an optimal \Ds-cover of $\eta\inv(s)$, it follows that $r \in \dopti{\eta\inv(s),\rho}$ which exactly says that $(s,r) \in  \popti{\Ds}{\eta}{\rho}$ as desired.
\end{proof}

%%% Local Variables:
%%% mode: latex
%%% TeX-master: "main"
%%% End:

%% file: cov.tex
We consider covering for the classes built with left/right polynomial closure. We prove that if \Cs is a \emph{finite} \vari and \Ds is a \vari \emph{with decidable covering} such that $\Cs \subseteq \Ds \subseteq \upol{\Cs}$, then covering is decidable for \ldet{\Ds} and \rdet{\Ds}. This can be lifted to \emph{all} levels \ldetn{\Ds} and \rdetn{\Ds} in the deterministic hierarchy of \Ds by induction.

The results are presented using \ratms and the framework introduced in Section~\ref{sec:ratms}: we give effective characterizations of \ldet{\Ds}- and \rdet{\Ds}-optimal \imprints. In particular, we rely on the additional notions designed to handle classes built from an arbitrary finite \vari \Cs. We work with \etac-pointed optimal \imprints where $\etac: A^* \to \canc$ is the canonical \Cs-morphism. Given a \mratm $\rho: 2^{A^*} \to R$, we characterize the subsets \pldetoptid and \prdetoptid of $\canc \times R$. Both characterizations are parameterized by the set $\popti{\Ds}{\etac}{\rho} \subseteq \canc \times R$ (this is how they depend on \Ds). When $\rho$ is \nice, they yield least fixpoint algorithms for computing the sets \pldetoptid and \prdetoptid from \popti{\Ds}{\etac}{\rho} (which is computable when \Ds-covering is decidable by Proposition~\ref{prop:covreduc}). Consequently, \ldet{\Ds}- and \rdet{\Ds}-covering are decidable in that case by Proposition~\ref{prop:breduc}. We first present the characterizations. The remainder of the section is then devoted to their proof.

\subsection{Statement}

Consider a morphism $\eta: A^* \to N$ and a \mratm $\rho: 2^{A^*}\to R$. For every set $P \subseteq N \times R$, we define the $(\ldeto,P)$-saturated subsets and the $(\rdeto,P)$-saturated subsets of $N \times R$ for $\eta$ and $\rho$. We fix $S \subseteq N \times R$ for the definition. We say that $S$ is $(\ldeto,P)$-saturated for $\eta$ and $\rho$ when it is saturated for $\eta$ and $\rho$, and satisfies the following additional property:
\begin{equation} \label{eq:detlc}
	\begin{array}{c}
		\text{for every pair of multiplicative idempotents $(e,f) \in S$ and every $(s,r) \in P$}\\
		\text{such that $e \Rord s$, we have $(es,fr) \in S$.}
	\end{array}
\end{equation}
Symmetrically, we say $S$ is $(\rdeto,P)$-saturated for $\eta$ and~$\rho$ when it is saturated for $\eta$ and $\rho$, and satisfies the following additional property:
\begin{equation} \label{eq:detrc}
	\begin{array}{c}
		\text{for every pair of multiplicative idempotents $(e,f) \in S$ and every $(s,r) \in P$}\\
		\text{such that $e \Lord s$, we have $(se,rf) \in S$.}
	\end{array}
\end{equation}
We are ready to state the characterization. We present it in the following theorem.

\begin{theorem} \label{thm:detlrc}
	Let \Cs be a finite \vari and \Ds a \vari such that $\Cs \subseteq \Ds \subseteq \upol{\Cs}$. Let $\rho: 2^{A^*} \to R$ be a \mratm and $P= \popti{\Ds}{\etac}{\rho}$. Then,
	\begin{itemize}
		\item \pldetoptid is the least $(\ldeto,P)$-saturated subset of $\canc \times R$ for \etac and $\rho$.
		\item \prdetoptid is the least $(\rdeto,P)$-saturated subset of $\canc \times R$ for \etac and $\rho$.
	\end{itemize}
\end{theorem}

Clearly, when $\rho: 2^{A^*} \to R$ is a \nice \mratm, Theorem~\ref{thm:detlrc} provides algorithms for computing the sets \pldetoptid and \prdetoptid from $P = \popti{\Ds}{\etac}{\rho}$. Indeed, the least $(\ldeto,P)$-saturated (resp. $(\rdeto,P)$-saturated) subset of $\canc\times R$ can be computed using a least fixpoint procedure. It starts from the set of trivial elements $(\etac(w),\rho(w)) \in\canc\times R$ and saturates it with the three operations in the definition: downset, multiplication and~\eqref{eq:detlc} (resp. \eqref{eq:detrc}). It is immediate that these operations can be implemented (for~\eqref{eq:detlc} and~\eqref{eq:detrc}, this is because we have the set $P$ in hand). Once \pldetoptid and \prdetoptid have been computed, it follows from Lemma~\ref{lem:pointgen} that the sets \opti{\ldet{\Ds}}{\rho} and \opti{\rdet{\Ds}}{\rho} can be computed as well. In view of Proposition~\ref{prop:breduc}, being able to compute these two sets is enough to decide covering for \ldet{\Ds} and \rdet{\Ds}. Thus, it follows that covering is decidable for \ldet{\Ds} and \rdet{\Ds} if one may compute the set $P = \popti{\Ds}{\etac}{\rho}$ from a \nice \mratm $\rho$. Finally, Proposition~\ref{prop:covreduc} implies that this set can be computed provided that \Ds-covering is decidable. %Altogether, we obtain the following corollary.

\begin{corollary} \label{cor:detlrc}
	Let \Cs be a finite \vari and \Ds a \vari with decidable covering such that $\Cs \subseteq \Ds \subseteq \upol{\Cs}$. Then, \ldet{\Ds}- and \rdet{\Ds}-covering are decidable.
\end{corollary}

Moreover, by definition of deterministic hierarchies, one may lift Corollary~\ref{cor:detlrc} to all levels \ldetn{\Ds} and \rdetn{\Ds} using induction. This yields the following corollary.

\begin{corollary} \label{cor:hdetlrc}
	Let \Cs be a finite \vari and \Ds a \vari with decidable covering such that $\Cs \subseteq \Ds \subseteq \upol{\Cs}$. Then, \ldetn{\Ds}- and \rdetn{\Ds}-covering are decidable for all $n \in \nat$.
\end{corollary}

An interesting application of Corollary~\ref{cor:hdetlrc} is the special case when $\Cs = \Ds$. Since \Cs is \emph{finite}, \Cs-covering is decidable (one may use a brute-force approach which consists in~testing all the finitely many possible \Cs-covers). Hence, we obtain that for every finite \vari \Cs, covering is decidable for all levels \ldetn{\Cs} and \rdetn{\Cs} for $n \in \nat$.

\smallskip
\noindent
{\bf A key application: the alphabet testable languages.} Let \at be the class containing the Boolean combinations of languages $B^*$ where $B \subseteq A$. One may verify that \at is a \vari. Moreover, it~is clearly finite by definition.   The class \at is particularly important in the literature because there are many operators $Op$ such that $Op(\at) = Op(\pt)$ where $\pt = \bpol{\stzer}$ is the class of \emph{piecewise testable languages}. For example, it is well-known~\cite{pin-straubing:upper} that $\pol{\at} = \pol{\pt}$ (see also~\cite{PZ:generic18} for a recent proof). This kind of result is important because finite \varis (such as \at) are often simpler to handle than infinite ones (such as \pt). This connection also holds for \ldeto, \rdeto and \upolo.

\begin{lemma} \label{lem:upolat}
	For every $n \in \nat$, we have $\upol{\at} = \upol{\pt}$, $\ldetn{\at} = \ldetn{\pt}$ and $\rdetn{\at} =\rdetn{\pt}$.
\end{lemma}

\begin{remark} \label{rem:upolat}
	On the other hand, Lemma~\ref{lem:upolat} fails for \mdeto: we have the strict inclusion $\mdet{\at} \subsetneq \mdet{\pt}$. This point will be important in Section~\ref{sec:covm}.
\end{remark}

\begin{proof}
	Clearly, it suffices to show that $\ldet{\at} = \ldet{\pt}$ and $\rdet{\at} = \rdet{\pt}$. That $\ldetn{\at} = \ldetn{\pt}$ and $\rdetn{\at}=\rdetn{\pt}$ for every $n \in \nat$, it then immediate by induction on $n$. Moreover, the equality  $\upol{\at} = \upol{\pt}$ also follows since \upol{\Cs} is exactly the union of all levels \ldetn{\Cs} (for every \vari \Cs) by Theorem~\ref{thm:apol}.
	
	By symmetry, we only prove that $\ldet{\at} = \ldet{\pt}$. Since $\at \subseteq \pt$ by definition, the left to right inclusion is immediate. We prove that $\pt \subseteq \ldet{\at}$. This will imply that $\ldet{\pt} \subseteq \ldet{\ldet{\at}} = \ldet{\at}$ as desired. Every language in \pt is a Boolean combination of marked products $A^*a_1A^* \cdots a_nA^*$. Therefore, since \ldet{\at} is a \vari by Theorem~\ref{thm:detclos}, it suffices to prove that every such marked product belongs to \ldet{\at}. Observe that $A^*a_1A^* \cdots a_nA^*$ is also defined by the marked product $(A \setminus \{a_1\})^*a_1(A \setminus \{a_2\})^*a_2 \cdots (A \setminus \{a_n\})^*a_nA^*$. One may verify that this a left deterministic marked product of languages in \at. Thus, $A^*a_1A^* \cdots a_nA^* \in \ldet{\at}$ as desired.
\end{proof}

Clearly, Corollary~\ref{cor:hdetlrc} implies that \ldetn{\at}- and \rdetn{\at}-covering are decidable for all $n \in \nat$. Hence, in view of Lemma~\ref{lem:upolat}, we obtain that \ldetn{\pt}- and \rdetn{\pt}-covering are decidable for all $n \in \nat$. Naturally, this extends to separation by Lemma~\ref{lem:covsep}.

\begin{corollary} \label{cor:detpt}
	For every level $n \in \nat$, \ldetn{\pt} and \rdetn{\pt} have decidable separation and covering.
\end{corollary}

Corollary~\ref{cor:detpt} is important since, as mentioned in Section~\ref{sec:deth}, the deterministic hierarchy associated to the class \pt is prominent in the literature.  Actually, there exists an alternate independent proof of the decidability of covering for all levels \ldetn{\pt} and \rdetn{\pt} by Henriksson and Kufleitner~\cite{kkufo2}. It is based on techniques tailored to this hierarchy.

\subsection{Proof argument}

We now prove Theorem~\ref{thm:detlrc}. It involves two independent statements which correspond respectively to soundness and completeness in the least fixpoint procedures computing \pldetoptid and \prdetoptid.  We first prove soundness.

\begin{proposition} \label{prop:detsound}
	Let \Cs be a finite \vari and \Ds a \vari such that $\Cs \subseteq \Ds \subseteq \upol{\Cs}$. Let $\rho: 2^{A^*} \to R$ be a \mratm and $P= \popti{\Ds}{\etac}{\rho}$. Then, \pldetoptid is $(\ldeto,P)$-saturated for \etac and $\rho$, and \prdetoptid is $(\rdeto,P)$-saturated for \etac and $\rho$.
\end{proposition}

\begin{proof}
	We prove that \pldetoptid is $(\ldeto,P)$-saturated. We leave the symmetrical argument for  \prdetoptid to the reader. By Theorem~\ref{thm:detclos}, \ldet{\Ds} is a \vari. Hence, Lemma~\ref{lem:genprin} implies that \pldetoptid is saturated for \etac and $\rho$. Let us prove~\eqref{eq:detlc}. We use Lemma~\ref{lem:moopt} which yields a \Cs-compatible \ldet{\Ds}-morphism $\alpha: A^* \to M$ such that for every $w \in A^*$ and $r \leq \rho(\atyp{w})$, we have $(\etac(w),r) \in \pldetoptid$. We may now prove~\eqref{eq:detlc}. Let $(e,f) \in \pldetoptid$ be a pair of multiplicative idempotents and $(s,r) \in P$ such that $e \Rord s$. We show that $(es,fr) \in \pldetoptid$. By definition of $\alpha$, it suffices to exhibit $w \in A^*$ such that $\etac(w) = es$ and $fr \leq \rho(\atyp{w})$. We write $k = \omega(M)$ for the proof.
	
	Since $(e,f) \in \pldetoptid$, Lemma~\ref{lem:optmo} yields $u \in A^*$ such that $\etac(u) = e$ and  $f \leq \rho(\atyp{u})$. Consider the congruence \caned on $M$ and let $\gamma = \dtype{\cdot} \circ \alpha: A^* \to {M}/{\caned}$ which is a \Ds-morphism by Lemma~\ref{lem:smult}. Thus, as  $(s,r) \in P= \popti{\Ds}{\etac}{\rho}$, Lemma~\ref{lem:optmo} yields $v \in A^*$ such that $\etac(v) = s$ and $r \leq \rho(\gtyp{v})$. We let $w =u^k v$. Since $e$ is an idempotent, we have $\etac(w) = es$ by definition. Let us show that $fr \leq \rho(\atyp{w})$. We prove that $(\atyp{u})^k\gtyp{v}\subseteq \atyp{w}$. Since $f \leq \rho(\atyp{u})$, $r \leq \rho(\gtyp{v})$ and $f$ is an idempotent, this yields $fr \leq  \rho(\atyp{w})$ as desired. 
	
	We fix $x \in(\atyp{u})^k \gtyp{v}$ and show that $\alpha(x) = \alpha(w)$. Let $g = (\alpha(u))^k$ which is idempotent by definition of $k$. We have $\alpha(w) = g\alpha(v)$ by definition. Moreover, the definition of $x$ yields $v'$ such that $\gamma(v) = \gamma(v')$ and $\alpha(x) = g\alpha(v')$. It remains to prove that $g\alpha(v) = g\alpha(v')$. By definition of $\gamma$, we have $\alpha(v) \caned \alpha(v')$. Moreover, recall that $\etac(u) = e$ which yields $\etac(u^k) = e$ and $\etac(v) = s$. Hence, since $\alpha$ is \Cs-compatible (which means that $\ctype{\cdot} \circ \alpha = \etac$), we have $\ctype{g} = e$ and $\ctype{\alpha(v)} = s$ which yields $\ctype{g} \Rord\ctype{\alpha(v)}$ by hypothesis on $e$ and $s$. Altogether, since $\alpha$ is an \ldet{\Ds}-morphism and $\Cs \subseteq \Ds \subseteq \upol{\Cs}$, it follows from the first assertion in Lemma~\ref{lem:lradv} that $g\alpha(v) = g\alpha(v')$ which completes the proof.
\end{proof}

We turn to completeness in Theorem~\ref{thm:detlrc}. We use the following proposition.

\begin{proposition} \label{prop:detcomp}
		Let \Cs be a finite \vari and \Ds a \vari such that $\Cs \subseteq \Ds \subseteq \upol{\Cs}$, $\eta: A^* \to N$ a \Cs-morphism, $\rho: 2^{A^*}\to R$ a \mratm and $P = \popti{\Ds}{\eta}{\rho}$.
	\begin{itemize}
		\item If $S \subseteq N \times R$ is $(\ldeto,P)$-saturated for $\eta$ and $\rho$, then, for each $s \in N$, there exists an \ldet{\Ds}-cover $\Kb_s$ of $\eta\inv(s)$ such that $(s,\rho(K)) \in S$ for every $K \in \Kb_s$.
		\item  If $S \subseteq N \times R$ is $(\rdeto,P)$-saturated for $\eta$ and $\rho$, then, for each $s \in N$, there exists an \rdet{\Ds}-cover $\Kb_s$ of $\eta\inv(s)$ such that $(s,\rho(K)) \in S$ for every $K \in \Kb_s$.
	\end{itemize}
\end{proposition}

\begin{proof}
	By symmetry, we only prove the first assertion. Hence, we consider $S \subseteq N \times R$ which is $(\ldeto,P)$-saturated for $\eta$ and $\rho$. Note that by closure under multiplication, we know that $S$ is a monoid for the componentwise multiplication (the neutral element is the trivial element $(1_N,1_R) = (\eta(\veps),\rho(\veps))$). The argument is based on the following lemma. We say that a cover \Kb of a language $L$ is \emph{tight} if $K \subseteq L$ for every $K \in \Kb$.
	
	\begin{lemma} \label{lem:detcov}
		Let $s \in N$ and $(t,q) \in S$. There exists a tight \ldet{\Ds}-cover of $\eta\inv(s)$ such that $(ts,q\rho(K))\in S$ for every $K \in \Kb$.
	\end{lemma}
	
	We first use Lemma~\ref{lem:detcov} to complete the main proof. Let $s \in N$ and $(t,q) = (1_N,1_R) \in S$. The lemma yields a tight \ldet{\Ds}-cover $\Kb_s$ of $\eta\inv(s)$ such that $(s,\rho(K)) \in S$ for every $K \in \Kb_s$ and the first assertion in Proposition~\ref{prop:detcomp} is proved.
	
	\smallskip
	
	It remains to prove Lemma~\ref{lem:detcov}. Let $s \in N$ and $(t,q) \in S$. We construct the tight \ldet{\Ds}-cover \Kb of $\eta\inv(s)$ by induction on two parameters which depend on the Green relations \Jrel and \Rrel of the monoids $N$ and $S$. They are as follows, listed by order of importance:
	\begin{enumerate}
		\item The \emph{\Jrel-rank of $s \in N$}: the number of elements $s' \in N$ such that $s \Jords s'$.
		\item The \emph{$\Rrel$-index of $(t,q) \in S$}: the number of pairs $(t',q') \in S$ such that $(t',q') \Rords (t,q)$.	
	\end{enumerate}
	We say that $(t,q) \in S$ is \emph{stabilized} by $s \in N$ to indicate that there exists $(t',q') \in S$ such that $t'\Rrel s$ and $(tt',qq')\Rrel (t,q)$. There are two cases depending on whether this holds.
	
	\smallskip
	\noindent
	{\bf Base case: $(t,q)$ is stabilized by $s$.} We define \Kb as an optimal \Ds-cover of $\eta\inv(s)$ for $\rho$. Note that we may assume without loss of generality that \Kb is tight as $\eta$ is a \Cs-morphism and $\Cs \subseteq \Ds$. It remains to prove that $(ts,q\rho(K))\in S$ for every $K \in \Kb$. We fix $K$ for the proof. Since $P = \popti{\Ds}{\eta}{\rho}$, and \Kb is an optimal \Ds-cover of $\eta\inv(s)$, we know that $(s,\rho(K)) \in P$.
	
	By hypothesis, there exists $(t',q'),(t'',q'') \in S$ such that $t'\Rrel s$ and $(tt't'',qq'q'') = (t,q)$. We define $(e,f) = ((t't'')^\omega,(q'q'')^\omega) \in S$ which is a pair of multiplicative idempotents. Clearly, $(te,qf) = (t,q)$. Moreover, since $t' \Rrel q$, it is immediate that $e \Rord s$. Hence, since $S$ is $(\ldeto,P)$-saturated, \eqref{eq:ldetcl} yields $(es,s\rho(K)) \in S$. Since $(t,q) \in S$, this yields $(tes,qf\rho(K)) \in S$. Finally, since $(qe,tf) = (q,t)$, we get $(ts,q\rho(K)) \in S$ as desired.
	
	\smallskip
	\noindent
	{\bf Inductive case: $(t,q)$ is not stabilized by $s$.} Let $T$ be the set of all $(s_1,a,s_2) \in  N\times A \times N$ such that $s_1 \eta(a) s_2 = s$ and $s \Rrel s_1\eta(a) \Rords s_1$. For every such triple $(s_1,a,s_2) \in T$, we use induction to build tight \ldet{\Ds}-covers of $\eta\inv(s_1)$ and $\eta\inv(s_2)$. We then combine them to construct \Kb. We fix a triple $(s_1,a,s_2) \in T$ for the definition.
	
	We have $s \Rords s_1$ by definition. This implies that $s \Jords s_1$ by Lemma~\ref{lem:jlr}. Hence, the \Jrel-rank of $s_1$ is strictly smaller than that of $s$. Hence, induction in  Lemma~\ref{lem:detcov} (for $s = s_1$ and $(t,q) = (1_N,1_R) \in S$) yields a tight \ldet{\Ds}-cover $\Ub_{s_1}$ of $\eta\inv(s_1)$ such that $(s_1,\rho(U)) \in S$ for every $U \in \Ub_{s_1}$. We now use our hypothesis in the inductive case to build several tight \ldet{\Ds}-covers of $\eta\inv(t_2)$: one for each $U \in \Ub_{s_1}$. We fix $U$ for the definition. We know that $s_1\eta(a) s_2 = s$ by definition. Hence, $s \Jord s_2$: the rank of $s_2$ is smaller than or equal to the one of $s$ (our first induction parameter has not increased). We also know that $(s_1,\rho(U)) \in S$ by definition of $\Ub_{s_1}$ and $(\eta(a),\rho(a)) \in S$ (this is a trivial element). Hence, $(s_1\eta(a),\rho(Ua))  \in S$. Moreover, $s \Rrel s_1\eta(a)$ by definition of $T$. Thus, since $(t,q)$ is not stabilized by $s$, we get $(ts_1\rho(a),q\rho(Ua)) \Rords (t,q)$. It follows that the \Rrel-index of $(ts_1\rho(a),q\rho(Ua))$ is strictly smaller than the one of $(t,q)$. Thus, induction on our second parameter in Lemma~\ref{lem:detcov} yields a tight \ldet{\Ds}-cover $\Vb_{(s_1,a,s_2),U}$ of $\eta\inv(s_2)$ such that $(ts_1\rho(a)s_2,q\rho(UaV)) \in S$ for every $V \in \Vb_{(s_1,a,s_2),U}$. We are ready to construct \Kb. We define,
	\[
	\Kb = \bigcup_{(s_1,a,s_2) \in T}  \{UaV \mid U \in \Ub_{s_1} \text{ and } V \in \Vb_{(s_1,a,s_2),U}\}.
	\]
	It remains to verify that \Kb is a tight \ldet{\Ds}-cover of $\eta\inv(s)$ and that $(ts,q\rho(K)) \in S$ for every $K \in \Kb$. We first show that \Kb is a cover of $\eta\inv(s)$. Let $w \in \eta\inv(s)$. We exhibit $K \in \Kb$ such that $w \in K$. Let $u' \in A^*$ be the \emph{least} prefix of $w$ such that $\eta(u') \Rrel \eta(w) = s$ and $v \in A^*$ the corresponding suffix: $w = u'v$. Observe that $u' \neq \veps$. Indeed, otherwise we have $1_N \Rrel s$ and since $(t1_N,q1_R) \Rrel (t,q)$ this contradicts the hypothesis that $(t,q)$ is not stabilized by $s$. Thus, we get $u \in A^*$ and $a \in A$ such that $u' = ua$. Moreover, $\eta(ua) \Rords \eta(u)$ by definition of $u' = ua$. Let $s_1 = \eta(u)$ and $s_2 = \eta(v)$. Clearly, $(s_1,a,s_2) \in T$: we have  $s_1 \eta(a) s_2 = \eta(uav) = \eta(w) = s$, $s \Rords s_1 = \eta(u)$ and $s \Rrel s_1\eta(a) = \eta(ua)$. Finally, since $\Ub_{s_1}$ and $\Vb_{(s_1,a,s_2),U}$ are covers of $\eta\inv(s_1)$ and $\eta\inv(s_2)$ respectively, we obtain $U \in \Ub_{s_1}$ and $V \in \Vb_{(s_1,a,s_2),U}$ such that $u \in U$ and $v \in V$. It follows that $w =uav \in UaV$ which is a language in \Kb by definition. Thus, \Kb is a cover of $\eta\inv(t)$. Moreover, it is simple to verify that that it is tight. If $K \in \Kb$ we have $K \subseteq  \eta\inv(s_1) a\eta\inv(s_2)$ for $(s_1,a,s_2)\in T$ by definition of \Kb. Since $s_1 \eta(a) s_2 = s$, this yields $K \subseteq \eta\inv(s)$.

	We now prove that every $K \in \Kb$ belongs to \ldet{\Ds} and satisfies $(ts,q\rho(K)) \in S$. By definition, $K = UaV$ for $U \in \Ub_{s_1}$ and $V\in\Vb_{(s_1,a,s_2),U}$ with $(s_1,a,s_2) \in T$. In particular, $U,V \in \ldet{\Ds}$. Hence, it suffices to show that $UaV$ is left deterministic. This is because $U \subseteq \eta\inv(s_1)$ since $\Ub_{s_1}$ is tight and $s_1\eta(a) \Rords s_1$ which implies that $UaA^* \cap U = \emptyset$. It remains to prove that $(ts,q\rho(K)) \in S$ for every $K \in \Kb$. This is immediate since $K = UaV$, $s = s_1\eta(a)s_2$ and $(ts_1\rho(a)s_2,q\rho(UaV)) \in S$ by definition of $\Vb_{(s_1,a,s_2),U}$. This concludes the proof of Lemma~\ref{lem:detcov}.	
\end{proof}

We are ready to prove Theorem~\ref{thm:detlrc}. The argument is standard: we merely combine Proposition~\ref{prop:detsound} and Proposition~\ref{prop:detcomp}.

\begin{proof}[Proof of Theorem~\ref{thm:detlrc}]
	Let \Cs be a finite \vari and \Ds a \vari which satisfies the inclusions $\Cs \subseteq \Ds \subseteq \upol{\Cs}$. Let $\rho: 2^{A^*} \to R$ be a \mratm and $P= \popti{\Ds}{\etac}{\rho}$. By symmetry, we only prove the first assertion: \pldetoptid is the least $(\ldeto,P)$-saturated subset of $\canc \times R$ for \etac and $\rho$. By Proposition~\ref{prop:detsound}, \pldetoptid is $(\ldeto,P)$-saturated for $\etac$ and $\rho$. It remains to show that it is the least such set. Hence, we let $S\subseteq \canc \times R$ which is $(\ldeto,P)$-saturated for \etac and $\rho$. We show that $\pldetoptid \subseteq S$. Let $(s,r) \in \pldetoptid$. Since $\etac$ is a \Cs-morphism, Proposition~\ref{prop:detcomp} yields an \ldet{\Ds}-cover $\Kb_s$ of $\eta\inv(s)$ such that $(s,\rho(K)) \in S$ for every $K \in \Kb_s$. Since $(s,r) \in \pldetoptid$, and $\Kb_s$ is a \ldet{\Ds}cover of $\eta\inv(s)$ we know that there exists $K \in \Kb_s$ such that $r \leq \rho(K)$. Hence, closure under downset for $S$ yields $(s,r) \in S$ as desired.
\end{proof}

%% file: covm.tex
We now consider covering for the classes built with mixed polynomial closure. In this case as well, we prove that if \Cs is a \emph{finite} \vari and \Ds is a \vari \emph{with decidable covering} such that $\Cs \subseteq \Ds \subseteq \upol{\Cs}$, then covering is decidable for \mdet{\Ds}. Using induction, this can be lifted to \emph{all} classes built from \Ds by applying \mdeto recursively. In particular, we use this result to show that covering is decidable for all levels $\fodb{n}(<)$ in the quantifier alternation hierarchy of \fodw (the link with \mdeto is established with Theorem~\ref{thm:mhiera}).

In this case as well, we rely on the framework of Section~\ref{sec:ratms}: we present an effective characterizations of \mdet{\Ds}-optimal \imprints. More precisely, given a \mratm $\rho: 2^{A^*} \to R$, we characterize the set $\pmdetoptid\subseteq \canc \times R$. The characterization is quite involved. In particular, it depends on \emph{three} auxiliary sets \popti{\Ds}{\etac}{\rho} (which can be computed when \Ds-covering is decidable by Proposition~\ref{prop:covreduc}) and the two sets \pldetoptid and \prdetoptid (which can also be computed if \Ds-covering is decidable by Theorem~\ref{thm:detlrc}).

\begin{remark}
	The characterization of \mdet{\Ds}-optimal \imprints is more involved than most of the typical results of this kind. Roughly, it directly describes the image under $\rho$ of the languages inside an optimal \mdet{\Ds}-cover. Intuitively, this can be explained by the discussion following Lemma~\ref{lem:dleast}: contrary to most of the operators that are typically considered, there exists no definition of \mdeto describing \mdet{\Ds} as the least class containing \Ds and closed under a list of operations involving  concatenation and union. 
\end{remark}

%We first state the characterization and discuss its consequences. The remainder of the section is then devoted to its proof.

\subsection{Statement}

We first present the property characterizing \mdet{\Ds}-optimal \imprints. We fix a morphism $\eta: A^* \to N$ into a finite monoid and a \mratm \mbox{$\rho: 2^{A^*}\to R$} for the definition. Moreover, we consider three subsets $P,P_1,P_2 \subseteq N \times R$ (in the characterization, they are \popti{\Ds}{\etac}{\rho}, \pldetoptid and \prdetoptid respectively). We define the $(\mdeto,P_1,P,P_2)$-saturated subsets of $N \times R$ for $\eta$ and $\rho$. First, we say that a pair $(s,r) \in N \times R$ is a \emph{$(P_1,P,P_2)$-block} when there exist $(s_1,r_1),(e_1,f_1) \in P_1$, $(s_3,r_3) \in P$ and $(s_2,r_2),(e_2,f_2) \in P_2$ such that $(e_1,f_1),(e_2,f_2)$ are pairs of multiplicative idempotents, $e_1 \Jrel e_2 \Jrel s$, $s = s_1 e_1 s_3 e_2 s_2$ and $r \leq r_1 f_1 r_3 f_2 r_2$. We may now define $(\mdeto,P_1,P,P_2)$-saturated sets. Consider a set $S \subseteq N \times R$. We say that $S$ is $(\mdeto,P_1,P,P_2)$-saturated for $\eta$ and $\rho$ when it is saturated for $\eta$ and $\rho$, and satisfies the following additional property:
\begin{equation} \label{eq:mdetop}
	\begin{array}{c}
		\text{for every $n \in \nat$, if the pairs $(s_0,r_0), \dots,(s_n,r_n) \in N \times R$ are $(P_1,P,P_2)$-blocks}\\
		\text{and $(s'_1,r'_1), \dots,(s'_n,r'_n) \in P$ satisfy $s_{i-1}s'_i \Jrel s_{i-1}$ and $s'_is_i \Jrel s_i$ for $1 \leq i \leq n$,}\\
		\text{then $(s_0s'_1s_1 \cdots s'_ns_n,r_0r'_1r_1 \cdots r'_nr_n) \in S$.}
	\end{array}
\end{equation}
Note that in particular, \eqref{eq:mdetop} implies that $S$ contains all $(P_1,P,P_2)$-blocks (this is the special case $n = 0$). We may now state the characterization of \mdet{\Ds}-optimal \imprints.

\begin{theorem} \label{thm:mainmpol}
	Let \Cs be a finite \vari and \Ds a \vari such that $\Cs \subseteq \Ds \subseteq \upol{\Cs}$. Let $\rho: 2^{A^*} \to R$ be a \mratm. Let $P\!=\! \popti{\Ds}{\etac}{\rho}$, $P_1\!=\! \pldetoptid$ and $P_2\! =\! \prdetoptid$. Then, \pmdetoptid is the least $(\mdeto,P_1,P,P_2)$-saturated subset of $\canc \times R$ for \etac and $\rho$.
\end{theorem}

Theorem~\ref{thm:mainmpol} yields an algorithm which computes the set \pmdetoptid associated a \nice \mratm $\rho: 2^{A^*} \to R$ provided that we have the sets $P =  \popti{\Ds}{\etac}{\rho}$, $P_1 = \pldetoptid$ and $P_2 = \prdetoptid$ in hand. Indeed, the least $(\mdeto,P)$-saturated subset of $\canc\times R$ can be computed using a least fixpoint procedure. It starts from the set of trivial elements $(\etac(w),\rho(w)) \in\canc\times R$ and saturates it with the operations in the definition: downset, multiplication and~\eqref{eq:mdetop}. It is simple to verify that these three operations can be implemented. In particular, this is possible for~\eqref{eq:mdetop} as we have $P$, $P_1$ and $P_2$ in hand (the number $n \in \nat$ in~\eqref{eq:mdetop} can be bounded using a standard pumping argument). Once \pmdetoptid has been computed, it follows from Lemma~\ref{lem:pointgen} that the set $\opti{\mdet{\Ds}}{\rho} \subseteq R$ can be computed as well. By Proposition~\ref{prop:breduc}, being able to compute this set is enough to decide \mdet{\Ds}-covering. Thus, it follows that covering is decidable for \mdet{\Ds}-covering are if one may compute the set $P = \popti{\Ds}{\etac}{\rho}$, $P_1 = \pldetoptid$ and $P_2 = \prdetoptid$ from a \nice \mratm $\rho$. It follows from Proposition~\ref{prop:covreduc} that $P$ can be computed provided that \Ds-covering is decidable. Moreover, we already proved with Theorem~\ref{thm:detlrc} that $P_1$ and $P_2$ can also be computed in this case.  Altogether, we obtain the following corollary.

\begin{corollary} \label{cor:mainmpol}
	Let \Cs be a finite \vari and \Ds a \vari with decidable covering such that $\Cs \subseteq \Ds \subseteq \upol{\Cs}$. Then, \mdet{\Ds}-covering is decidable.
\end{corollary}

An immediate induction  implies that Corollary~\ref{cor:mainmpol} extends to all classes that can be built from \Ds by applying \mdeto recursively. In this context, a key application is the quantifier alternation hierarchy of two-variable first-order logic equipped with only the linear ordering (\fodw). It follows from Theorem~\ref{thm:level1} and Lemma~\ref{lem:gensig} that the first level (\emph{i.e.}, $\fodb{1}(<)$) is the class $\pt =\bpol{\stzer}$ of piecewise testable languages. Moreover, we proved in Theorem~\ref{thm:mhiera} that the quantifier alternation hierarchy can then be climbed with mixed polynomial closure: $\fodb{n+1}(<) = \mdet{\fodb{n}(<)}$ for every $n \in \nat$. Yet, the situation is slightly more complicated than what happened in Section~\ref{sec:cov} for the operators \ldeto and \rdeto. In this case, the class \mdet{\pt} is \emph{strictly larger} than \mdet{\at} where \at is the finite \vari of alphabet testable languages (see Remark~\ref{rem:upolat}). However, $\at \subseteq \pt \subseteq \upol{\pt}$ and we have $\upol{\pt}=\upol{\at}$ by Lemma~\ref{lem:upolat}. Moreover, it is well-known that \pt has decidable covering (see~\cite{pzcovering} for a proof). Altogether, we obtain the following result from Corollary~\ref{cor:mainmpol} and a simple induction.

\begin{corollary}\label{cor:fodhiera}
	For all $n \in \nat$, covering and separation are decidable for $\fodb{n}(<)$.
\end{corollary}

\begin{remark}
	There exists an alternate specialized proof of the decidability of covering for all levels $\fodb{n}(<)$ by Henriksson and Kufleitner~\cite{kkufo2}. 
\end{remark}

\begin{remark}
	Corollary~\ref{cor:fodhiera} can be lifted to the levels $\fodb{n}(\! <,+1)$ and $\fodb{n}(\! <,+1,MOD)$ in the hierarchies of \fodws and \fodwsm using independent techniques. It is known that $\fodb{n}(<)$, $\fodb{n}(<,+1)$ and $\fodb{n}(<,+1,MOD)$ are connected by another operator called ``enrichment'' or ``wreath product'' which is used to combine \emph{two} classes into a larger one. First, we have $\fodb{n}(<,+1) = \fodb{n}(<) \circ \su$ with \su as the class of ``suffix languages'' (the Boolean combinations of languages $A^*w$ with $w \in A^*$). A proof is available in~\cite{Lauser2014FormalLT}. Moreover, $\fodb{n}(<,+1,MOD) = \fodb{n}(<,+1) \circ \md$ (this is a standard property which holds for many fragments of first-order logic, see~\cite{prwmodulo} for example). Finally, it is known that the operators $\Cs \mapsto \Cs \circ \su$ and $\Cs \mapsto \Cs \circ \su \circ \md$ preserve the decidability of separation~\cite{PlaceZ20,prwmodulo}. Therefore, Corollary~\ref{cor:fodhiera} also implies that for every $n \in \nat$, separation is decidable for both $\fodb{n}(<,+1)$ and $\fodb{n}(<,+1,MOD)$.
\end{remark}
%
%Note that Corollary~\ref{cor:mainmpol} also applies to the levels \jdetn{\Cs} and \idetn{\Cs} within the determinsitic hierarchy of basis \Cs when \Cs is a finite \vari. %TODO

\subsection{Proof argument}

We now concentrate on the proof of Theorem~\ref{thm:mainmpol}. In this case as well the argument involves two independent directions corresponding respectively to soundness and completeness. We first handle the former.

\begin{proposition} \label{prop:msound}
	Let \Cs be a finite \vari and \Ds a \vari such that \mbox{$\Cs\! \subseteq\! \Ds\! \subseteq\! \upol{\Cs}$}. Let $\rho: 2^{A^*} \to R$ be a \mratm. Let $P= \popti{\Ds}{\etac}{\rho}$, $P_1= \pldetoptid$ and $P_2 = \prdetoptid$. Then, \pmdetoptid is $(\mdeto,P_1,P,P_2)$-saturated for \etac and $\rho$.
\end{proposition}

\begin{proof}
	Since \Ds is a \vari, Theorem~\ref{thm:detclos} implies that \mdet{\Ds} is a \vari as well. Hence, Lemma~\ref{lem:genprin} yields that \pmdetoptid is saturated for $\etac$ and $\rho$. It remains to prove that it satisfies~\eqref{eq:mdetop}. We use Lemma~\ref{lem:moopt} which yields a \Cs-compatible \mdet{\Ds}-morphism $\alpha: A^* \to M$ such that for every $w \in A^*$ and $r \leq \rho(\atyp{w})$, we have $(\eta(w),r) \in \popti{\Ds}{\eta}{\rho}$. Note that since $\alpha$ is \Cs-compatible, we have $\ctype{\cdot} \circ \alpha = \etac$ by definition.
	
	We start with a preliminary lemma concerning $(P_1,P,P_2)$-blocks. We say that a word $w \in A^*$ is \emph{good} if there exists an idempotent $g \in E(M)$ such that $\alpha(w) \Jord g$ and $\etac(w) \Jrel \ctype{g}$.
	
	\newcommand{\ldtype}[1]{\typ{#1}{\ldet{\Ds}}}
	\newcommand{\rdtype}[1]{\typ{#1}{\rdet{\Ds}}}
	
	\begin{lemma} \label{lem:msound}
		Let $(s,r) \in \canc \times R$ be a $(P_1,P,P_2)$-block. There exists a good word $w \in A^*$ such that $\etac(w) = s$ and $r \leq  \rho(\atyp{w})$.
	\end{lemma}
	
	\begin{proof}
		We write  $Q = {M}/{\caned}$,  $Q_1 = {M}/{\caneld}$ and $Q_2 = {M}/{\canerd}$. Lemma~\ref{lem:smult} implies that $\gamma = \dtype{\cdot} \circ \alpha: A^* \to Q$ is a \Ds-morphism, that $\gamma_1= \ldtype{\cdot} \circ \alpha: A^* \to Q_1$ is an \ldet{\Ds}-morphism and that $\gamma_2= \rdtype{\cdot} \circ \alpha: A^* \to Q_2$ is a \rdet{\Ds}-morphism. Moreover, one may verify that $\gamma$, $\gamma_1$ and $\gamma_2$ remain \Cs-compatible since $\Cs \subseteq \Ds$.
		
		By definition of $(P_1,P,P_2)$-blocks, we know that $s = s_1 e_1 s_3 e_2 s_2$ and $r \leq  r_1 f_1 r_3 f_2 r_2$ where $(s_1,r_1),(e_1,f_1) \in P_1$, $(s_2,r_2),(e_2,f_2) \in P_2$, $(s_3,r_3) \in P$, $(e_1,f_1),(e_2,f_2)$ are pairs of multiplicative idempotents and $e_1 \Jrel e_2 \Jrel s$.  We use these pairs to exhibit elements in $Q_1$, $Q_2$ and $Q$. First, since we have $(s_1,r_1),(e_1,f_1) \in P_1 = \pldetoptid$ and $\gamma_1: A^* \to Q_1$ is an \ldet{\Ds}-morphism, it follows from Lemma~\ref{lem:optmo} that there are $u_1,v_1 \in A^*$ which satisfy  $\etac(u_1) = s_1$, $\etac(v_1) = e_1$,  $r_1 \leq \rho(\ftyp{u_1}{\gamma_1})$ and $f_1 \leq \rho(\ftyp{v_1}{\gamma_1})$. Symmetrically, we have $(s_2,r_2),(e_2,f_2) \in P_2 = \prdetoptid$. Hence, since $\gamma_2: A^* \to Q_2$ is an \rdet{\Ds}-morphism, Lemma~\ref{lem:optmo} yields $u_2,v_2 \in A^*$ such that $\etac(u_2) = s_2$, $\etac(v_2) = e_2$, $r_2 \leq \rho(\ftyp{u_2}{\gamma_2})$ and $f_2 \leq \rho(\ftyp{v_2}{\gamma_2})$. Finally, since $(s_3,r_3) \in P = \popti{\Ds}{\etac}{\rho}$ and $\gamma: A^* \to Q$ is a \Ds-morphism, Lemma~\ref{lem:optmo} yields $u_3 \in A^*$  such that $\etac(u_3) = s_3$ and $r_3 \leq \rho(\gtyp{u_3})$. 
		
		Let $k = \omega(M)$ (by definition, $k$ is a multiple of $\omega(Q)$, $\omega(Q_1)$ and $\omega(Q_2)$). We define $w = u_1v_1^ku_3v_2^ku_2$. Clearly, $\etac(w) = s_1e_1^ks_3e_2^ks_2 = s_1e_1s_3e_2s_2 = s$. Let us now verify that $w$ is good. Let $g = \alpha(v_1^k) \in E(M)$. Since $v_1^k$ is a factor of $w$, we have $\alpha(w) \Jord g$. Finally, since $\etac(v_1^k) = e_1$ and $\alpha$ is \Cs-compatible, we have $\ctype{g} = e_1$. Thus, since $e_1 \Jrel s = \etac(w)$, we have $\etac(w) \Jrel \ctype{g}$. It remains to prove that $r \leq  \rho(\atyp{w})$. The argument is based on Lemma~\ref{lem:lradv}. We use it to prove the following inclusion for $m = |M|$:
		\begin{equation} \label{eq:hdet:sminc}
			\ftyp{u_1}{\gamma_1} (\ftyp{v_1}{\gamma_1})^{km} \gtyp{u_3} (\ftyp{v_2}{\gamma_2})^{km} 	\ftyp{u_2}{\gamma_2} \subseteq \atyp{w}.
		\end{equation}
		Recall that by definition, we have $r_1 \leq \rho(\ftyp{u_1}{\gamma_1})$, $f_1 \leq \rho(\ftyp{v_1}{\gamma_1})$, $r_2 \leq \rho(\ftyp{u_2}{\gamma_2})$, $f_2 \leq \rho(\ftyp{v_2}{\gamma_2})$ and $r_3 \leq \rho(\gtyp{u_3})$. Hence, it follows from~\eqref{eq:hdet:sminc} that $r_1f_1^{nk}f_1r_3f_2^{nk}r_2 \leq \rho(\atyp{w})$. Since $f_1,f_2 \in R$ are idempotents and $r = r_1f_1r_3f_2r_2$, this yields $r \leq  \rho(\atyp{w})$ as desired.
		
		\smallskip
		
		We now prove~\eqref{eq:hdet:sminc}. We fix a word $w' \in \ftyp{u_1}{\gamma_1}(\ftyp{v_1}{\gamma_1})^{kn} \gtyp{u_3}(\ftyp{v_2}{\gamma_2})^{kn} 	\ftyp{u_2}{\gamma_2}$ and show that $\alpha(w') =  \alpha(w)$. By definition, we have $w' = u'_1v'_1u'_3v'_2v'_2$ with $u'_1 \in \ftyp{u_1}{\gamma_1}$, $v'_1 \in (\ftyp{v_1}{\gamma_1})^{km}$, $u'_3 \in \gtyp{u_3}$, $v'_2 \in (\ftyp{v_2}{\gamma_2})^{km}$ and	$u'_2 \in \ftyp{u_2}{\gamma_2}$. Recall that $e_1 \Jrel e_2 \Jrel s = s_1 e_1 s_3 e_2 s_2$. Hence, it follows from  Lemma~\ref{lem:jlr} that $e_1 \Rrel e_1s_3e_2s_2$ and $e_2 \Lrel s_1e_1s_3e_2$. We have the following fact.
		
		\begin{fct} \label{fct:msound}
			We have $\alpha(u_3v^k_2u_2) \canerd \alpha(u'_3v'_2u'_2)$.
		\end{fct}
	
		\begin{proof}
			By definition of $\gamma_2$ this boils down to proving that $\gamma_2(u_3v_2^ku_2) = \gamma_2(u'_3v'_2u'_2)$. Moreover, since the definitions of $u'_2$ and $v'_2$ imply that $\gamma_2(u_2) = \gamma_2(u'_2)$ and $\gamma_2(v^k_2) = \gamma_2(v'_2)$, it suffices to show that $\gamma_2(u_3v^k_2) = \gamma_2(u'_3v^k_2)$. By definition of $k$, we know that $\gamma_2(v^k_2) \in E(Q_2)$. Moreover, $\gamma(u_3) = \gamma(u'_3)$ by definition of $u'_3$ and it follows that $\gamma_2(u_3) \caned \gamma_2(u'_3)$ by definition of $\gamma$. Finally, since $e_2 \Lrel s_1e_1s_3e_2$, $\etac(v_2) = e_2$ and $\etac(u_3) = s_3$, we know that $\gamma_2(u_3v^k_2) \canec \gamma_2(v^k_2)$. Altogether, since $\Cs \subseteq \Ds \subseteq \upol{\Cs}$ and $\gamma_2$ is an \rdet{\Ds}-morphism by definition, the second assertion in Lemma~\ref{lem:lradv} yields $\gamma_2(u_3v^k_2) = \gamma_2(u'_3v^k_2)$ as desired.
		\end{proof}
		
		We may now prove that $\alpha(w) = \alpha(w')$. This involves two steps: we prove independently that $\alpha(w) = \alpha(u_1v_1^ku'_3v'_2u'_2)$ and $\alpha(u_1v_1^ku'_3v'_2u'_2) = \alpha(w')$. Let us start with the former. By Fact~\ref{fct:msound}, we have $\alpha(u_3v^k_2u_2) \canerd \alpha(u'_3v'_2u'_2)$. Moreover, since $e_1 \Rrel e_1s_3e_2s_2$, we know that $\ctype{\alpha(v_1^k)}  \Rrel \ctype{\alpha(v_1^ku_3v^k_2u_2)}$. Also, $\alpha(v_1^k)$ is an idempotent of $M$. Finally,  since $\mdet{\Ds} \subseteq \ldet{\rdet{\Ds}}$, we know that $\alpha$ is an \ldet{\rdet{\Ds}}-morphism. Altogether, it follows from the first assertion in Lemma~\ref{lem:lradv} that $\alpha(v_1^ku_3v_2^ku_2) = \alpha(v_1^ku'_3v'_2u'_2)$. Hence, multiplying by $\alpha(u_1)$ on the right yields $\alpha(w) = \alpha(u_1v_1^ku'_3v'_2u'_2)$.
		
		It remains to show that $\alpha(u_1v_1^ku'_3v'_2u'_2) = \alpha(w')$. Recall that by definition, we have $v'_2 \in (\ftyp{v_2}{\gamma_2})^{km}$ for $m = |M|$. Hence, since $(\gamma_2(v_2))^k$ is an idempotent, it follows from a pumping argument that $v'_2$ admits a decomposition $v'_2 = xyz$ where $x,y,z \in (\ftyp{v_2}{\gamma_2})^k$ and $\alpha(yz) = \alpha(z)$. Let $g = (\alpha(y))^\omega \in E(M)$. Since $\etac(u_3) = s_3$, $\etac(v_2)= e_2$ and $\alpha$ is \Cs-compatible, we have $\ctype{\alpha(u'_3)} = s_3$, $\ctype{\alpha(x)} = e_2$ and $\ctype{g} = e_2$. Thus, as $e_2 \Lrel s_1e_1s_3e_2$, we get $\ctype{g} \Lrel \ctype{\alpha(u_1v_1^ku'_3x)g}$. Moreover, $\gamma_1(u_1v_1^k) = \gamma_1(u'_1v'_1)$ by definition which yields $\gamma_1(u_1v_1^ku'_3x) = \gamma_1(u'_1v'_1u'_3x)$. Hence, we get $\alpha(u_1v_1^ku'_3x)g \caneld \alpha(u'_1v'_1u'_3x)g$ by definition of $\gamma_1$. Finally, since $\mdet{\Ds} \subseteq \rdet{\ldet{\Ds}}$, $\alpha$ is an \rdet{\ldet{\Ds}}-morphism. Altogether,  the second assertion in Lemma~\ref{lem:lradv} yields $\alpha(u_1v_1^ku'_3x)g = \alpha(u'_1v'_1u'_3x)g$. We may now multiply by $\alpha(yu_2)$ on the right to get $\alpha(u_1v_1^ku'_3v'_2u'_2) = \alpha(u'_1v'_1u'_3v'_2u'_2)$. This exactly says that $\alpha(u_1v_1^ku'_3v'_2u'_2) = \alpha(w')$, completing the proof.
	\end{proof}
	
	We may now prove~\eqref{eq:mdetop}. We fix $n \in \nat$, $n+1$ $(P_1,P,P_2)$-blocks $(s_0,r_0), \dots,(s_n,r_n)$ and $(s'_1,r'_1),\dots,(s'_n,r'_n) \in P$ such that $s_{i-1}s'_i \Jrel s_{i-1}$ and $s'_is_i \Jrel s_i$ for $1 \leq i \leq n$. \mbox{Finally, we define} $(s,r) =  (s_0s'_1s_1 \cdots s'_ns_n,r_0r'_1r_1 \cdots r'_nr_n)$ and prove that $(s,r) \in \pmdetoptid$. By definition of $\alpha$, it suffices to exhibit $w \in A^*$ such that  $\etac(w) = s$ such that $r \leq \rho(\atyp{w})$.

	It follows from Lemma~\ref{lem:msound} that for every $i$ such that $0 \leq i \leq n$, there exists a good word $w_i \in A^*$ such that $\etac(w_i) = s_i$ and $r_i \leq \rho(\atyp{w_i})$. Moreover, let  $Q = {M}/{\caned}$ and $\gamma = \dtype{\cdot} \circ \alpha: A^* \to Q$ which is a \Ds-morphism by Lemma~\ref{lem:smult}. For $1 \leq i \leq n$, we have $(s'_i,r'_i) \in P = \popti{\Ds}{\etac}{\rho}$ by definition. Hence, Lemma~\ref{lem:optmo} yields $u_i \in A^*$ such that $\etac(u_i) = s'_i$ and $r'_i \leq \rho(\gtyp{u_i})$. We define $w = w_0u_1w_1 \cdots u_nw_n$. By definition $\etac(w) = s_0s'_1s_1 \cdots s'_ns_n = s$. It remains to show that $r \leq \rho(\atyp{w})$. The argument is based on the following inclusion:
	\begin{equation} \label{eq:hdet:minc}
		\atyp{w_0}\gtyp{u_1}\atyp{w_1} \ \cdots\  \gtyp{u_1}\atyp{w_n} \subseteq \atyp{w}.
	\end{equation}
	By definition, we have $r_i \leq  \rho(\atyp{w_i})$ for $0 \leq i \leq n$ and $r'_i \leq \rho(\gtyp{u_i})$ for $1 \leq i \leq n$. Hence, it is immediate from~\eqref{eq:hdet:minc} that $r = r_0r'_1r_1 \cdots r'_nr_n \leq \rho(\atyp{w})$ as desired. 
	
	We now concentrate on proving~\eqref{eq:hdet:minc}. Let $w' \in 			\atyp{w_0}\gtyp{u_1}\atyp{w_1} \ \cdots\  \gtyp{u_n}\atyp{w_n}$. We have to show that $\alpha(w') = \alpha(w)$. By definition, for $1 \leq i \leq n$, there exists $u'_i \in A^*$ such that $\gamma(u'_i) = \gamma(u_i)$ and $\alpha(w') = \alpha(w_0u'_1w_1 \cdots u'_nw_n)$. Moreover,  $\alpha(w) = \alpha(w_0u_1w_1 \cdots u_nw_n)$ by definition. Consequently, it now suffices to show that $\alpha(w_{i-1}u_iw_i) = \alpha(w_{i-1}u'_iw_i)$ for $1 \leq i \leq n$. This will imply that $\alpha(w) = \alpha(w')$ as desired. We fix $i$ and write $t_{i-1} = \alpha(w_{i-1})$, $t_i = \alpha(w_i)$, $p_i = \alpha(u_i)$ and $p'_i = \alpha(u'_i)$ for the proof. We have to show that  $t_{i-1}p_it_i =  t_{i-1}p'_it_i$. 
	
	\begin{lemma} \label{lem:hdet:soundfinal}
		There exist $x_i,y_i \in M$ such that $t_{i-1}p_ix_i = t_{i-1}$ and $y_ip_it_i = t_i$.
	\end{lemma}
	
	\begin{proof}
		By symmetry, we only prove the existence of $y_i \in M$ such that $y_ip_it_i = t_i$. By definition, $\etac(u_iw_i) = s'_is_i$ and $\alpha(u_iw_i) = p_it_i$. Moreover, we have $s'_is_i \Jrel s_i$ by hypothesis which means that $\etac(u_iw_i) \Jrel \etac(w_i)$. Since $\alpha$ is \Cs-compatible, this exactly says that $\ctype{p_{i}t_i} \Jrel \ctype{t_{i}}$. Moreover, $w_i$ is good by definition. This yields an idempotent $g \in E(M)$ such that $t_i \Jord g$ and $\ctype{t_i} \Jrel \ctype{g}$. The former yields $z,z' \in M$ such that $t_{i} = zgz'$. Moreover, since $\ctype{p_{i}t_i} \Jrel \ctype{t_{i}}$, we obtain  $\ctype{p_{i}t_i} \Jrel \ctype{g}$. Altogether, we get $\ctype{p_izgz'} \Jrel \ctype{g}$ which implies that $\ctype{p_izg} \Jrel \ctype{g}$. By Lemma~\ref{lem:jlr}, this yields $\ctype{p_izg} \Lrel \ctype{g}$. We get $z'' \in M$ such that $\ctype{z''p_izg} = \ctype{g}$. By definition, $\alpha$ is an \mdet{\Ds}-morphism and therefore a \upol{\Cs}-morphism as well since $\Ds \subseteq \upol{\Cs}$. Thus, it follows from Theorem~\ref{thm:cupol} that $g = gz''p_izg$. We obtain, $t_i = zgz' = zgz''p_izgz' = zgz''p_it_i$. Therefore, we have $y_ip_it_i = t_i$ for $y_i = zgz''$ which completes the proof.
	\end{proof}
	
	We now prove that $t_{i-1}p_it_i =  t_{i-1}p'_it_i$. Let $x_i,y_i \in M$ be as defined in Lemma~\ref{lem:hdet:soundfinal}. Recall that $p_i = \alpha(u_i)$ and $p'_i = \alpha(u'_i)$ where $\gamma(u_i) = \gamma(u'_i)$. In particular, since $\gamma = \dtype{\cdot} \circ \alpha$, it follows that $p_i \caned p'_i$. Hence, since $\alpha$ is an \mdet{\Ds}-morphism, Theorem~\ref{thm:cmdet} yields,
	\[
	(p_ix_i)^\omega p_i (y_ip_i)^\omega =  (p_ix_i)^\omega p'_i (y_ip_i)^\omega.
	\]
	We may now multiply by $t_{i-1}$ on the left and $t_i$ on the right. Since $t_{i-1}p_ix_i = t_{i-1}$ and $y_ip_it_i = t_i$ by Lemma~\ref{lem:hdet:soundfinal}, this yields $t_{i-1}p_it_i =  t_{i-1}p'_it_i$ as desired, concluding the proof.
\end{proof}

It remains to handle completeness in Theorem~\ref{thm:mainmpol}. 

\begin{proposition}\label{prop:hdet:mcov}
	Let \Cs be a finite \vari and \Ds a \vari such that \mbox{$\Cs\!\! \subseteq\!\! \Ds\!\! \subseteq\!\! \upol{\Cs}$}. Let $\eta: A^* \to N$ be a \Cs-morphism and $\rho: 2^{A^*} \to R$ a \mratm. Let $P= \popti{\Ds}{\eta}{\rho}$, $P_1= \popti{\ldet{\Ds}}{\eta}{\rho}$ and $P_2 = \popti{\rdet{\Ds}}{\eta}{\rho}$. If $S \subseteq N \times R$ is $(\mdeto,P_1,P,P_2)$-saturated for $\eta$ and $\rho$, then, for each $s \in N$, there exists an \mdet{\Ds}-cover $\Kb_s$ of $\eta\inv(s)$ such that $(s,\rho(K)) \in S$ for every $K \in \Kb_s$.
\end{proposition}

\begin{proof}
	We first use the sets $P,P_1$ and $P_2$ to construct a special \Ds-morphism $\alpha: A^*\to M$. All languages in \mdet{\Ds} that we build in the proof will be \eqmak-classes for some $k \in \nat$.
	
		\begin{fct} \label{fct:propm}
		There exists a \Ds-morphism $\alpha: A^* \to M$, a morphism $\delta : M \to N$ and $m \in \nat$ such that $\eta = \delta \circ \alpha$ and the three following properties hold:
		\begin{itemize}
			\item For every $w \in A^*$, we have $(\eta(w),\rho(\atyp{w})) \in P$.
			\item For every $w \in A^*$ and $k \geq m$, we have $(\eta(w),\rho(\lhtyp{w}{\alpha,k})) \in P_1$.
			\item For every $w \in A^*$ and $k \geq m$, we have $(\eta(w),\rho(\rhtyp{w}{\alpha,k})) \in P_2$.
		\end{itemize}
	\end{fct}
	\begin{proof}
		For every element $s \in N$, we let $\Hb_{1,s}$ be an optimal \ldet{\Ds}-cover of $\eta\inv(s)$ for $\rho$,  $\Hb_{2,s}$ be an optimal \rdet{\Ds}-cover of $\eta\inv(s)$ for $\rho$ and $\Hb_s$ be an optimal \Ds-cover of $\eta\inv(s)$ for $\rho$. Corollary~\ref{cor:opcar} yields two \Ds-morphisms $\alpha_1: A^*\to M_1$ and $\alpha_2: A^*\to M_2$ and $k_1,k_2 \in \nat$ such that for each $s \in N$, every $H\in\Hb_{1,s}$ is a union of \eqlp{\alpha_1,k_1}-classes and every $H\in\Hb_{2,s}$ is a union of \eqlp{\alpha_2,k_2}-classes. Finally, Lemma~\ref{lem:moopt} yields a \Cs-compatible \Ds-morphism $\alpha_3: A^* \to M_3$ such that $(\eta(w),\rho(\typ{w}{\alpha_3})) \in P$ for every $w \in A^*$. Let $Q = M_1 \times M_2 \times M$ be the monoid equipped with the componentwise multiplication and $\gamma: A^* \to Q$ the morphism defined by $\gamma(w) = (\alpha_1(w),\alpha_2(w),\alpha(w))$ for every $w \in A^*$. Finally, let $\alpha: A^* \to M$ be the surjective restriction of $\gamma$. Since \Ds is a \vari, one may verify that $\alpha$ remains a \Ds-morphism. Moreover, one may also verify that $\alpha$ is \Cs-compatible since this was the case for $\alpha_3$. As $\eta$ is a \Cs-morphism, this yields a morphism $\delta : M \to N$ such that $\eta = \delta \circ \alpha$ by Lemma~\ref{lem:cmdiv}. Finally, we let $m = max(k_1,k_2)$. It remains to prove the three assertions.
		
		First, if $w \in A^*$, it is immediate by definition that $\atyp{w}\subseteq\typ{w}{\alpha_3}$. Thus, since $(\eta(w),\rho(\typ{w}{\alpha_3})) \in P$ by hypothesis and $P =\popti{\Ds}{\eta}{\rho}$ is closed under downset, we get $(\eta(w),\rho(\atyp{w})) \in P$. We turn to the last two assertions. By symmetry, we only prove the second one. Let $w \in A^*$ and $k \geq m$. We show that $(\eta(w),\rho(\lhtyp{w}{\alpha,k})) \in P_1$. Let $s = \eta(w)$. By construction $\Hb_{1,s}$ is a cover of $\eta\inv(s)$ which yields $H \in \Hb_{s,1}$ such that $w \in H$. Moreover, since  $\Hb_{1,s}$ is an optimal \ldet{\Ds}-cover of $\eta\inv(s)$, we know that $(\eta(w),\rho(H)) \in \popti{\ldet{\Ds}}{\eta}{\rho} = P_1$. Moreover, $H$ is a union of  \eqlp{\alpha_1,k_1} by definition which yields $\lhtyp{w_1}{\alpha_1,k_1} \subseteq H$. Finally, we have $k \geq m \geq k_1$ by hypothesis and one may verify from the definition of $\alpha$ that \eqlp{\alpha,k} is finer than \eqlp{\alpha_1,k_1}. Thus, $\lhtyp{w_1}{\alpha,k} \subseteq \lhtyp{w_1}{\alpha_1,k_1} \subseteq H$ and closure under downset  now implies that $(\eta(w),\rho(\lhtyp{w}{\alpha,k})) \in P_1$ as desired.
	\end{proof}
	
	We fix the \Ds-morphism $\alpha: A^* \to M$ described in Fact~\ref{fct:propm} for the remainder of the proof. The argument is now based on the following key lemma.
		
	\begin{lemma} \label{lem:mcov}
		There exists $k \in \nat$ such that $(\eta(w),\rho(\mhtyp{w}{\alpha,k})) \in S$ for all $w \in A^*$.
	\end{lemma}
	
	Before we prove Lemma~\ref{lem:mcov}, let us apply it to complete the main proof. Given $s \in N$, we exhibit an appropriate \mdet{\Ds}-cover $\Kb_s$ of $\eta\inv(s)$. We let $\Kb_s = \{\mhtyp{w}{\alpha,k} \mid w \in \eta\inv(s)\}$ where $k \in \nat$ is the number given by Lemma~\ref{lem:mcov}. Proposition~\ref{prop:opcar} implies that $\Kb_s$ is an \mdet{\Ds}-cover of $\eta\inv(s)$. Finally, Lemma~\ref{lem:mcov} yields $(s,\rho(K)) \in S$ for every $K \in \Kb_s$.
	
	\medskip
	
	We now concentrate on proving Lemma~\ref{lem:mcov}. Let us start with preliminary terminology that we shall use to decompose arbitrary words in $A^*$. Let $p\in\nat$. A \emph{$p$-iteration} is a word $u \in A^*$ which admits a decomposition $u = xu_1 \cdots u_py$ with $x,y,u_1,\dots,u_p \in A^*$ such that $\eta(u_i)\Jrel\eta(u)$ for every $i\leq p$.  We have the following key lemma concerning $p$-iterations.
	
		\begin{lemma} \label{lem:iterdec}
		There exist $p,h \in \nat$ such that for all $p$-iterations $u \in A^*$,  the pair $(\eta(u),\rho(\mhtyp{u}{\alpha,h}))$ is a $(P_1,P,P_2)$-block.
	\end{lemma}
	
	\begin{proof}
		We use induction to prove a slightly more general property. By Lemma~\ref{lem:genprin}, the sets $P_1 =\popti{\ldet{\Ds}}{\eta}{\rho}$ and $P_2 =\popti{\rdet{\Ds}}{\eta}{\rho}$ are sub-monoids of $N \times R$ for the componentwise multiplication. For each $i \in \{1,2\}$, if $(s,r) \in P_i$, we define the \emph{\Jrel-depth of $(s,r)$} as the number of pairs $(t,q) \in P_i$ such that $(t,q) \Jords (s,r)$ (note that here, we are considering the Green relation \Jrel of the monoid $P_i$).
		
		Consider $(s_1,r_1) \in P_1$, $(s_2,r_2) \in P_2$ of \Jrel-depths $d_1$ and $d_2$, and $t \in N$ such that $t \Jrel s_1 t s_2$. We use induction on $d_1$ and $d_2$ (in any order) to prove that if $p \geq d_1+d_2$ and $h \geq d_1+d_2+m$, then for every $p$-iteration $u \in \eta\inv(t)$, the pair $(s_1\eta(u)s_2 ,r_1\rho(\mhtyp{u}{\alpha,h}))r_2$ is a $(P_1,P,P_2)$-block. Clearly, the lemma follows from the special case when $(s_1,r_1) = (s_1,r_1) = (1_M,1_R)$ (which is an element of $P_1$ and $P_2$ by Lemma~\ref{lem:genprin}). There are two cases.
		
		First, assume that there exist $(t_1,q_1) \in P_1$ and such that $t \Jrel t_1$ and $(s_1t_1,r_1q_1) \Jrel (s_1,r_1)$, and $(t_2,q_2) \in P_2$  such that $t \Jrel t_2$ $(t_2s_1,q_2r_2) \Jrel (s_2,r_2)$. We prove that $(s_1\eta(u)s_2,r_1\rho((\mhtyp{u}{\alpha,h})r_2)$ is a $(P_1,P,P_2)$-block directly. Lemma~\ref{lem:jlr} yields $(s_1t_1,r_1q_1) \Rrel (s_1,r_1)$. We get $(t'_1,q'_1) \in P_1$ such that $(s_1,r_1) = (s_1t_1t'_1,s_1r_1r'_1)$. Let $(e_1,f_1) = ((t_1t'_1))^\omega, (r_1r'_1)^\omega) \in P_1$. By definition, $(s_1,r_1) = (s_1e_1,s_1f_1)$. Moreover, since $\eta(u) = t \Jrel t_1$ and $t \Jrel s_1ts_2$, we have $s_1e_1\eta(u)e_2s_2 \Jrel e_1$. A symmetrical 	argument yields a pair of multiplicative idempotents $(e_2,f_2)\in P_2$ such that $(s_2,r_2) = (e_2s_2,f_2r_2)$ and $s_1e_1\eta(u)e_2s_2 \Jrel e_2$. Finally, Fact~\ref{fct:propm} yields $(\eta(u),\rho(\atyp{u})) \in P$. Moreover, $\mhtyp{u}{\alpha,h} \subseteq \atyp{u}$ by definition and since $P = \popti{\Ds}{\eta}{\rho}$ is closed under downset by Lemma~\ref{lem:genprin}, we get $(\eta(u),\rho(\mhtyp{u}{\alpha,h})) \in P$. Hence, since we have $e_1 \Jrel e_2 \Jrel s_1e_1\eta(u)e_2s_2$, it follows that $(s_1e_1\eta(u)e_2s_2,r_1f_1\rho(\mhtyp{u}{\alpha,h})f_2r_2)$ is a $(P_1,P,P_2)$-block. By hypothesis on $(e_1,f_1)$ and $(e_2,f_2)$, it follows that $(s_1\eta(u)s_2,r_1\rho(\mhtyp{u}{\alpha,h})r_2)$ is a $(P_1,P,P_2)$-block as desired.
		
		We turn to the inductive case. We assume that either $(s_1t_1,r_1q_1) \Jords (s_1,r_1)$ for every  $(t_1,q_1) \in P_1$ such that $t \Jrel t_1$, or $(t_2s_1,q_2r_2) \Jords (s_2,r_2)$ for every $(t_2,q_2) \in P_2$  such that $t \Jrel t_2$. We only treat the case when $(s_1t_1,r_1q_1) \Jords (s_1,r_1)$ for every $(t_1,q_1)\in P_1$ such that $t \Jrel t_1$ (the converse case is symmetrical). Since $u$ is a $p$-iteration, one may verify that $u$ admits a decomposition $u = vau'$ where $u'$ is a $(p-1)$-iteration, $\alpha(u') \Jrel \alpha(u)$ and $\eta(u) \Jrel \eta(va) \Jords \eta(v)$ (in other words, $va$ is the least prefix of $u$ such that $\eta(u) \Jrel \eta(va)$). Let $i \in \posc{u}$ be the position carrying the highlighted letter `$a$' in $u = vau'$. Since $\eta(va) \Jords \eta(v)$, we have $\eta(va) \Rords \eta(v)$ by Lemma~\ref{lem:jlr} which yields $\alpha(va) \Rords \alpha(v)$ by Fact~\ref{fct:propm}. Hence, $i\in\poslp{\alpha}{1}{u}$ by definition and one may verify from the definition of \eqmep{\alpha,h} that,
		\begin{equation} \label{eq:theblock}
		\mhtyp{u}{\alpha,h} \subseteq \lhtyp{v}{\alpha,h}\ a\ \mhtyp{u'}{\alpha,h-1} \subseteq  \lhtyp{va}{\alpha,h}\ \mhtyp{u'}{\alpha,h-1}.
		\end{equation}
		Let $(s'_1,r'_1) = (s_1\eta(va),r_1\rho(\lhtyp{va}{\alpha,h}))$. We have $h \geq m$, which yields $(\eta(va),\rho(\lhtyp{va}{\alpha,h})) \in P_1$ by the second assertion in Fact~\ref{fct:propm}. Hence, our hypothesis yields $(s'_1,r'_1) \Jords (s_1,r_1)$ which implies that the \Jrel-depth $d'_1$ of $(s'_1,r'_1)$ is strictly smaller than the \Jrel-depth $d_1$ of $(s_1,r_1)$. by definition, it follows that $p-1 \geq d_1+d_2-1 \geq d'_1+d_2$ and $h-1 \geq d_1+d_2+m-1 \geq d'_1+d_2+m$. Consequently, since $u'$ is a $(p-1)$-iteration, induction on the \Jrel-depth of $(s_1,r_1)$ yields that $(s'_1\eta(u')s_2 ,r'_1\rho(\mhtyp{u'}{\alpha,h-1}))r_2)$ is a $(P_1,P,P_2)$-block. By definition of $(s'_1,r'_1)$, this exactly says that $(s_1\eta(u)s_2 ,r_1\rho(r_1\rho(\lhtyp{va}{\alpha,h}\ \mhtyp{u'}{\alpha,h-1}))r_2)$ is a $(P_1,P,P_2)$-block. In view of~\eqref{eq:theblock} and since the set of $(P_1,P,P_2)$-blocks is closed under downset by definition, it follows that $(s_1\eta(u)s_2,r_1\rho(\mhtyp{u}{\alpha,h})r_2)$ is a $(P_1,P,P_2)$-block as desired.
	\end{proof}
	
	Unfortunately, given a fixed $p \in \nat$, not all words are $p$-iterations. We deal with arbitrary words using the following notion. Let $p,\ell \in \nat$ and $w \in A^*$. A \emph{$p$-decomposition of length $\ell$} for $w$ is a decomposition $w = w_0a_1w_1  \cdots a_{\ell} w_\ell$ where $a_1,\dots,a_{\ell} \in A$, every factor $w_i \in A^*$ for $0 \leq i \leq \ell$ is a $(p+1)$-iteration, $\eta(w_{i-1}a_{i}) \Rords \eta(w_{i-1})$ and  $\eta(w_{i-1}a_{i}w_{i}) \Lords \eta(w_{i})$ for $1 \leq i \leq \ell$. The proof of Lemma~\ref{lem:mcov} is not based on the two following statements.
	
	\begin{lemma} \label{lem:pdec}
		Let $p \in \nat$. Each $w \in A^*$ admits a $p$-decomposition of length  $\ell \leq (p+1)^{|N|}-1$.
	\end{lemma}
	
	\begin{proof}
		For every $w \in A^*$, we define $d(w) \in \nat$ as the number of elements $s \in N$ such that $\eta(w) \Jords s$. Clearly, $d(w) \leq |N|$ for every $w \in A^*$. Hence, it suffices to prove that every $w \in A^*$ admits a  $p$-decomposition of length at most $(p+2)^{d(w)}-1$. We proceed by induction on $d(w)$. If $d(w) = 0$, then $\eta(w) \Jrel 1_N$ and $w = \veps \veps^{p+1} w$ is a $(p+1)$-iteration. In particular, $w$ admits a  $p$-decomposition of length $0 = (p+1)^{0}-1$ which concludes this case. Assume now that $d(w) \geq 1$. In that case, $\eta(w) \Jords 1_N$. This yields $n \geq 1$, $u_0,\dots,u_{n} \in A^*$ and $b_1,\dots,b_n\in A$ such that $w=u_0b_1u_1\cdots b_n u_n$ and for all $i \leq n$, we have $\eta(w) \Jrel \eta(u_{i-1}b_i) \Jords \eta(u_{i-1})$ and $\eta(w) \Jords \eta(u_n)$. We consider two independent cases. First, assume that $n \geq p+1$. In that case, since $\eta(u_{i-1}b_i) \Jrel \eta(w)$ for all $i \leq n$, it is clear that $w$ is a $(p+1)$-iteration. In particular, $w$ admits a  $p$-decomposition of length $0 \leq  (p+1)^{d(w)} -1$ and we are finished.  Conversely, assume that $n < p+1$. Since $\eta(w) \Jords \eta(u_{i})$ for every $i \leq \ell$, we have $d(u_i) \leq d(w)-1$ by definition. Hence, induction yields that each word $u_i$ admits a  $p$-decomposition of length at most $(p+1)^{d(w)-1}-1$. We may now replace each factor $u_i$ in $w=u_0b_1u_1\cdots b_n u_n$ by its  $p$-decomposition to obtain a new decomposition $w = v_0c_1v_1  \cdots c_{\ell} v_\ell$ where each factor $v_i$ for $i \leq \ell$ is a $(p+1)$-iteration, $\eta(v_{i-1}c_{i}) \Rords \eta(v_{i-1})$ for $1 \leq i \leq \ell$ and $\ell \leq (p+1)^{d(w)-1} -1 + p \times (p+1)^{d(w)-1}  =  (p+1)^{d(w)}-1$. However, it may happen that  $\eta(v_{i-1}c_{i}v_{i}) \Lrel \eta(v_{i})$ for some $i$. Yet, it is immediate that in this case $v_{i-1}c_{i}v_{i}$ is a $(p+1)$-iteration and $v_{i-1}c_{i}v_{i}c_{i+1} \Rords v_{i-1}c_{i}v_{i}$. Hence, we may reduce  the decomposition by making $v_{i-1}c_{i}v_{i}$ a single factor. Doing so recursively eventually yields the desired $p$-decomposition of length at most $(p+1)^{d(w)}-1$ for $w$.
	\end{proof}

	We are ready to prove  Lemma~\ref{lem:mcov}. Let $p,h \in \nat$ be the numbers defined in Lemma~\ref{lem:iterdec}. We now use induction on $\ell$ to prove that for every $\ell \in \nat$, if $k \geq h + \ell$ and $w \in A^*$ admitting a $p$-decomposition of length $\ell$, then $(\eta(w),\rho(\mhtyp{w}{\alpha,k})) \in S$. By Lemma~\ref{lem:pdec}, it will then follow that Lemma~\ref{lem:mcov} holds for $k = h + (p+2)^{|N|}-1$. We now fix $\ell$ and $k \geq h + \ell$. Let $w \in A^*$ admitting a $p$-decomposition $w = w_0a_1w_1  \cdots a_{\ell} w_\ell$ of length $\ell$. There are two cases. 
	
	First, assume that $\eta(a_{g}w_{g}) \Lrel \eta(w_{g})$ for all $g$ such that $1 \leq g \leq \ell$. This is the base case: we use~\eqref{eq:mdetop} to prove that $(\eta(w),\rho(\mhtyp{w}{\alpha,k})) \in S$ directly. Consider an index $g$ such that $1 \leq g \leq \ell$. By definition of $p$-decompositions, we have $\eta(w_{g-1}a_{g}w_{g}) \Lords \eta(w_{g})$ and our hypothesis states that $\eta(a_{g}w_{g})\Lrel\eta(w_{g})$. Hence, there exists a decomposition $w_{g-1}=u_{g-1}b_gv_g$ of $w_{g-1}$ with $u_{g-1},v_g\in A^*$ such that $\eta(b_gv_ga_gw_g) \Lords \eta(v_ga_gw_g) \Lrel \eta(w_g)$ (\emph{i.e.}, $v_ga_gw_g$ is the greatest suffix of $w_{g-1}a_{g}w_{g}$ whose image under $\eta$ is \Lrel-equivalent to $\eta(w_g)$). Since $w_{g-1}$ is a $(p+1)$-iteration (this is by definition of $p$-decompositions), one may verify that $u_{g-1}$ is a $p$-iteration and $\eta(u_{g-1}) \Rrel \eta(w_{g-1})$. We write $u'_0 = u_0b_1$, $u'_g = a_gu_gb_{g+1}$ for $1 \leq g \leq \ell-1$ and $u'_\ell = a_\ell u_\ell$. We have the following fact.
	
	\begin{fct} \label{fct:mproof}
		For all $g$ such that $0 \leq g \leq \ell$, the pair $(\eta(u'_g),\rho(\mhtyp{u'_g}{\alpha,h}))$ is a $(P_1,P,P_2)$-block. Moreover, for all $g$ such that $1 \leq g \leq \ell$, we have $\eta(u'_{g-1}v_g) \Jrel \eta(u'_{g-1})$ and $\eta(v_gu'_g) \Jrel \eta(u'_{g})$.
	\end{fct}

	\begin{proof}
		We first fix $g$ such that $0 \leq g \leq \ell$ and prove that $u'_g$ is a $p$-iteration: since $h$ and $p$ are the numbers given by Lemma~\ref{lem:iterdec}, this implies as desired that $(\eta(u'_g),\rho(\mhtyp{u'_g}{\alpha,h}))$ is a $(P_1,P,P_2)$-block. We show that $\eta(u_g) \Jrel \eta(u'_g)$. Since $u_g$ is a $p$-iteration and an infix of $u_g$, this implies as desired that $u'_g$ is a $p$-iteration as well. We only detail the case when $1 \leq g \leq \ell-1$ (the cases $g = 0$ and $g = \ell$ are similar). By definition, $u'_g = a_gu_gb_{g+1}$ and $\eta(v_ga_gw_g) \Lrel \eta(w_g)$ and since $u'_g$ is an infix of $v_ga_gw_g$, this yields $\eta(w_g) \Jord \eta(u'_g)$. Since we also know that $\eta(u_{g}) \Rrel \eta(w_{g})$ (by definition of $u_g$), this yields $\eta(u_g) \Jord \eta(u'_g)$ and since the converse inequality is trivial, we get $\eta(u_g) \Jrel \eta(u'_g)$.
		
		We now fix $g$ such that $1 \leq g \leq \ell$. By definition $\eta(v_ga_gw_g) \Lrel \eta(w_g)$ which implies that $\eta(v_ga_gu_g) \Lrel \eta(a_gu_g)$ since $\eta(w_g) \Rrel \eta(u_g)$. By definition of $u'_g$, this yields $\eta(v_gu'_g) \Lrel \eta(u'_g)$. Moreover, $w_{g-1}=u_{g-1}b_gv_g$ and $\eta(u_{g-1}) \Rrel \eta(w_{g-1})$ which means that  $\eta(u_{g-1}) \Rrel \eta(u_{g-1}b_gv_g)$. By definition of $u'_{g-1}$, this yields  $\eta(u'_{g-1}v_g) \Rrel \eta(u'_{g-1})$, concluding the proof.
	\end{proof}
	
	By definition, $w = u_0b_1v_1a_1u_1 \cdots b_\ell v_\ell a_\ell w_\ell = u'_0v_1u'_1 \cdots v_\ell u'_\ell$.  We write $i_1,\dots,i_n \in \pos{w}$ for the positions carrying the letters $a_1,\cdots,a_n$ and $j_1,\dots,j_n \in \pos{w}$ for the positions carrying the letters $b_1,\cdots,b_n$. By definition of $p$-decomposition, $\eta(w_{g-1}a_{g}) \Rords \eta(w_{g-1})$ for $1 \leq g \leq \ell$. This yields $\eta(u_{g-1}b_gv_ga_{g}) \Rords \eta(u_{g-1}b_gv_g)$ and by Fact~\ref{fct:propm}, this implies that $\alpha(u_{g-1}b_gv_ga_{g}) \Rords \alpha(u_{g-1}b_gv_g)$. Thus, $i_1,\dots,i_n \in \poslp{\alpha}{\ell}{w}$. Conversely, we know that $\eta(b_gv_ga_gw_g) \Lords \eta(v_ga_gw_g)$ and $u_{g} \Rrel w_{g}$  for $1 \leq g \leq \ell$ by definition. Hence, one may then verify that  $\eta(b_gv_ga_gu_g) \Lords \eta(v_ga_gu_g)$ and Fact~\ref{fct:propm} yields $\alpha(b_gv_ga_{g}u_g) \Lords \alpha(v_ga_gu_g)$ for  $1 \leq g \leq \ell$. Thus, $j_1,\dots,j_n \in \posrp{\alpha}{\ell}{w}$ by definition. Since $k \geq h + \ell$, one may now verify that $\mhtyp{w}{\alpha,k} \subseteq \mhtyp{u_0}{\alpha,h}\ b_1\ \atyp{v_1}\ a_1\ \mhtyp{u_1}{\alpha,h}\ \cdots \ b_n\ \atyp{v_n}\ a_n\  \mhtyp{u_\ell}{\alpha,h}$ by definition of \eqmp{\alpha,k}. Since \eqmp{\alpha,k} is a congruence, this yields,
	\begin{equation} \label{eq:mcase1}
			\mhtyp{w}{\alpha,k} \subseteq \mhtyp{u'_0}{\alpha,h}\ \atyp{v_1}\ \mhtyp{u'_1}{\alpha,h}\ \cdots \ \atyp{v_n}\ \mhtyp{u'_\ell}{\alpha,h}.
	\end{equation}
	By Fact~\ref{fct:mproof}, $(\eta(u'_g),\rho(\mhtyp{u'_g}{\alpha,h}))$ is a $(P_1,P,P_2)$-block for $0 \leq g \leq \ell$, and $\eta(u'_{g-1}v_g) \Jrel \eta(u'_{g-1})$ and $\eta(v_gu'_g) \Jrel \eta(u'_{g})$ for $1 \leq g \leq \ell$. Finally, Fact~\ref{fct:propm} yields $(\alpha(v_g),\rho{(\atyp{v_g})}) \in P$ for $1 \leq g \leq \ell$. Hence, \eqref{eq:mdetop} in the definition of $(\mdeto,P_1,P,P_2)$-saturated sets yields,
	\[
	(\eta(u'_0v_1u'_1 \cdots v_\ell u'_\ell), \rho(\mhtyp{u'_0}{\alpha,h}\ \atyp{v_1}\ \mhtyp{u'_1}{\alpha,h}\ \cdots \ \atyp{v_n}\ \mhtyp{u'_\ell}{\alpha,h})) \in S.
	\]
	It then follows from closure under downset and~\eqref{eq:mcase1} that $(\eta(w),\rho(\mhtyp{w}{\alpha,k})) \in S$ as desired.
	
	\medskip
	
	It remains to handle the converse case. We assume that there exists $g$ such that  $1 \leq g \leq \ell$ and $\eta(a_{g}w_{g}) \Lords \eta(w_{g})$. Let $i \in \pos{w}$ be the position carrying the letter $a_g$ in the decomposition $w = w_0a_1w_1  \cdots a_{\ell} w_\ell$. By definition of $p$-decompositions, we have $\eta(w_{q-1}a_{q}w_{q}) \Lords \eta(w_{q})$ for $1 \leq q \leq \ell$. Hence, since $\eta(a_{g}w_{g}) \Lords \eta(w_{g})$, one may verify that $i \in \posrp{\eta}{\ell -(g-1)}{w}$. Symmetrically, $\eta(w_{q-1}a_{q}) \Rords \eta(w_{q-1})$ for $1 \leq q \leq \ell$ which implies that $i \in \poslp{\eta}{g}{w}$. Let $w' = w_0a_1w_1  \cdots a_{g-1} w_{g-1}$ and $w'' = w_ga_{g+1}w_{g+1} \cdots a_\ell w_\ell$ (in particular, $w = w'a_gw''$). Since $i \in \poslp{\eta}{g}{w} \cap \posrp{\eta}{\ell -(g-1)}{w}$, one may now verify from the definition of \eqmp{\alpha,k} that,
	\begin{equation} \label{eq:mcase2}
		\mhtyp{w}{\alpha,k} \subseteq \mhtyp{w'}{\alpha,k-\ell+(g-1)}\ a_g\ \mhtyp{w''}{\alpha,k-g}.
	\end{equation}
	By definition $w'$ admits a $p$-decomposition of length $g-1 < \ell$. Moreover, since $k \geq h + \ell$, we have $k-\ell+(g-1) \geq h +(g-1)$. Hence, induction yields $(\eta(w'),\rho(\mhtyp{w'}{\alpha,k-\ell+(g-1)})) \in S$. Symmetrically, $w''$ admits a $p$-decomposition of length $\ell-g < \ell$. Moreover, since $k \geq h + \ell$, we have $k-g \geq h + (\ell-g)$. Hence, induction yields $(\eta(w''),\rho(\mhtyp{w''}{\alpha,k-g})) \in S$. Finally, we have $(\eta(a_g),\rho(a_g)) \in S$ since $S$ is saturated. Hence, since $w = w'a_gw''$, closure under multiplication yields $(\eta(w),\rho(\mhtyp{w'}{\alpha,k-\ell+(g-1)}\ a_g\ \mhtyp{w''}{\alpha,k-g})) \in S$. It then follows from~\eqref{eq:mcase2} and closure under downset that $(\eta(w),\rho(\mhtyp{w}{\alpha,k})) \in S$ which complete the proof of Lemma~\ref{lem:mcov}.
\end{proof}

We are ready to prove Theorem~\ref{thm:mainmpol}. This is now straightforward: we merely combine Proposition~\ref{prop:msound} and Proposition~\ref{prop:hdet:mcov}.

\begin{proof}[Proof of Theorem~\ref{thm:mainmpol}]
	Let \Cs be a finite \vari and \Ds a \vari such that we have the inclusions $\Cs \subseteq \Ds \subseteq \upol{\Cs}$. Let $\rho: 2^{A^*} \to R$ be a \mratm. We define $P= \popti{\Ds}{\etac}{\rho}$, $P_1= \pldetoptid$ and $P_2 = \prdetoptid$. We prove that \pmdetoptid is the least $(\mdeto,P_1,P,P_2)$-saturated subset of $\canc \times R$ for \etac and $\rho$. It is immediate from  Proposition~\ref{prop:msound} that \pmdetoptid is $(\mdeto,P_1,P,P_2)$-saturated for $\etac$ and $\rho$. It remains to show that it is the least such set. Let $S\subseteq \canc \times R$ which is $(\mdeto,P_1,P,P_2)$-saturated for \etac and $\rho$. We show that $\pmdetoptid \subseteq S$. Let $(s,r) \in \pmdetoptid$, \emph{i.e.}, $r \in \opti{\mdet{\Ds}}{\etac\inv(s)}{\rho}$. Proposition~\ref{prop:hdet:mcov} yields an \mdet{\Ds}-cover $\Kb$ of $\etac\inv(s)$ such that $(s,\rho(K)) \in S$ for every $K \in \Kb$. By definition, $r \in \prin{\rho}{\Kb}$ which yields $K \in \Kb$ such that $r \leq \rho(K)$. Since $(s,\rho(K)) \in S$ and $S$ is saturated, closure under downset yields $(s,r) \in S$ which completes the proof.
\end{proof}

%% file: conc.tex
We investigated the operators \ldeto, \rdeto and \mdeto, and the associated deterministic hierarchies. We proved that these three operators preserve the decidability of membership. Moreover, we used \mdeto to characterize the quantifier alternation hierarchies of the variants $\fod(<,\prefsigg)$ and $\fod(<,+1,\prefsigg)$ of \fod for a group \vari \Gs. They imply the decidability of membership for all levels when \emph{separation} is decidable for \Gs. Finally, we looked at separation and covering for our operators and used the results to show that all levels in the quantifier alternation hierarchy of \fodw have decidable separation. In particular, \mdeto is the linchpin upon which most of our results are based.

There are several follow-up questions. A first point concerns membership for the levels \jdetn{\Cs} of the hierarchies introduced in Section~\ref{sec:deth}. These are the only levels which we are not able to handle in a generic manner. Indeed, it follows from Theorems~\ref{thm:cmdet} and~\ref{thm:joincarn} that membership is decidable for all these levels as soon as this is the case for the first one: \jdet{\Cs}. Yet, we do not have a generic result for handling this initial level. Another question is whether our covering results for the levels $\fodbn(<)$ can be generalized to the variants $\fodbn(<,\prefsigg)$ and $\fodbn(<,+1,\prefsigg)$ for arbitrary group \varis \Gs. Such a result is proved in~\cite{pzconcagroup} for the first level: if \Gs has decidable separation, then so  $\fodb{1}(<,\prefsigg)$ has decidable covering (the proof considers \bpol{\Gs} which characterizes $\fodb{1}(<,\prefsigg)$ by Theorem~\ref{thm:level1}) Finally, one may also look at the other variants of \fod: the classes $\fod(\infsigc)$ for an \emph{arbitrary} \vari \Cs. Unfortunately, our results fail in the general case. An example is considered in~\cite{between}: \fod with ``between relations''. It is simple to verify from the definition that this class is exactly $\fod(\infsig{\at})$. The results of~\cite{between} imply that $\fod(\infsig{\at})$ is distinct from \upol{\bpol{\at}} which means that Corollary~\ref{cor:fod} fails in this case.